%% file: main.tex
\documentclass[12pt]{article}

\usepackage{mdframed}

\usepackage[most]{tcolorbox}

\usepackage{amsmath, amssymb, amsthm}
\usepackage{booktabs}
\usepackage{multirow}
\usepackage{graphicx}
\usepackage{float}
\usepackage{hyperref}
\usepackage[capitalize]{cleveref}
\usepackage{natbib}
\usepackage{xcolor}
\usepackage{geometry}
\usepackage{microtype}               
\usepackage{enumitem}                 
\usepackage{titlesec}                 

\definecolor{morandiGray}{RGB}{181, 181, 170}
\definecolor{morandiSage}{RGB}{163, 177, 138}
\definecolor{morandiBlue}{RGB}{165, 180, 196}
\definecolor{morandiRed}{RGB}{180, 120, 120}
\definecolor{morandiTaupe}{RGB}{192, 170, 145}
\definecolor{reasoningColor}{RGB}{70, 130, 180}

\titleformat{\section}
  {\Large\bfseries}{\thesection}{1em}{}
\titlespacing*{\section}{0pt}{2.5ex plus 1ex minus .2ex}{1.5ex plus .2ex}

\titleformat{\subsection}
  {\large\bfseries}{\thesubsection}{1em}{}
\titlespacing*{\subsection}{0pt}{2ex plus 1ex minus .2ex}{1ex plus .2ex}

\titleformat{\subsubsection}
  {\normalsize\bfseries}{\thesubsubsection}{1em}{}
\titlespacing*{\subsubsection}{0pt}{1.5ex plus .5ex minus .2ex}{0.8ex plus .2ex}

\setlist{itemsep=2pt, topsep=4pt, parsep=0pt}

\newmdenv[
    topline=false,
    bottomline=false,
    rightline=false,
    leftline=true,
    linewidth=2pt,
    linecolor=morandiGray,
    backgroundcolor=morandiGray!8,
    innerleftmargin=10pt,
    innerrightmargin=10pt,
    innertopmargin=8pt,
    innerbottommargin=8pt,
    skipabove=10pt,
    skipbelow=10pt
]{defenv}
\theoremstyle{definition}
\newtheorem{definition}{Definition}

\newtcolorbox{paperquote}[1][]{
    enhanced,
    colback=morandiBlue!12,
    colframe=morandiBlue!70,
    arc=2mm,
    boxrule=0.5pt,
    left=8pt,
    right=8pt,
    top=10pt,
    bottom=6pt,
    fonttitle=\bfseries\small,
    title={#1},
    coltitle=black,
    attach boxed title to top left={yshift=-3mm, xshift=4mm},
    boxed title style={colback=morandiBlue!50, arc=1mm, boxrule=0pt, size=small}
}

\title{Training Language Models for Bilateral Trade
\newline with Private Information}
\author{
    Dirk Bergemann\thanks{Department of Economics, Yale University. Email: dirk.bergemann@yale.edu} \quad 
    Soheil Ghili\thanks{Yale School of Management, Yale University. Email: soheil.ghili@yale.edu} \quad 
    Xinyang Hu\thanks{Department of Statistics and Data Science, Yale University. Email: \{xinyang.hu, zhuoran.yang\}@yale.edu} \quad \\
    Chuanhao Li\thanks{Department of Industrial Engineering, Tsinghua University. Email: chuanhao-li@mail.tsinghua.edu.cn} \quad 
    Zhuoran Yang\footnotemark[3]
}
\begin{document}
\maketitle

\begin{abstract}
Bilateral bargaining under incomplete information provides a controlled
testbed for evaluating large language model (LLM) agent capabilities. Bilateral trade  demands individual rationality, strategic surplus maximization, and
cooperation to realize gains from trade. We develop a structured
bargaining environment in which LLMs negotiate via tool calls within an
event-driven simulator, separating binding offers from natural-language
messages to enable automated evaluation. The environment serves two
purposes: as a benchmark for frontier models and as a training environment
for open-weight models via reinforcement learning.

In benchmark experiments, a round-robin tournament among five frontier
models (15,000 negotiations) reveals that effective strategies implement
price discrimination through sequential offers. Aggressive anchoring,
calibrated concession, and temporal patience are associated with both the
highest surplus share and the highest deal rate. Accommodating strategies
that concede quickly disable price discrimination in the buyer role,
yielding the lowest surplus capture and deal completion. Strategically
competent models scale their behavior proportionally to item value,
maintaining consistent performance across price tiers; weaker models
perform well only when wide zones of possible agreement compensate for
suboptimal strategies.

In training experiments, we fine-tune Qwen3 (8B, 14B) via supervised
fine-tuning (SFT) followed by Group Relative Policy Optimization (GRPO)
against a fixed frontier opponent. The two stages optimize competing
objectives: SFT approximately doubles surplus share but reduces deal
rates, while RL recovers deal rates but erodes surplus gains---a tension
traceable to the reward structure. SFT also compresses surplus variation
across price tiers, and this compression generalizes to opponents unseen
during training, suggesting that behavioral cloning instills proportional
strategies rather than memorized price points.
\end{abstract}
\noindent \textbf{JEL Classification:} D82, D83, D84. \\
\noindent \textbf{Keywords:} Bilateral Bargaining, Two-Sided Private Information, Agentic Bargaining.
\input{sections/sec1}
\input{sections/sec2}

\input{sections/sec3}
\input{sections/sec4}
\input{sections/sec5}

\input{sections/sec6}

\bibliographystyle{plainnat}  
\bibliography{references}

\input{sections/appendix}

\end{document}

%% file: sections/sec1.tex

\section{Introduction}
\label{sec:intro}

Bilateral bargaining under incomplete information is a foundational problem in
economics and operations research. A buyer and seller, each constrained by a
private reservation price, must discover through strategic interaction whether
mutually beneficial trade is possible---and if so, at what price. The formal
study of this problem spans decades, from Nash's axiomatic bargaining
solution~\citep{nash1950} and Rubinstein's alternating-offers
model~\citep{rubinstein1982} to the impossibility results of
\citet{myerson1983} and the bilateral trade formulation of
\citet{chatterjee1983}. Applications range from procurement and supply chain
contracting to dispute resolution and online marketplaces, where
alternating-offer bargaining accounts for a substantial share of
transaction volume \citep{backus2018sequential}.

Large language models (LLMs) are increasingly deployed as autonomous
agents in multi-turn, goal-directed settings---customer service, code
generation, and tool-augmented reasoning. Bilateral bargaining provides a
controlled testbed for evaluating whether these models can function as
rational, effective decision-making agents. An effective bargaining agent must
simultaneously (i)~respect hard numeric constraints across many rounds of
counterpart pressure, (ii)~maximize surplus through strategic offers and
communication, and (iii)~coordinate with a counterpart to realize gains from
trade when they exist. These three capabilities---individual rationality,
strategic effectiveness, and allocative efficiency---map directly to
requirements for any agentic LLM deployment that involves multi-turn
instruction fidelity, objective optimization, and coordination under
uncertainty.

Two limitations prevent direct use of general-purpose LLMs for this task.
First, frontier models exhibit systematic behavioral failures: they accept
deals yielding negative utility and fail to walk away from
scenarios with no-gains-from-trade. These
failures reflect a mismatch between pre-training objectives---which reward
fluency and agreeableness---and the requirements of strategic negotiation.
Second, free-form dialogue creates a measurement problem: it is unclear
whether a mentioned price constitutes a binding offer, a hypothetical, or a
market reference, making automated evaluation of utilities, concession rates,
and agreement status impossible at scale.

This paper addresses both limitations. We make three contributions:

\begin{enumerate}[leftmargin=*]

\item \textbf{A structured bilateral bargaining environment with a
three-dimensional evaluation framework (\Cref{sec:bargaining,sec:agent_system}).}
We construct an agent system in which LLMs negotiate through structured tool
calls (\texttt{make\_offer}, \texttt{respond\_to\_offer},
\texttt{send\_message}, etc.) within an event-driven simulation engine.
Separating formal offers from natural-language messages resolves the
measurement problem: every offer, acceptance, and rejection is a
machine-parseable event. We define three evaluation dimensions---individual
rationality, strategic effectiveness (surplus share),
and allocative efficiency (deal rate)---each with formal metrics that can be
computed automatically from the negotiation trace. A price-tier decomposition
disaggregates all metrics by item value, revealing whether models maintain
consistent performance across the full price range.

\item \textbf{A systematic benchmark of frontier LLMs on strategic
negotiation (\Cref{sec:benchmark}).}
We evaluate five frontier models in a round-robin tournament where every model
plays as both buyer and seller against every model, yielding 15,000 total
negotiations. Three empirical findings emerge.
(a)~Aggressive anchoring with calibrated concession is the most effective
observed strategy: the top-performing model achieves both the highest surplus
share and the highest deal rate by opening far above cost and then conceding
to close, contradicting the naive expectation of a surplus--efficiency
tradeoff.
(b)~Accommodating strategies are associated with weak buyer performance: a
cautious, high-concession profile produces both the lowest buyer surplus share
and the lowest buyer deal rate---cooperativeness without strategic backbone
hurts both surplus capture and deal completion.
(c)~Strategic competence is associated with consistency across price tiers:
strong models maintain narrow surplus spreads across value quintiles, while
weaker models improve only on high-value items where larger zones of possible
agreement buffer suboptimal strategies.

\item \textbf{A training pipeline for open-weight negotiation agents, with a
diagnosis of the SFT--RL tension (\Cref{sec:training}).}
We fine-tune Qwen3 models (8B and 14B parameters) via a two-stage pipeline:
supervised fine-tuning (SFT) on synthetic self-play trajectories, followed by
reinforcement learning via Group Relative Policy Optimization (GRPO) against a
fixed frontier opponent. 
The two stages optimize competing objectives: SFT teaches ``hold out for good
terms''; RL teaches ``close the deal.'' This tension, driven by a reward
structure that provides no signal to distinguish rational walk-aways from
irrational acceptances, identifies concrete design changes for future work:
explicit IR penalties, diverse training opponents, and reward curricula that
stage constraint compliance before surplus optimization.

\end{enumerate}

The remainder of the paper is organized as follows. \Cref{sec:related}
reviews related work. \Cref{sec:bargaining} formalizes the bargaining game,
evaluation dimensions, and agent system. \Cref{sec:benchmark} presents the
frontier model benchmark. \Cref{sec:training} describes the training pipeline
and results. \Cref{sec:conclusion} concludes with discussions of limitations and future work.

%% file: sections/sec2.tex

\section{Related Work}
\label{sec:related}

We organize the literature along four streams that correspond to the
contributions of this paper: the game-theoretic foundations of bilateral
trade, automated negotiation systems, LLM evaluation in negotiation and
agentic settings, and reinforcement learning for language models.

\subsection{Bilateral Trade and Negotiation Theory}
\label{sec:related_theory}

The bilateral bargaining problem studied here is rooted in a well-established
theoretical tradition. \citet{nash1950} introduces the axiomatic bargaining
solution, predicting equal surplus division under symmetric disagreement
points. \citet{rubinstein1982} models alternating-offers bargaining with
discounting, deriving the unique subgame-perfect equilibrium.
\citet{chatterjee1983} analyze bilateral trade under incomplete information
with simultaneous offers, characterizing the set of equilibria when
reservation prices are private. \citet{myerson1983} establish the
impossibility of achieving ex-post efficiency in bilateral trade under any
incentive-compatible, individually rational, budget-balanced mechanism---a
result that provides theoretical context for our empirical finding that even
the strongest agents leave some gains unrealized.

Large-scale empirical evidence on sequential bargaining has only recently
become available. \citet{backus2018sequential} study over 25~million
alternating-offer negotiations on eBay's Best Offer platform---the same
protocol our simulator implements---and document several patterns that
inform our evaluation framework: more patient buyers obtain lower prices,
experienced players capture more surplus, prices converge through Coasian
dynamics (seller offers decline, buyer offers increase), and bargaining
behavior differs systematically between expensive and inexpensive items
due to fixed costs of negotiation. Our price-tier decomposition
(\Cref{sec:benchmark_price_tier}) tests whether LLMs exhibit the same
value-dependent patterns, and our behavioral metrics (concession rate,
temporal patience) directly measure the mechanisms
\citeauthor{backus2018sequential} identify as determinants of bargaining
outcomes.

The behavioral negotiation literature documents systematic deviations from
game-theoretic predictions. \citet{galinsky2001} demonstrate the
first-offer anchoring effect: the opening offer shifts the counterpart's
reference point and predicts final settlement prices.
\citet{bazerman2000negotiation} survey cognitive biases in negotiation,
including anchoring, the fixed-pie assumption, and reactive devaluation.
\citet{thompson1990negotiation} show that negotiators routinely fail to
identify integrative potential, leaving surplus on the table even in
cooperative settings. These findings inform our evaluation framework:
initial aggressiveness, concession rate, and temporal patience
(\Cref{sec:metrics}) are designed to capture the behavioral mechanisms
that the psychology literature identifies as determinants of negotiation
outcomes.

\subsection{Automated Negotiation Systems}
\label{sec:related_automated}

Pre-LLM automated negotiation systems operate in structured, language-free
environments. \citet{jennings2001automated} survey game-theoretic, heuristic,
and argumentation-based approaches for bilateral agent negotiation.
The Automated Negotiating Agents Competition (ANAC) provides a standardized
platform for evaluating agent strategies in multi-issue bargaining with
predefined utility functions \citep{baarslag2011first,baarslag2015anac}.
\citet{chang2020multi} train actor--critic networks via self-play for
multi-issue bargaining, learning adaptive bidding and acceptance strategies.
\citet{green2022science} model eBay's Best Offer protocol as a Markov
decision process and train neural-network agents on historical transaction
data, achieving near-optimal buyer utility. Their agents outperform humans
by anchoring more aggressively and bargaining longer---a pattern consistent
with the behavioral drivers we measure in \Cref{sec:benchmark_results}.

These systems demonstrate that RL can produce effective bargaining strategies
when the action space is well-defined and rewards are directly observable. Our
work builds on this insight by preserving a structured action space
(\Cref{sec:action_space}) while adding natural-language communication as a
parallel channel---enabling both automated metric computation and the richer
strategic signaling that language affords.

A separate line of work combines language with strategic reasoning in more
complex multi-party settings. \citet{meta2022human} achieve human-level play
in Diplomacy by coupling a language model with a planning module that
proposes joint actions, demonstrating that language and strategy can be
integrated when the game structure provides sufficient scaffolding.
\citet{lewis2017deal} introduce end-to-end learning for negotiation dialogues
in the DealOrNoDeal task, training models that negotiate over item bundles
using both language and discrete choices.
\citet{he2018decoupling} decouple strategy and generation in
CraigslistBargains, using a dialogue manager to select actions and a
separate module to generate utterances. These architectures rely on
template retrieval or action-conditioned generation
\citep{keizer2017evaluating,yang2020improving}, limiting linguistic
flexibility. LLMs remove this bottleneck but introduce the behavioral
failures we document in \Cref{sec:motivation}.

\subsection{LLM Evaluation in Negotiation and Agentic Settings}
\label{sec:related_eval}

Several concurrent benchmarks evaluate LLM negotiation capabilities.
\citet{davidson2024evaluating} evaluate LLM agency through multi-issue
negotiations, finding that models frequently accept dominated offers and fail
to maintain consistent goals. \citet{xia2024measuring} measure bargaining
abilities and susceptibility to adversarial tactics, quantifying violations of
negotiation rationality. \citet{bianchi2024negotiationarena} introduce
NegotiationArena, a multi-domain evaluation platform that reveals limited
strategic diversity across models. \citet{abdelnabi2024cooperation} study
cooperation and deception in multi-agent LLM games, finding that models
exhibit inconsistent strategic behavior across repeated interactions.

Our benchmark differs from these efforts in three respects. First, the
structured action space separates binding offers from natural-language
messages, enabling unambiguous utility computation without human annotation or
LLM-based parsing. Second, we evaluate along three formally defined
dimensions (individual rationality, strategic effectiveness, allocative
efficiency) with automated metrics, rather than relying on aggregate scores or
qualitative assessments. Third, the round-robin tournament design---every
model against every model in both roles---reveals pairwise behavioral dynamics
(e.g., which models induce opponent IR violations) that single-opponent
evaluations cannot capture.

More broadly, LLM agent evaluation has expanded beyond negotiation.
\citet{liu2024agentbench} benchmark LLMs across diverse agentic tasks
including web browsing, tool use, and database interaction.
\citet{yao2023react} introduce the ReAct framework, interleaving reasoning
and action for tool-augmented tasks.
\citet{xi2025rise} survey LLM-based agents, identifying planning, memory,
and tool use as core capabilities. Our work contributes to this literature by
providing a testbed where multi-turn instruction fidelity, numeric constraint
adherence, and strategic optimization can be measured simultaneously---capabilities
that general-purpose agent benchmarks typically evaluate in isolation.

\subsection{LLM-Based Negotiation Agents}
\label{sec:related_agents}

Recent work on improving LLM negotiation performance falls into two
categories.

\paragraph{Prompt-based and modular approaches.}
\citet{fu2023improving} use self-play with a critic to select strong
negotiation transcripts, appending them as in-context examples to improve
pricing outcomes. \citet{oh2025llm} augment prompts with utility-based
feedback, helping models anticipate opponent behavior.
\citet{li2024dialogue} introduce Dialogue Action Tokens---planning signals
that guide a frozen LLaMA through multi-turn negotiation acts.
\citet{kong2025fishbargain} deploy FishBargain, a modular agent combining
price extraction, planning, and generation, tested with real sellers on
Xianyu. These approaches improve performance without modifying model
weights, but their gains are bounded by the base model's strategic
reasoning capacity.

\paragraph{Fine-tuning approaches.}
\citet{chatterjee2024agreemate} generate synthetic dialogues via
chain-of-thought prompting and fine-tune LLaMA-based models, improving
fairness and agreement rate.
\citet{verma2022chai} apply conservative Q-learning on offline negotiation
logs. \citet{ahmad2023ina} construct a mixed human/GPT-J dataset and
train agents using PPO with reward signals for rational behavior.
These fine-tuning efforts use small datasets, offline RL, or single-metric
optimization. None diagnoses the interaction between supervised and
reinforcement learning stages---specifically, the tension between the
selectivity instilled by behavioral cloning and the deal-closing bias
introduced by online RL---which our training experiments identify as a
central design challenge (\Cref{sec:training_discussion}).

\subsection{Reinforcement Learning for Language Models}
\label{sec:related_rl}

RL from human feedback (RLHF) has become standard for aligning LLMs with
human preferences \citep{ouyang2022training,bai2022training}. The dominant
algorithm, Proximal Policy Optimization (PPO) \citep{ppo}, has been
applied to instruction following, summarization, and dialogue.
\citet{grpo} introduce Group Relative Policy Optimization (GRPO), which
normalizes advantages within groups of rollouts for the same prompt,
reducing variance in reward estimation. We adopt GRPO for negotiation
training because its within-context normalization is well-suited to the
high context-dependence of negotiation outcomes (\Cref{sec:rl}).

Recent work applies RL to verifiable reasoning tasks where correctness can be
checked automatically. \citet{deepseekr1} train DeepSeek-R1 via large-scale
RL on mathematical and coding problems, demonstrating that RL can
substantially improve reasoning capabilities.
\citet{luong2024reft} apply RL fine-tuning to improve generalization on
multi-step reasoning tasks. These settings feature single-turn interactions
with bounded output length and binary or near-binary rewards (correct vs.\
incorrect). Multi-turn negotiation introduces three complications absent from
these tasks: (i)~context grows unboundedly over rounds, risking overflow
during training; (ii)~the opponent is a non-stationary part of the
environment, creating distribution shift; and (iii)~the reward must balance
protocol compliance with outcome quality, as a model that generates
well-formed tool calls but accepts irrational deals has failed.
\Cref{sec:rl_challenges} addresses these challenges explicitly.

%% file: sections/sec3.tex

\section{Bilateral Bargaining as a Testbed for LLMs}
\label{sec:bargaining}

\subsection{The Bargaining Game}
\label{sec:game}
Consider a bilateral negotiation where a buyer and seller seek to trade an item under conditions of incomplete information. Each agent is constrained by a private reservation price\footnote{A \emph{reservation price is the highest a buyer will pay or the lowest a seller will accept. We
use this term throughout; it is equivalent to ``private value'' in the auction
literature.}} that is hidden from their counterpart. The two
parties exchange offers and natural-language messages over multiple rounds, each
trying to reach a deal that maximizes their own surplus while respecting their
constraint. Neither party knows whether a mutually beneficial agreement even
exists; they must discover this through strategic interaction.

This setup, rooted in the bilateral trade formulations of \citet{chatterjee1983}
and \citet{myerson1983}, creates a rich testbed for LLM capabilities. An effective
agent must simultaneously (i)~respect hard constraints across many rounds,
(ii)~maximize surplus through strategic offers and communication, and
(iii)~coordinate with a counterpart through natural-language communication to
realize gains from trade when they exist. Bilateral bargaining under incomplete
information is a foundational problem in operations research and economics, with
applications to procurement, supply chain contracting, and dispute resolution.

We now formalize the game and define the three evaluation dimensions.

\begin{definition}[Bilateral Bargaining Game]
\label{def:game}
A buyer with reservation price $b$ and a seller with reservation price $s$
negotiate over a single indivisible item through multi-round alternating offers
\citep{rubinstein1982}. Each round combines natural-language communication with
formal price proposals. Both parties observe the item description, historical
prices, and the full negotiation history (all offers and messages). Neither party
observes the other's reservation price. The utility functions are:
\begin{equation}
\label{eq:utility}
U_B = \begin{cases}
b - p & \text{if agreement at price } p \leq b, \\
0     & \text{if no agreement,}\\
-\infty & \text{if agreement at price } p > b;
\end{cases}
\quad
U_S = \begin{cases}
p - s & \text{if agreement at price } p \geq s, \\
0     & \text{if no agreement.}\\
-\infty & \text{if agreement at price } p < s.
\end{cases}
\end{equation}
Accepting a price that violates the reservation constraint ($p > b$ for the buyer
or $p < s$ for the seller) yields negative utility---an irrational outcome.
\end{definition}

\begin{definition}[Zone of Possible Agreement]
\label{def:zopa}
The zone of possible agreement (ZOPA) is the interval $[s, b]$ when $b \geq s$.
The total available surplus (gains from trade) is $b - s$. Because reservation
prices are private, neither party knows whether a ZOPA exists or how large it is;
this must be inferred through the negotiation process.
\end{definition}

\begin{definition}[GFT and NGFT Scenarios]
\label{def:gft}
A scenario with $b \geq s$ is a \emph{gains-from-trade} (GFT) scenario: a
mutually beneficial deal exists, and any price $p \in [s, b]$ yields non-negative
utility for both parties. A scenario with $b < s$ is a
\emph{no-gains-from-trade} (NGFT) scenario: no price yields non-negative utility
for both parties simultaneously, and the rational outcome is no deal.
\end{definition}

\begin{figure}[H]
    \centering
    \includegraphics[width = 0.85\textwidth]{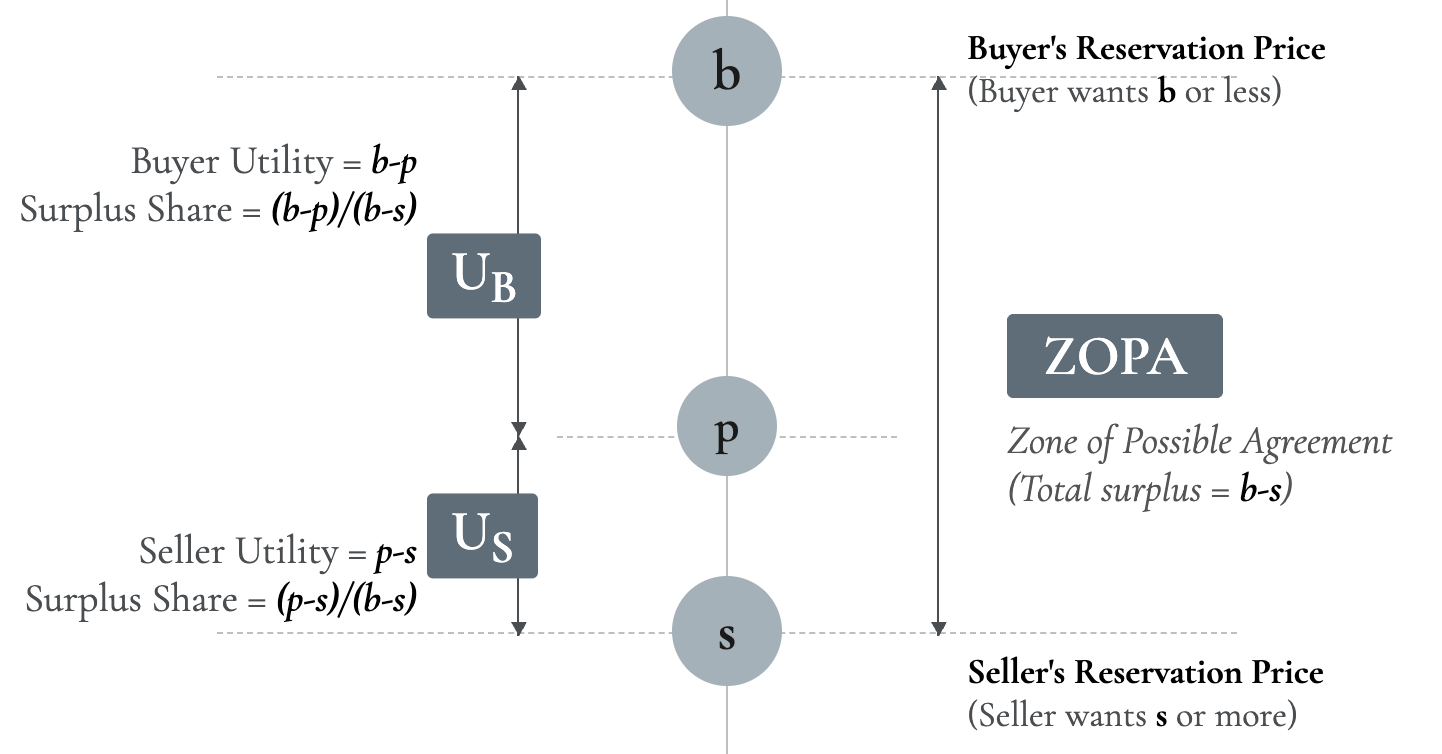}
    \caption{The diagram illustrates the bilateral bargaining space defined by the overlap between the seller’s reservation price ($s$) and the buyer’s reservation price ($b$). The ZOPA represents the total available economic surplus, calculated as $b - s$. Within this range, any agreed settlement price ($p$) facilitates allocative efficiency by distributing the surplus between the two parties.}
    \label{fig:zopa}
\end{figure}

\medskip\noindent\textbf{Three evaluation dimensions.}\quad
The game tests three distinct capabilities. Together, they provide a comprehensive
assessment of whether an LLM can function as a rational, effective negotiating
agent.

\medskip\noindent\textbf{Dimension 1: Individual rationality.}\quad
Can the agent respect hard constraints? An agent that accepts a deal yielding
negative utility---paying more than its maximum or selling below its minimum---has
failed at the most basic level.

\begin{definition}[Individual Rationality]
\label{def:ir}
An agent satisfies individual rationality (IR) if it never accepts a price yielding
negative utility: the buyer rejects any $p > b$, and the seller rejects any
$p < s$. This is the standard IR constraint in mechanism design
\citep{myerson1981}.
\end{definition}

IR is a necessary condition for credible negotiation. An
agent that violates it cannot be considered a rational participant regardless of
its average surplus.

\medskip\noindent\textbf{Dimension 2: Strategic effectiveness (surplus maximization).}\quad
Given that the agent respects its constraints, how much surplus does it capture?
Two individually rational agents can differ substantially in the utility they
obtain. We measure strategic effectiveness through average utility (across all
negotiations and conditional on deals) and the surplus share each agent captures.

\begin{definition}[Surplus Share]
\label{def:surplus_share}
For GFT scenarios reaching agreement at price $p$, the buyer's surplus share is
$(b - p) / (b - s)$ and the seller's surplus share is $(p - s) / (b - s)$.
The two shares sum to one: buyer and seller are dividing the total available
surplus $b - s$. A larger surplus share indicates a more strategically effective
agent.
\end{definition}

Because the seller posts the initial listing price, the protocol introduces a
structural role asymmetry that may advantage the seller through anchoring. We
examine this asymmetry empirically in \Cref{sec:benchmark_results}. While IR is a
necessary condition (floor), strategic effectiveness is an optimization objective
(ceiling).

\medskip\noindent\textbf{Dimension 3: Allocative efficiency (gains-from-trade realization).}\quad
Can the buyer--seller pair discover and realize the available gains from trade
through dialogue? This is a \emph{pair-level} property: it depends on both agents'
strategies jointly, not on either agent alone.

\begin{definition}[Deal Rate]
\label{def:deal_rate}
Given a set of $N$ negotiations, the deal rate is
\begin{equation}
\label{eq:deal_rate}
\text{Deal Rate} = \frac{1}{N} \sum_{i=1}^{N} \mathbf{1}\{\text{negotiation } i \text{ reaches agreement}\},
\end{equation}
where $\mathbf{1}\{\cdot\}$ is the indicator function. The deal rate counts all
agreements regardless of whether the agreed price yields positive or negative
utility. 
\end{definition}

In GFT scenarios, higher deal rates indicate better allocative
efficiency (surplus exists and should be realized). In NGFT scenarios, deal rate coincides with violation rate. Lower deal rates indicate better rationality (any deal violates IR for at least one
party).
Beyond the deal rate itself, allocative efficiency is shaped by behavioral drivers:
\emph{initial aggressiveness} of opening offers \citep{galinsky2001}, the
\emph{concession rate} (the pace of price adjustments across rounds), and
\emph{temporal patience} (willingness to sustain dialogue rather than terminating
prematurely). These behavioral metrics jointly determine whether a pair reaches
agreement and how quickly.


\medskip\noindent\textbf{Instruction adherence under multi-turn pressure.}\quad
The three dimensions share a common prerequisite: the agent must faithfully
execute the instructions encoded in its system prompt\footnote{A \emph{system
prompt} is a set of instructions provided to the LLM at the beginning of a
conversation, specifying its role, objectives, and constraints.}---reservation
price constraints, tool-use protocols, and behavioral rules---across many
rounds in which the counterpart actively attempts to persuade and extract
concessions. Each capability failure admits a dual interpretation: an IR
violation is also a failure to follow the reservation price constraint; a
an NGFT
agreement is also a violation of the surplus-maximization objective. If a
model cannot maintain adherence to a numeric constraint under sustained
counterpart pressure, it is likely to exhibit similar failures in any agentic
deployment requiring multi-turn instruction fidelity.

\subsection{Motivation: Why General-Purpose LLMs Are Insufficient}
\label{sec:motivation}

The most straightforward approach is to prompt a general-purpose LLM with the
game setup and let it negotiate in free-form dialogue. Two fundamental
limitations make this insufficient.

\begin{enumerate}

\item \textbf{Systematic behavioral failures.}
General-purpose LLMs exhibit failures across all three evaluation
dimensions. \citet{davidson2024evaluating} find that models frequently
accept dominated offers and fail to maintain consistent goals across
rounds. \citet{xia2024measuring} quantify violations of negotiation
rationality and show that models are susceptible to adversarial tactics
that pressure them into irrational concessions.
\citet{bianchi2024negotiationarena} reveal limited strategic diversity:
models converge on similar negotiation patterns regardless of context.
\citet{abdelnabi2024cooperation} document inconsistent strategic behavior
across repeated interactions in multi-agent games. These findings span
our three dimensions: IR violations (accepting deals at negative utility),
weak strategic effectiveness (failing to maximize surplus), and
allocative inefficiency (inability to close deals when gains from trade
exist or walking away too easily). The common root is a mismatch between
pre-training objectives---which reward fluency and agreeableness---and the
requirements of strategic negotiation: sustained goal-directed reasoning,
adherence to hard numeric constraints, and multi-round planning.

\item \textbf{The measurement problem.}
In free-form dialogue, it is unclear whether a mentioned price constitutes a
binding offer, a hypothetical suggestion, or a reference to market data.
This makes it impossible to compute utilities, track concession rates, or
determine whether agreement was reached. Without structured actions, every
evaluation requires human annotation or unreliable LLM-based parsing,
precluding both large-scale benchmarking and the use of outcome quality as a
training signal for RL.

\end{enumerate}

These two limitations motivate a domain-specific agent system that provides
structured actions for precise, automated measurement while preserving
natural-language communication.

\subsection{The Agent System}
\label{sec:agent_system}

We construct a simulation environment and structured agent interface whose design
choices directly address the three evaluation dimensions (\Cref{sec:game}) and two
limitations (\Cref{sec:motivation}). The system is modular: the negotiation domain
(item catalog, reservation price distributions, behavioral rules) is configurable
for adaptation to different application domains.

\subsubsection{Structured Action Space}
\label{sec:action_space}

We define the agent's action space via structured tool calls:\footnote{A
\emph{tool call} is a structured function invocation that the LLM generates as part
of its output; the simulation engine parses and executes each call, returning the
result to the agent's context for the next turn.}
\begin{table}[H]
\centering
\small
\begin{tabular}{@{}p{8cm}p{8cm}@{}}
\toprule
\textbf{Tool call} & \textbf{Description} \\
\midrule
\texttt{make\_offer(agent, price, side\_offer)}
  & Propose a price for the item. \\[3pt]
\texttt{respond\_to\_offer(agent, response)}
  & Accept or reject opponent's offer. \\[3pt]
\texttt{send\_message(agent, content)}
  & Send a negotiation message to opponent. \\[3pt]
\texttt{search\_price()}
  & Retrieve historical market reference prices. \\[3pt]
\texttt{quit\_negotiation(agent)}
  & Terminal tool; end negotiation without deal. \\[3pt]
\texttt{wait\_for\_response(agent)}
  & Terminal tool; wait for opponent's action. \\
\bottomrule
\end{tabular}
\caption{Structured action space: tool calls available to each agent.}
\label{tab:action_space}
\end{table}

Separating formal offers from messages enables unambiguous utility computation and
automated detection of IR violations (\Cref{def:ir}). Without structured actions, it is often unclear from free-form
text whether a price constitutes a binding offer. Structured actions also produce a
machine-readable negotiation trace from which surplus share, concession
rates, and initial aggressiveness can be computed automatically. The
\texttt{send\_message} tool preserves the communicative richness of natural
language---agents can persuade, signal flexibility, and extract information---while
the \texttt{make\_offer}/\texttt{respond\_to\_offer} pair provides a formal
mechanism for registering agreement.

The structured action space resolves the measurement problem identified in
\Cref{sec:motivation}. Offers, acceptances, rejections, and outside-option
exercises are all explicit, machine-parseable events. This enables fully automated,
reproducible evaluation without reliance on LLM-as-judge. The tool-based design
also mirrors real-world negotiation platforms, where interactions occur through
structured forms or APIs; each simulator action maps naturally to backend
components used in commercial systems.

Each agent receives a system prompt specifying its role, reservation price
constraint, utility formula, item description, and available tools. The prompt
also includes expert-crafted negotiation strategies (e.g., anchoring with
aggressive opening offers, pacing concessions, exercising the outside option when
appropriate). \Cref{fig:prompt} shows the hierarchical structure of the system
prompt. These strategies provide a behavioral baseline; fine-tuning can
subsequently improve upon them (\Cref{sec:training}).

\begin{figure}[H]
\centering
\begin{paperquote}[System Prompt Skeleton]
\small
\begin{description}
    \item[Role and Objective.] You are a skilled bargaining agent negotiating on behalf of...
    \item[Critical Rule: Constraint Compliance.] Mandatory adherence to role-specific constraints...
    \item[Variable Definitions.] Definition of \texttt{HARD\_LIMITS} and \texttt{UTILITY.} formulas...
    \item[Constraint Rules.] Explicit prohibition of deals resulting in negative utility...
    \item[Mandatory Decision Process.] Step-by-step verification: Identify $\rightarrow$ Recall $\rightarrow$ Calculate $\rightarrow$ Decide...
    \item[Negotiation Strategy Guidelines.] \hfill
    \begin{itemize}
        \item \textit{General Approach:} Strategic initialization and utility maximization...
        \item \textit{Anchoring and Positioning:} Use of precise numbers and pressure tactics...
        \item \textit{Concessions:} Tapered movements to signal limits...
        \item \textit{Non-Monetary Negotiation:} Exploration of bonuses and bundles...
    \end{itemize}
    \item[Rules You Must Follow] Requirements for persuasiveness, professionalism, and firm exit strategies...
    \item[How You Operate (CoT)] Use of \textbf{Thought} (private reasoning) and \textbf{Code} (tool calls)...
    \item[Tool Space and API Definitions] Specifications for \texttt{make\_offer}, \texttt{send\_message}, etc...
    \item[Critical Tool Usage Patterns] Mandatory utility pre-calculation patterns for API calls...
    \item[Code Rules] Syntax strictness, state persistence, and environmental constraints...
\end{description}
\end{paperquote}
\caption{Hierarchical structure of the agent system prompt. The prompt integrates
private economic constraints ($s, b$), strategic bargaining directives, and the
``Thought-Action'' protocol governing agent behavior within the simulation.}
\label{fig:prompt}
\end{figure}

\subsubsection{Event-Driven Simulation Engine}
\label{sec:simulation}

The simulation engine is a discrete-event system with a priority queue; events are
processed chronologically. Each event comprises a timestamp, type, acting agent, and
content. The negotiation protocol proceeds as: initialization $\to$ seller posts
listing price $\to$ buyer opens $\to$ alternating turns $\to$ termination.
A negotiation terminates when either party accepts an offer, exercises the outside
option (\texttt{quit\_negotiation}), or the round limit is reached. The maximum
number of rounds is set to $N = 10$; each turn permits at most 3 tool calls. Each agent's output
consists of a \texttt{Thought:} block (internal reasoning, not visible to the
counterpart) followed by a \texttt{Code:} block (tool calls), which the simulator
parses and executes.

\begin{figure}[H]
    \centering
\includegraphics[width=0.9\textwidth]{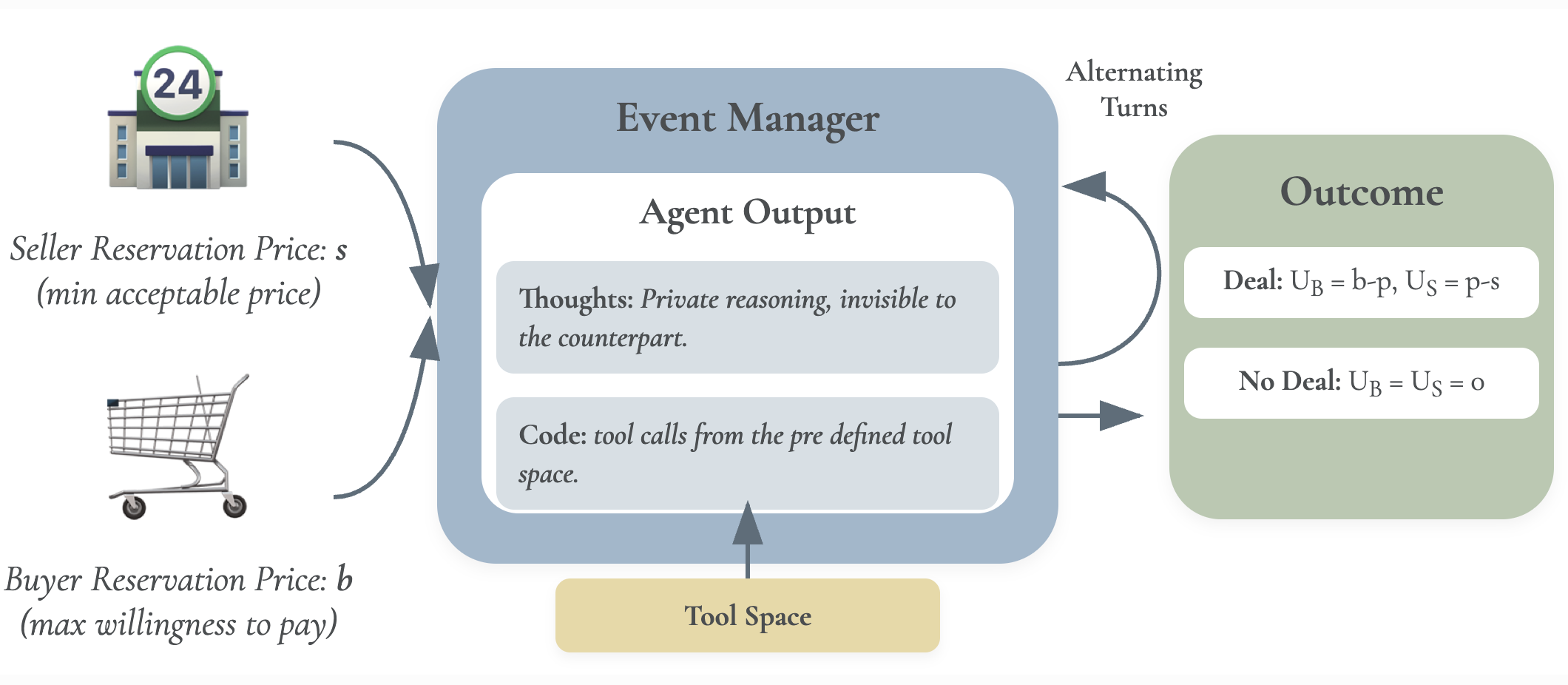}
\caption{This is the system architecture of the bilateral bargaining simulation environment. Agents are initialized with reservation prices ($s$ for the seller, $b$ for the buyer). The negotiation phase is governed by an Event Manager that enforces a protocol of alternating turns, a limit of max rounds, and a maximum of three structured tool calls per turn from the predefined Tool Space. Agent outputs have private "Thoughts" (internal reasoning) and "Code" (external tool-based actions). The cycle concludes in the Outcome phase, where the environment evaluates deal status and calculates resulting utilities ($U_B = b - p$; $U_S = p - s$) based on the final settlement price ($p$).}
\label{fig:system_architecture}
\end{figure}

\subsubsection{Evaluation Metrics}
\label{sec:metrics}

The structured action space enables a comprehensive suite of automated metrics,
organized by the three evaluation dimensions.

\begin{table}[H]
\centering
\small
\begin{tabular}{@{}p{4.5cm}p{10.5cm}@{}}
\toprule
\textbf{Metric} & \textbf{Description} \\
\midrule
Violation rate
  & Fraction of negotiations where an agent accepts a price yielding
    negative utility ($p > b$ for the buyer or $p < s$ for the seller).
    A no-deal outcome (utility $= 0$) is \emph{not} a violation.
    In GFT scenarios, violations arise when a deal is struck outside the
    ZOPA; in NGFT scenarios, every deal violates IR for at least one
    party, so the violation rate coincides with the NGFT deal rate
    (see \Cref{tab:ae_metrics}). The two metrics are otherwise
    distinct: violation rate measures individual constraint adherence,
    while deal rate measures pair-level agreement realization. \\
\bottomrule
\end{tabular}
\caption{Dimension~1 metrics: individual rationality (IR).}
\label{tab:ir_metrics}
\end{table}

\begin{table}[H]
\centering
\small
\begin{tabular}{@{}p{4.5cm}p{10.5cm}@{}}
\toprule
\textbf{Metric} & \textbf{Description} \\
\midrule
Average utility (all)
  & Mean $U_B$ and $U_S$ across all negotiations in a given regime
    (GFT or NGFT), including no-deal outcomes (utility $= 0$). \\[4pt]
Average utility (deals)
  & Mean $U_B$ and $U_S$ restricted to negotiations reaching agreement.
    Isolates agreement quality from agreement frequency. \\[4pt]
Surplus share
  & Share of surplus captured in both-valid deals: $(b - p)/(b - s)$
    for the buyer, $(p - s)/(b - s)$ for the seller
    (\Cref{def:surplus_share}). Defined only for GFT scenarios. \\[4pt]
\bottomrule
\end{tabular}
\caption{Dimension~2 metrics: strategic effectiveness (surplus maximization).}
\label{tab:se_metrics}
\end{table}

\begin{table}[H]
\centering
\small
\begin{tabular}{@{}p{4.5cm}p{10.5cm}@{}}
\toprule
\textbf{Metric} & \textbf{Description} \\
\midrule
Deal rate
  & Fraction of negotiations reaching agreement (\Cref{def:deal_rate}).
    Higher is better in GFT; lower is better in NGFT. \\
\bottomrule
\end{tabular}
\caption{Dimension~3 metrics: allocative efficiency (gains-from-trade
realization).}
\label{tab:ae_metrics}
\end{table}

\begin{table}[H]
\centering
\small
\begin{tabular}{@{}p{4.5cm}p{10.5cm}@{}}
\toprule
\textbf{Metric} & \textbf{Description} \\
\midrule
Seller initial aggressiveness
  & Ratio of the seller's opening offer $p_0^{\text{seller}}$ to the seller's
    reservation price:
    $\alpha^{\text{seller}} = p_0^{\text{seller}} \big/ s$.
    Higher values indicate a more aggressive opening position
    (asking further above cost). \\[4pt]
Buyer initial aggressiveness
  & Measured in two variants (higher $=$ more aggressive):
    \begin{itemize}[nosep,leftmargin=*]
      \item \emph{Gap closure:}
        $\alpha^{\text{buyer},1}
         = \bigl(p_0^{\text{seller}} - p_0^{\text{buyer}}\bigr)
           \big/ p_0^{\text{seller}},$
        fraction of the seller's asking price cut by the buyer's
        first counter-offer.
      \item \emph{Reservation ratio:}
        $\alpha^{\text{buyer},2}
         = \bigl(b - p_0^{\text{buyer}}\bigr) \big/ b,$
        fraction of the buyer's reservation price retained as
        surplus in the opening bid.
    \end{itemize}
    \vspace{-4pt} \\
Concession rate
  & Fraction of remaining surplus conceded per round (computed for valid
    deals only).
    \begin{itemize}[nosep,leftmargin=*]
      \item \emph{Seller:}
        $c_t^{\text{seller}} = \bigl(p_{t}^{\text{seller}} - p_{t+1}^{\text{seller}}\bigr)
         \big/ \bigl(p_{t}^{\text{seller}} - s\bigr).$
      \item \emph{Buyer:}
        $c_t^{\text{buyer}} = \bigl(p_{t+1}^{\text{buyer}} - p_{t}^{\text{buyer}}\bigr)
         \big/ \bigl(b - p_{t}^{\text{buyer}}\bigr).$
    \end{itemize}
    Higher values indicate aggressive bargaining. The sign is used as a
    Dimension~1 diagnostic (\Cref{tab:ir_metrics}).
    \vspace{-4pt} \\
Temporal patience
  & Mean number of rounds to termination. Measures willingness to
    sustain dialogue rather than terminating prematurely. \\
\bottomrule
\end{tabular}
\caption{Behavioral driver metrics: mechanisms underlying strategic
effectiveness and allocative efficiency outcomes.}
\label{tab:behavioral_metrics}
\end{table}

\paragraph{Price-tier decomposition.}
Aggregate metrics can mask systematic variation across item price ranges. Human
negotiation behavior differs by item value: buyers tend to bargain more tightly on
inexpensive items (where a fixed dollar concession is proportionally larger) and
more flexibly on expensive items. To capture this heterogeneity, we decompose all
metrics by item value.

For each buyer--seller model pair, we partition the item catalog into five quintiles
based on each item's reference price (midpoint of the historical price range).
Within each quintile, we compute all IR, strategic effectiveness, and allocative
efficiency metrics separately. This decomposition reveals whether models exhibit
price-sensitivity patterns analogous to human behavior, whether certain models
systematically under-perform in specific price tiers, and whether IR violations
concentrate at the extremes of the price distribution. It also serves as a
diagnostic for training: if fine-tuning improves performance on expensive items but
degrades it on inexpensive ones, the price-tier decomposition makes this visible.

\subsection{Item Catalog and Reservation Price Sampling}
\label{sec:catalog}

The negotiation scenarios are constructed from a product catalog and a reservation
price sampling procedure that together determine the input to the agent system.

\paragraph{Product catalog.}
Listings are sourced from the CraigslistBargains dataset \citep{craigslist}
(1,402~items) and Amazon Price History data (930~items), totaling 2,332~products.
Each listing includes a title, description, category, and historical price range.
For Amazon listings, historical price bounds are taken directly from the dataset.
Craigslist listings contain only a posted price; we set the highest price to the
posted price and the lowest price to 50\% of that value. An example listing is
in \Cref{fig:ex_listing} (\Cref{app:example_listing}). The catalog is partitioned
into training (80\%), validation (8.5\%), and test (11.5\%) splits.

\paragraph{Reservation price sampling.}
The \texttt{search\_price()} tool returns historical high and low market reference prices for each
item, providing reference information for anchoring. Seller reservation prices are sampled as
\begin{equation}
\label{eq:seller_res}
s \sim \mathrm{Uniform}[0.6 \times p_{\min},\; 0.9 \times p_{\min}],
\end{equation}
and buyer reservation prices as
\begin{equation}
\label{eq:buyer_res}
b \sim \mathrm{Uniform}[0.4 \times p_{\min},\; p_{\max} + 0.1(p_{\max} - p_{\min})],
\end{equation}
where $p_{\min}$ and $p_{\max}$ are the historical low and high prices from the
item's catalog listing. This produces a mix of GFT and NGFT scenarios
(${\sim}65\%$ GFT, ${\sim}35\%$ NGFT), testing whether agents both realize surplus
when it exists and avoid value destruction when it does not.

\paragraph{Connection to evaluation dimensions.}
Reservation prices define each agent's hard constraints; the sampling range creates
scenarios from wide ZOPA to no ZOPA, testing whether agents maintain IR across
varying difficulty. The range of ZOPA widths creates variation in available surplus,
enabling meaningful comparison of surplus share across agents. The GFT/NGFT split
directly operationalizes allocative efficiency---do agents reach agreement when (and
only when) gains from trade exist?

\subsection{Two Uses of the System}
\label{sec:two_uses}

The agent system serves two complementary purposes, corresponding to the next two
sections:
\begin{enumerate}
    \item \textbf{Benchmark for evaluating frontier LLMs (\Cref{sec:benchmark}).}
      Running models in the simulator and measuring the metrics above (including
      price-tier decomposition) yields a systematic evaluation of each model across
      the three dimensions. This extends prior LLM negotiation evaluations by
      providing structured, dimension-specific diagnostics rather than aggregate
      scores.
    \item \textbf{Environment for training negotiation agents
      (\Cref{sec:training}).}
      The simulator enables large-scale synthetic trajectory generation via LLM
      self-play. The behavioral metrics serve as reward signals for reinforcement
      learning. Together, these support a two-stage training pipeline (supervised
      fine-tuning\footnote{\emph{Supervised fine-tuning} (SFT) is additional
      training of a pre-trained LLM on task-specific demonstrations, analogous to
      behavioral cloning in the RL literature.} on filtered synthetic data, then RL
      with rule-based rewards) to fine-tune open-weight models for the negotiation
      task.
\end{enumerate}

%% file: sections/sec4.tex

\section{Benchmarking Frontier LLMs}
\label{sec:benchmark}

\subsection{Evaluation Protocol}
\label{sec:eval_protocol}

We use the agent system (\Cref{sec:agent_system}) as a standardized benchmark for
evaluating frontier LLMs on bilateral price negotiation.

\paragraph{Models.}
We evaluate five frontier models: DeepSeek-V3-0324 \citep{deepseekv3},
Gemini-2.5-Flash \citep{gemini25_flash_pro}, Gemini-2.5-Pro
\citep{gemini25_flash_pro}, GPT-4.1 \citep{gpt4}, and o3 \citep{o3}. All
models receive the same system prompt template, tool definitions, and item
information. See \Cref{tab:models} for model details and costs.

\begin{table}[H]
\centering
\small
\begin{tabular}{@{}llr@{\,/\,}l@{}}
\toprule
\textbf{Model} & \textbf{Provider}
  & \multicolumn{2}{c}{\textbf{Cost (\$/1M tokens, in\,/\,out)}} \\
\midrule
DeepSeek-V3-0324 & DeepSeek & \$0.27  & \$1.10 \\
Gemini-2.5-Flash & Google   & \$0.30  & \$2.50 \\
Gemini-2.5-Pro   & Google   & \$1.25  & \$10.00 \\
GPT-4.1          & OpenAI   & \$2.00  & \$8.00 \\
o3$^\dagger$     & OpenAI   & \$2.00  & \$8.00 \\
\bottomrule
\end{tabular}
\caption{Frontier models evaluated in the round-robin benchmark. Prices
reflect API list rates at the time of experiments. $^\dagger$o3 is a
reasoning model; internal reasoning tokens are billed as output tokens,
increasing effective cost.}
\label{tab:models}
\end{table}

\paragraph{Round-robin tournament.}
Each model plays as both buyer and seller against every model (including itself),
producing $5 \times 5 = 25$ distinct (buyer, seller) pairings. All 25 pairings
negotiate over the same 600 product listings (400 GFT + 200 NGFT), ensuring that
differences across pairings reflect model behavior rather than item-level variance.
This yields a total of 15,000 negotiations. Self-play pairings (model vs.\ itself) are included to examine
whether a model's strategies lead to efficient outcomes or deadlock when facing
its own behavioral profile.

\subsection{Results}
\label{sec:benchmark_results}

A unifying lens for the results that follow is price discrimination through
negotiation. The economics literature recognizes bargaining as a mechanism for
first-degree price discrimination: rather than posting a single price, the
seller uses sequential offers to discover each buyer's willingness to pay and
extract surplus accordingly \citep{riley1983optimal,sobel1983multistage}. Three findings from the
benchmark align with this interpretation. First, the top-performing seller
strategy---anchoring far above cost and conceding calibratedly over multiple rounds---creates maximum
room for price adaptation: different buyers end up paying different prices based
on their strategic resistance, yet deals still close at near-maximal rates
(\Cref{sec:benchmark_summary}, Finding~1). Second, accommodating buyer
strategies that concede quickly and uniformly represent the failure mode: they
reveal willingness to pay immediately, eliminating the seller's need to
discriminate and leaving surplus on the table for the buyer role (Finding~2).
Third, consistency across price tiers reflects the ability to discriminate
proportionally: strong models scale their anchoring and concession to item
value, extracting a similar fraction of surplus whether the item costs \$20 or
\$2{,}000; weak models succeed only when wide ZOPAs mask their inability to
adapt (Finding~3).

\subsubsection{Model-Level Overview}
\label{sec:benchmark_three_dim}

We evaluate models along the three dimensions defined in \Cref{sec:game}:
(1)~individual rationality, measured by violation
rates; (2)~strategic effectiveness, measured by surplus share; and
(3)~allocative efficiency, measured by deal rate. We also report behavioral
drivers---initial aggressiveness and concession rate---that characterize
the mechanisms associated with outcome differences.
We discuss each dimension below with detailed tables; pairwise heatmaps
for all metrics are in \Cref{app:pairwise_heatmaps}.

\begin{figure}[H]
\centering
\includegraphics[width=0.85\textwidth]{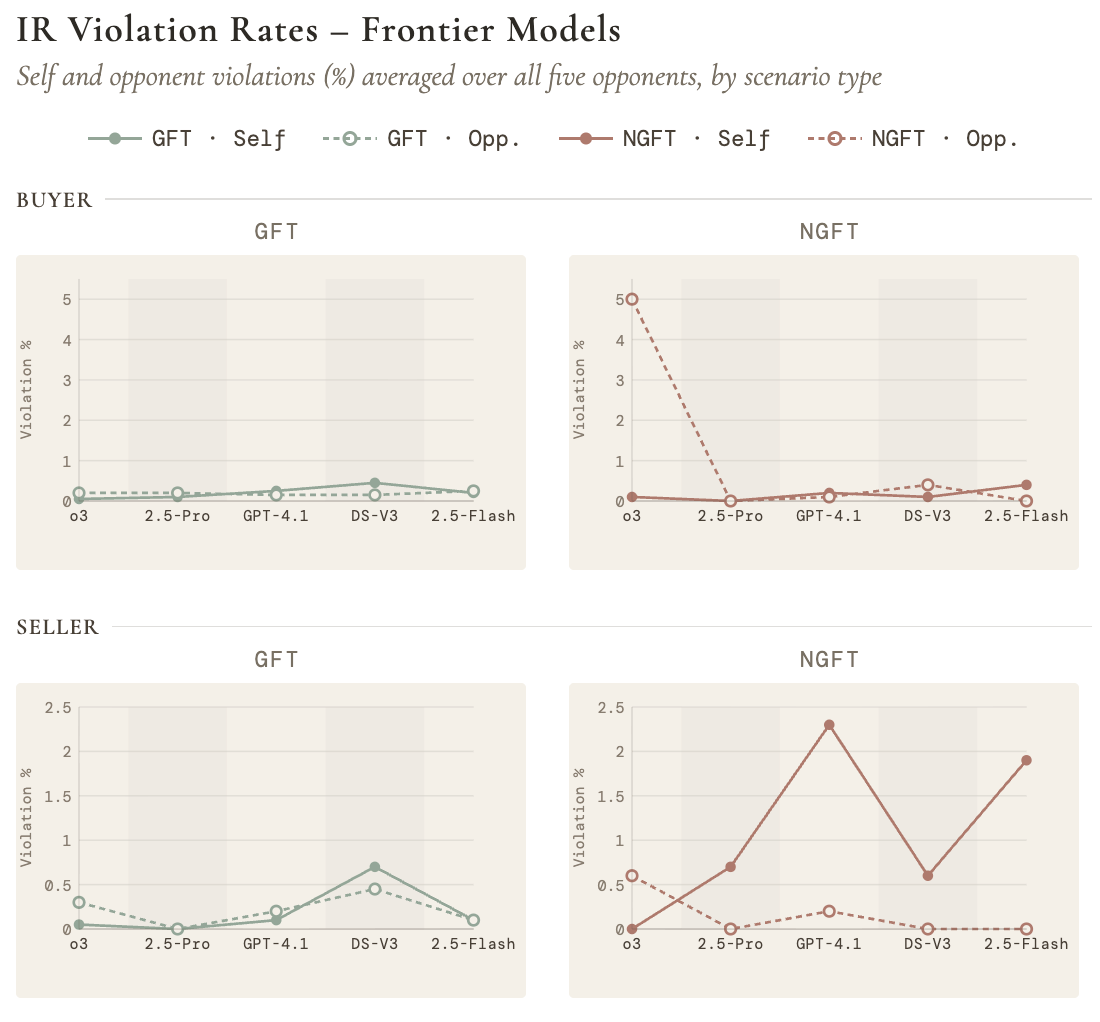}
\caption{IR compliance for frontier models. Self: the model's own violation
rate (deals yielding negative utility for itself). Opp.: the opponent's
violation rate when facing this model. All values averaged over all five
opponents. Green/red borders highlight best/worst values in each column. In NGFT, any deal is a
violation for at least one party; o3 as buyer induces 5.00\% opponent
violations despite only 0.10\% self-violations. Full table in \Cref{tab:frontier_ir}.}
\label{fig:frontier_ir}
\end{figure}

\paragraph{Dimension 1: Individual rationality.}
\Cref{fig:frontier_ir} reports violation rates by model and role for both GFT
and NGFT scenarios, distinguishing between a model's own violations (self) and
violations it induces in opponents (opp.). Full numerical values are in
\Cref{tab:frontier_ir}.

\textbf{GFT violation rates are uniformly low.} All GFT self-violation
rates are below 1\%, and opponent-violation rates are below 0.5\%.
At these magnitudes, differences across models are not reliably
distinguishable from sampling variability.

\textbf{NGFT reveals a seller-side vulnerability.} Buyer NGFT
self-violations are all ${\leq}0.40\%$, while seller NGFT
self-violations range from 0.00\% (o3) to 2.30\% (GPT-4.1).
GPT-4.1 is the most vulnerable seller (2.30\%), followed by
Gemini-2.5-Flash (1.90\%), Gemini-2.5-Pro (0.70\%), and
DeepSeek-V3-0324 (0.60\%). This buyer--seller asymmetry is consistent
across all five models, suggesting a systematic difficulty with the
seller role in NGFT scenarios rather than a model-specific failure.

\begin{figure}[H]
\centering
\includegraphics[width=0.85\textwidth]{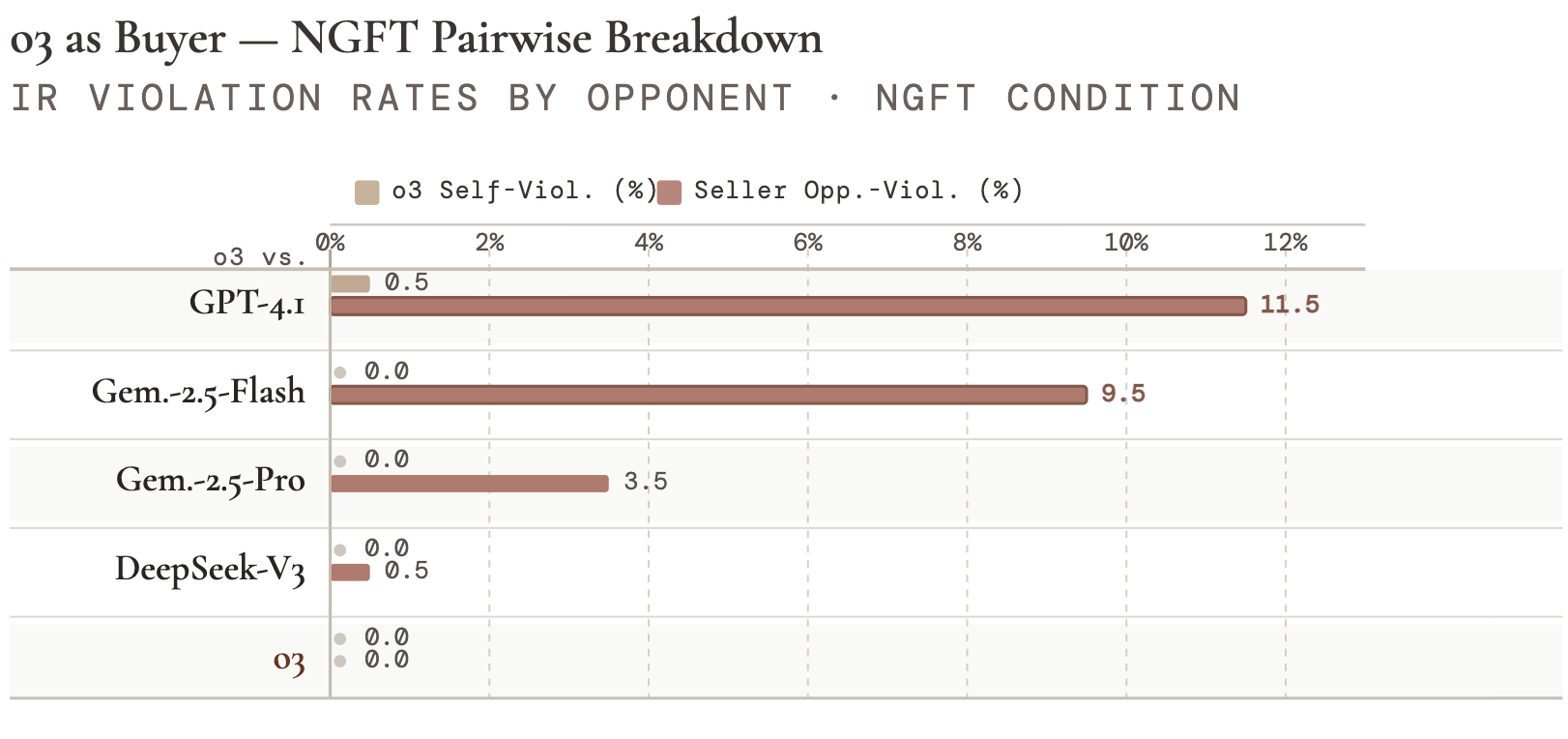}
\caption{NGFT pairwise breakdown for o3 as buyer. The irrationality comes
almost entirely from opponents: GPT-4.1 and Gemini-2.5-Flash violate at
11.5\% and 9.5\%, while o3 itself violates at most 0.5\%. Full table in
\Cref{tab:o3_buyer_ngft}.}
\label{fig:o3_buyer_ngft}
\end{figure}

\textbf{o3 as buyer induces substantially higher NGFT opponent
violations than any other model.} o3 buyer NGFT opponent-violation
averages 5.00\% across all opponents, far above the next highest
(DeepSeek buyer at 0.40\%). \Cref{fig:o3_buyer_ngft} decomposes this
by opponent (\Cref{tab:o3_buyer_ngft}): GPT-4.1 as seller violates
at 11.5\% when facing o3; Gemini-2.5-Flash at 9.5\%.
o3 self-play produces 0.0\% opponent violations. The pattern is consistent
with opponent strength: weaker sellers (GPT-4.1, Flash) are pressured
into accepting irrational deals, while stronger opponents (Pro,
DeepSeek, o3 itself) resist. Pairwise violation rate heatmaps are in
\Cref{app:violation_rate}.

\begin{figure}[H]
\centering
\includegraphics[width=0.85\textwidth]{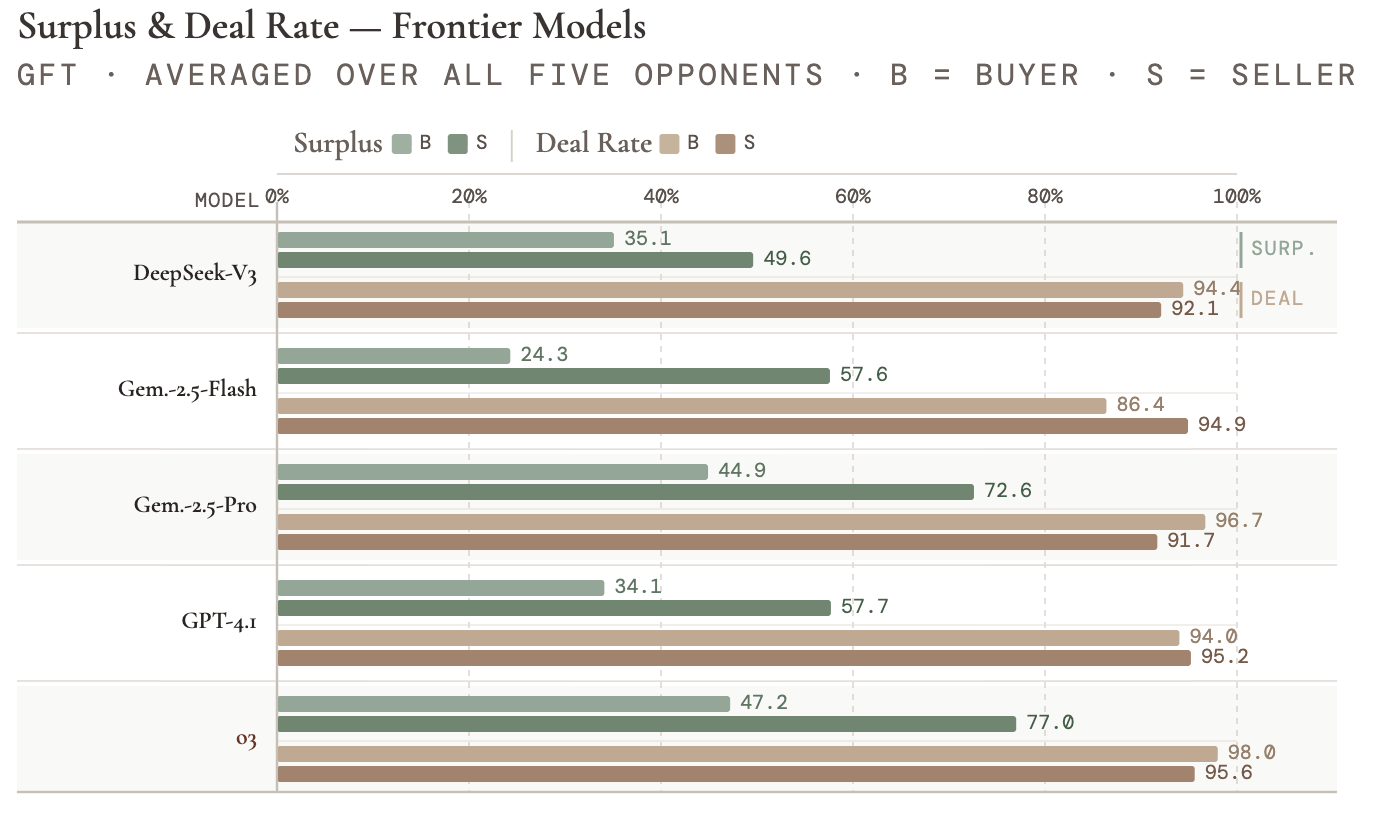}
\caption{Surplus share and deal rate for frontier models (GFT scenarios),
averaged over all five opponents. B = buyer, S = seller. Surplus: fraction of
ZOPA captured conditional on agreement. Deal rate: fraction of negotiations
reaching agreement. Full table in \Cref{tab:frontier_surplus}.}
\label{fig:frontier_surplus}
\end{figure}

\paragraph{Dimension 2: Strategic effectiveness---surplus maximization.}
\Cref{fig:frontier_surplus} reports surplus share and deal rate by model and
role (\Cref{tab:frontier_surplus}). A clear hierarchy emerges at the top:
\textbf{o3 captures the most surplus in both roles} (47.2\% buyer, 77.0\%
seller), followed by Gemini-2.5-Pro (44.9\%/72.6\%). The remaining three
models are closer together, and their relative ordering shifts across roles:
as buyers, DeepSeek-V3-0324 (35.1\%) edges GPT-4.1 (34.1\%), with
Gemini-2.5-Flash last (24.3\%); as sellers, GPT-4.1 (57.7\%) and
Gemini-2.5-Flash (57.6\%) are nearly tied, while DeepSeek-V3-0324 is last
(49.6\%). \textbf{The top two models rank consistently across roles; the
remaining three swap positions.}

Sellers capture the majority of surplus in nearly every pairing, reflecting the
structural \textbf{first-mover advantage}. Self-play isolates this effect:
DeepSeek-V3-0324 self-play is nearly symmetric (48.5\%/51.5\%), while o3
self-play is heavily seller-skewed (32.3\%/67.7\%), revealing that o3's
strategy amplifies the structural advantage far beyond the baseline. The
50 percentage point (pp) spread between the best buyer surplus share (o3 vs.\
DeepSeek, 61.8\%) and the worst (Gemini-2.5-Flash vs.\ o3, 11.6\%)
underscores that \textbf{model choice matters as much as role assignment}. Pairwise utility and
surplus share heatmaps are in \Cref{app:utility,app:surplus_share}.

\begin{figure}[H]
\centering
\includegraphics[width=0.85\textwidth]{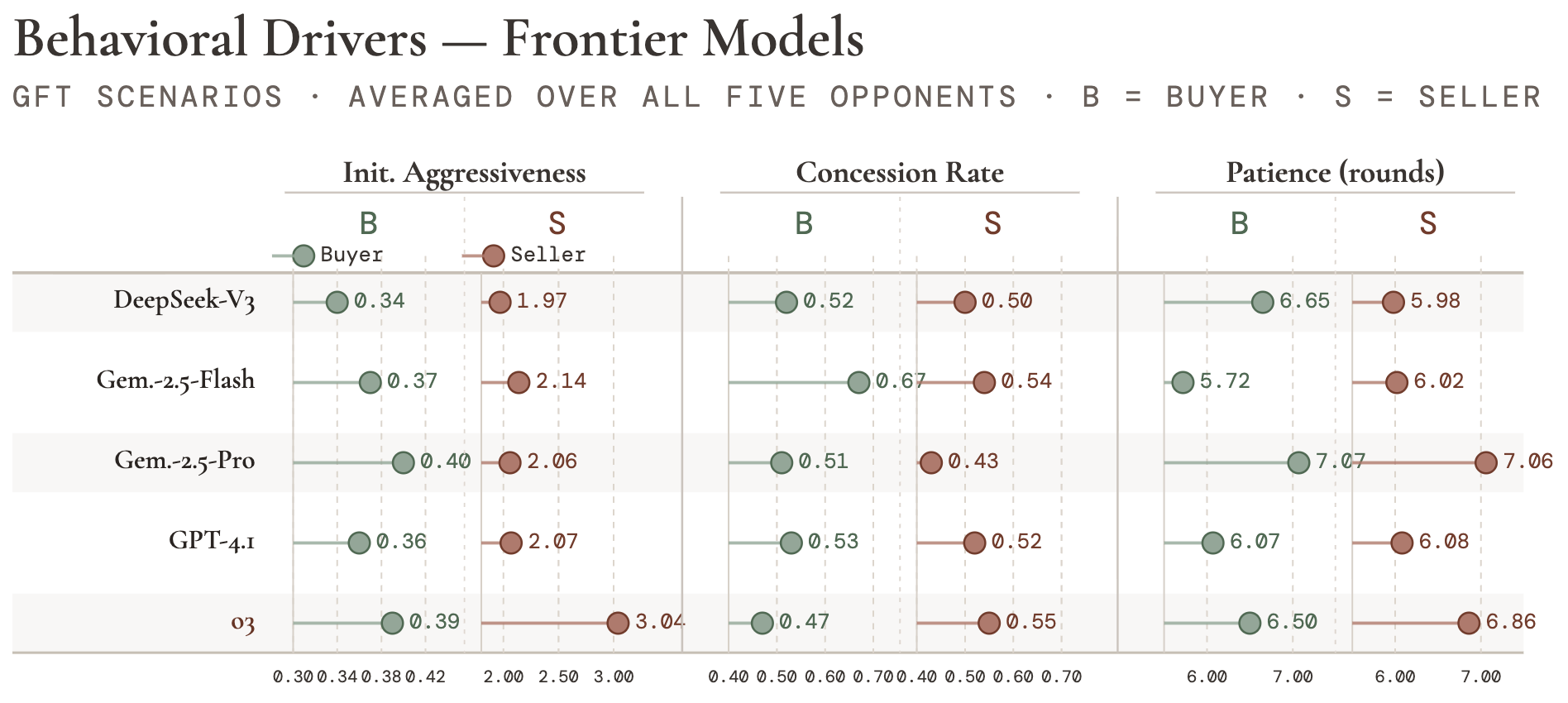}
\caption{Behavioral drivers for frontier models (GFT scenarios), averaged
over all five opponents. B = buyer, S = seller. Init.\ Aggr.\ (buyer):
fraction of the seller’s asking price cut by the buyer’s first counter-offer
(higher = more aggressive). Init.\ Aggr.\ (seller): ratio of first offer to
reservation price (higher = more aggressive). Conc.\ Rate: fraction of
remaining surplus conceded per round. Patience: mean number of rounds to
termination (max 10). Full table in \Cref{tab:frontier_surplus}.}
\label{fig:frontier_behavior}
\end{figure}

\paragraph{Dimension 3: Allocative efficiency---deal rate.}
\Cref{fig:frontier_surplus} reports GFT deal rates by model and role.
\textbf{o3 achieves the highest rates in both roles} (98.0\% buyer, 95.6\%
seller), demonstrating that high surplus capture can coexist with high deal
completion. GPT-4.1 shows balanced performance across roles
(94.0\%/95.2\%). Gemini-2.5-Flash as buyer is the clear outlier (86.4\%):
its cautious profile is associated with more walk-aways, compounding its
strategic disadvantage---though it recovers as seller (94.9\%).
Gemini-2.5-Pro shows the reverse asymmetry: strong as buyer (96.7\%) but the
lowest seller deal rate (91.7\%), associated with its low concession rate as
seller. Pairwise deal rate heatmaps are in \Cref{app:deal_rate}.

\paragraph{Behavioral drivers.}
The preceding three dimensions document \emph{what} models achieve; the
\Cref{fig:frontier_behavior} reveals \emph{how}.

\emph{Initial aggressiveness.} Seller aggressiveness varies substantially:
o3 opens at $3.04\times$ its cost, far above every other model
(1.97--2.14$\times$). \textbf{This aggressive anchoring is the most
distinctive behavioral difference} across models, consistent with the
first-offer anchoring effect \citep{galinsky2001}. The remaining four
sellers cluster tightly, with DeepSeek-V3-0324 the most moderate
($1.97\times$). On the buyer side, aggressiveness is more compressed
(0.34--0.40), with Gemini-2.5-Pro cutting the most from the seller's
asking price (0.40) and DeepSeek-V3-0324 the least (0.34).

\emph{Concession rates.} o3 employs opposite strategies by role: as buyer it
concedes the least (0.47), while as seller it concedes the most (0.55). This
\textbf{``anchor high, concede to close'' pattern is consistent with o3's
combined high surplus share and high deal rate}. Gemini-2.5-Flash shows the
opposite profile: the highest buyer concession rate (0.67) is associated with
the weakest buyer surplus share (24.3\%), while Gemini-2.5-Pro's low seller
concession rate (0.43) is associated with its strong seller surplus share
(72.6\%) and the lowest seller deal rate (91.7\%).

\emph{Temporal patience.} Gemini-2.5-Flash's impatience (5.72 rounds as
buyer) correlates with its weak bargaining power: it closes deals quickly but
captures less surplus. \textbf{Gemini-2.5-Pro's patience (7.07 rounds) is
associated with its strong surplus capture} and low concession rate. Pairwise
heatmaps for initial aggressiveness, concession rate, and temporal patience
are in
\Cref{app:initial_aggressiveness,app:concession_rate,app:temporal_patience}.

\subsubsection{Price-Tier Decomposition}
\label{sec:benchmark_price_tier}

Human negotiators exhibit price-sensitive behavior: they bargain more
tightly on inexpensive items, where a fixed dollar concession is
proportionally larger. Do frontier LLMs show similar patterns? We
disaggregate all three evaluation dimensions by reservation price quintile
(Q1\,=\,lowest, Q5\,=\,highest), averaging across all five opponents per
model. Buyer-side metrics are broken down by the buyer's reservation price
$b$; seller-side metrics by the seller's reservation price $s$. This
decomposition reflects that each agent's own reservation price determines
how much room it has to negotiate: a buyer with low $b$ faces thin ZOPAs
and must bargain tightly, while a buyer with high $b$ has more margin for
suboptimal play. The results split sharply: surplus capture and deal rates are
value-sensitive for weaker models but stable for stronger ones, while
violation rates show no systematic price-tier trend for any model.

\begin{figure}[H]
\centering
\includegraphics[width=\textwidth]{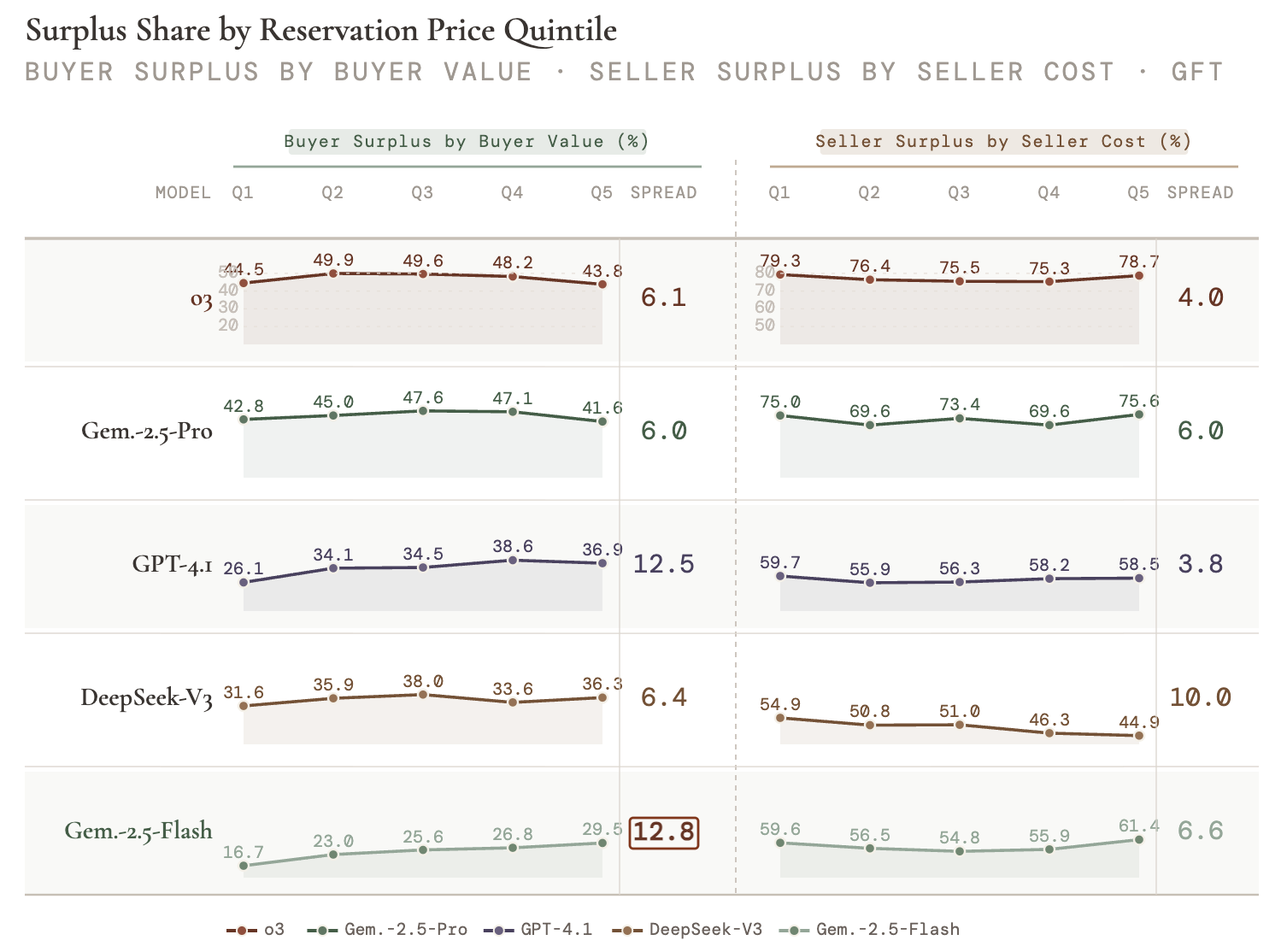}
\caption{Surplus share by reservation price quintile, averaged across all 5
opponents (GFT). Left panel: buyer surplus by buyer-value bracket. Right
panel: seller surplus by seller-cost bracket. Strong models (o3, Gem.-Pro)
show 4--6\,pp spreads; weak buyers (Gem.-Flash) show 13\,pp.
Full table in \Cref{tab:frontier_bracket_surplus}.}
\label{fig:frontier_bracket_surplus}
\end{figure}

\textbf{Surplus capture: stronger models show smaller bracket spreads.}
\Cref{fig:frontier_bracket_surplus} reports each model's surplus share
across value quintiles (\Cref{tab:frontier_bracket_surplus}). o3 maintains
44--50\% buyer surplus across all brackets (6.1\,pp spread);
Gemini-2.5-Pro is similarly stable at 42--48\% (6.0\,pp).
Gemini-2.5-Flash improves from 17\% at Q1 to 30\% at Q5---a 12.8\,pp
spread---because high-value brackets have larger ZOPAs that buffer
suboptimal strategies. DeepSeek-V3 as seller shows an approximately
declining pattern, from 55\% (Q1) to 45\% (Q5), suggesting difficulty
maintaining margins on more expensive items. The top two models (o3,
Gemini-2.5-Pro) and the bottom one (Gemini-2.5-Flash as buyer) maintain
their relative positions across all quintiles; GPT-4.1 and DeepSeek-V3
swap at Q4 and Q5.

\begin{figure}[H]
\centering
\includegraphics[width=\textwidth]{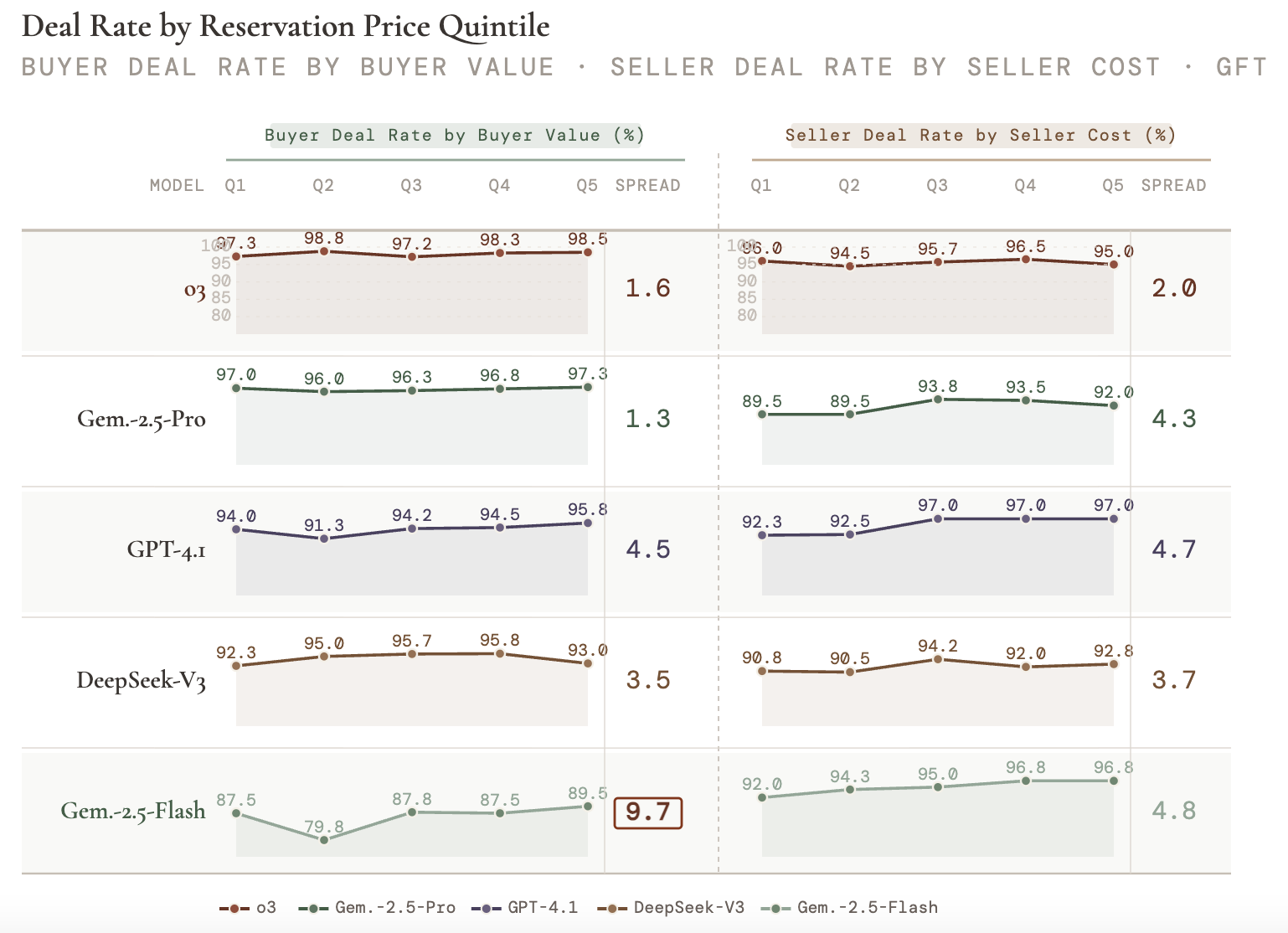}
\caption{Deal rates by reservation price quintile, averaged across all 5
opponents (GFT). Left panel: buyer deal rate by buyer-value bracket. Right
panel: seller deal rate by seller-cost bracket. Flash as buyer shows the
largest spread (9.7\,pp); all other models stay within 1--5\,pp.
Full table in \Cref{tab:frontier_bracket_dealrate}.}
\label{fig:frontier_bracket_dealrate}
\end{figure}

\textbf{Deal rates: weaker buyers show more bracket variation.}
\Cref{fig:frontier_bracket_dealrate} shows that deal rates follow a
similar pattern (\Cref{tab:frontier_bracket_dealrate}). o3 closes
97--99\% of deals regardless of bracket (1.6\,pp spread).
Gemini-2.5-Flash as buyer drops to 80\% at Q2---its minimum---and
reaches 90\% at Q5, producing the largest spread of any model
(9.7\,pp). As sellers, all models are more stable: spreads range from
2.0\,pp (o3) to 4.8\,pp (Gemini-2.5-Flash). Buyers exhibit more
value-sensitivity than sellers, consistent with sellers' structural
advantage in setting initial asking prices.

\begin{figure}[H]
\centering
\includegraphics[width=\textwidth]{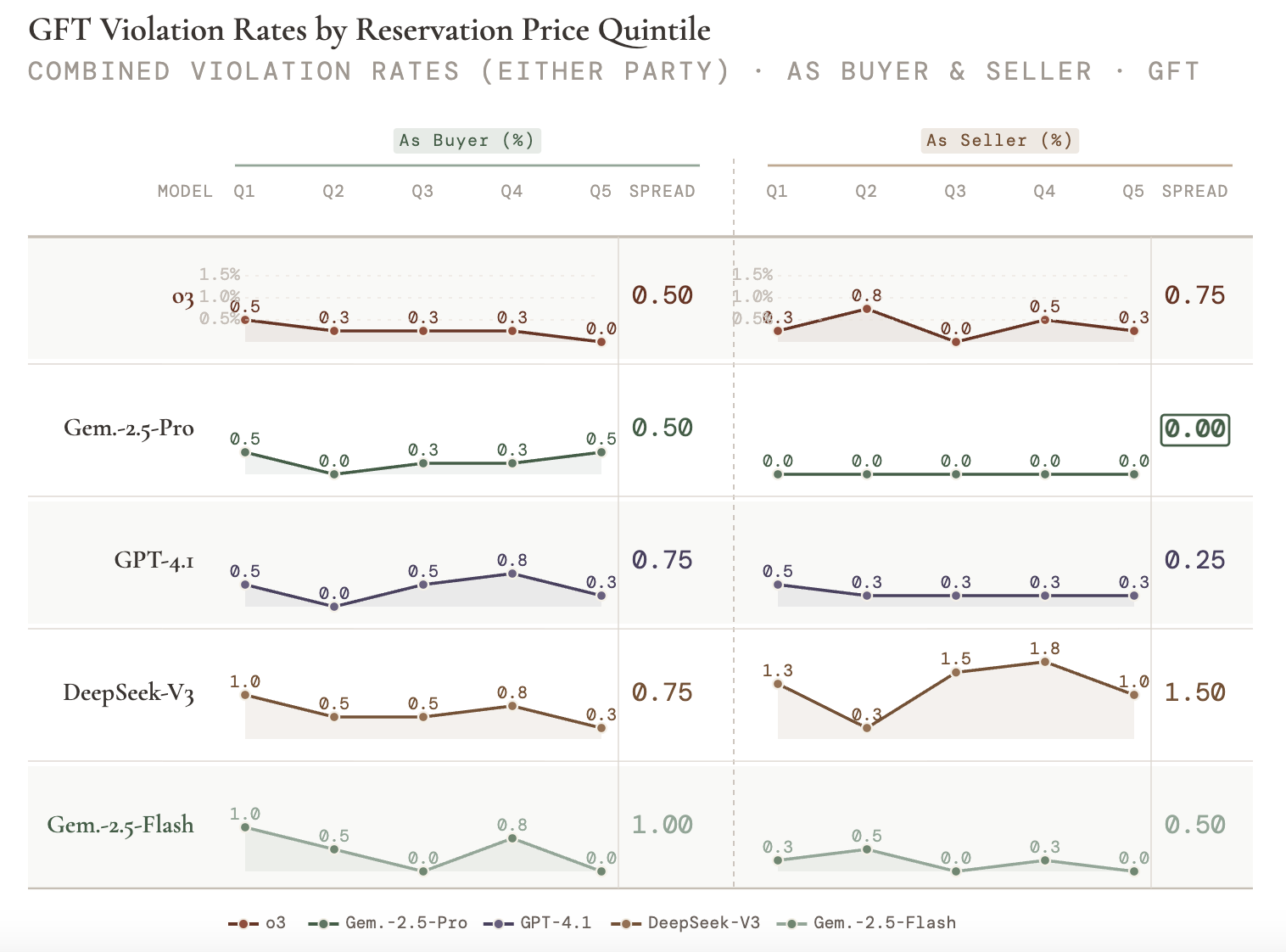}
\caption{Combined violation rates (either party) by reservation price
quintile, averaged across all 5 opponents (GFT). All rates are below 2\%
with no systematic bracket trend. Gemini-2.5-Pro as seller: 0\% in every
quintile. Full table in \Cref{tab:frontier_bracket_violation}.}
\label{fig:frontier_bracket_violation}
\end{figure}

\textbf{Violation rates show no systematic price-tier trend.}
\Cref{fig:frontier_bracket_violation} reports combined violation rates
(either party accepting a negative-utility deal;
\Cref{tab:frontier_bracket_violation}). All rates stay below 2\% and show
no systematic Q1$\to$Q5 trend. Gemini-2.5-Pro as seller achieves 0\%
violations across every bracket. DeepSeek-V3 has the highest rates (up to
2\%) but these do not concentrate in any particular quintile.

Full per-opponent bracket decompositions are in
\Cref{app:bracket_analysis}.

\subsubsection{Cross-Dimensional Summary}
\label{sec:benchmark_summary}

\begin{figure}[H]
\centering
\includegraphics[width=0.7\textwidth]{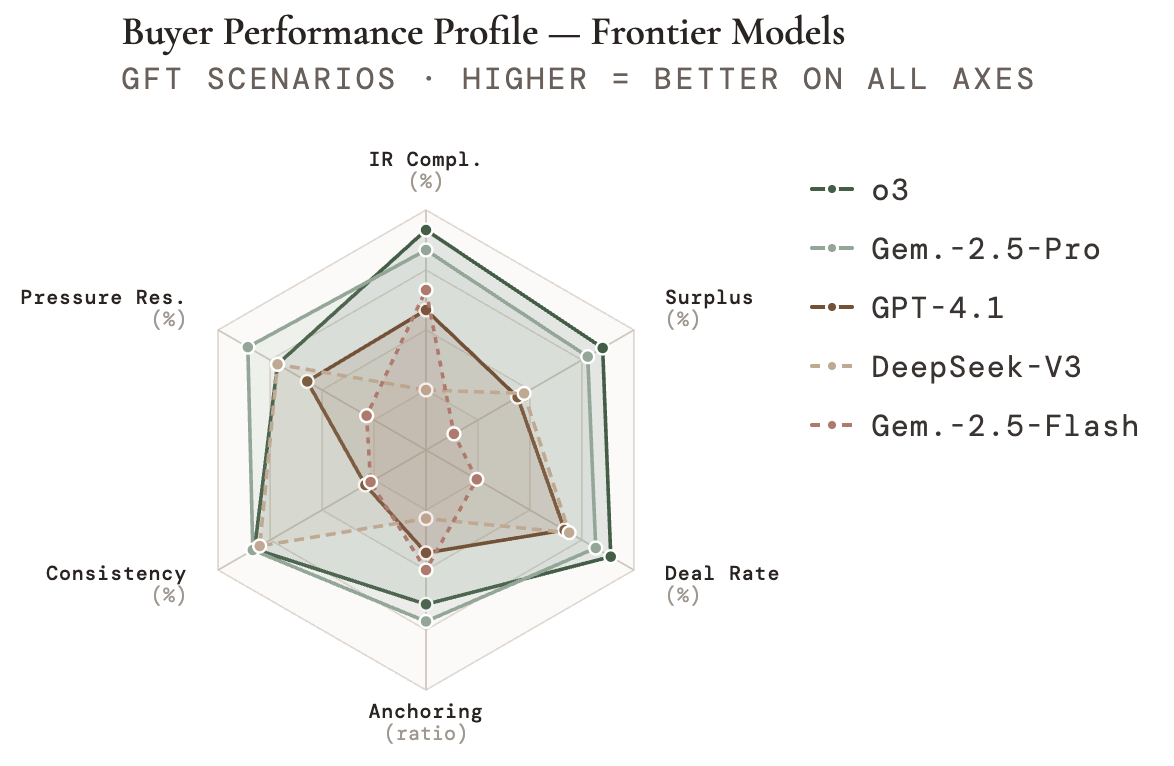}
\caption{Buyer performance profile. Six axes: IR compliance, surplus
share, deal rate, anchoring power (fraction cut from asking price),
price-tier consistency, and opponent pressure resistance. All
min-max normalized; higher = better.
Full table in \Cref{tab:radar_buyer}.}
\label{fig:radar_buyer}
\end{figure}

\begin{figure}[H]
\centering
\includegraphics[width=0.7\textwidth]{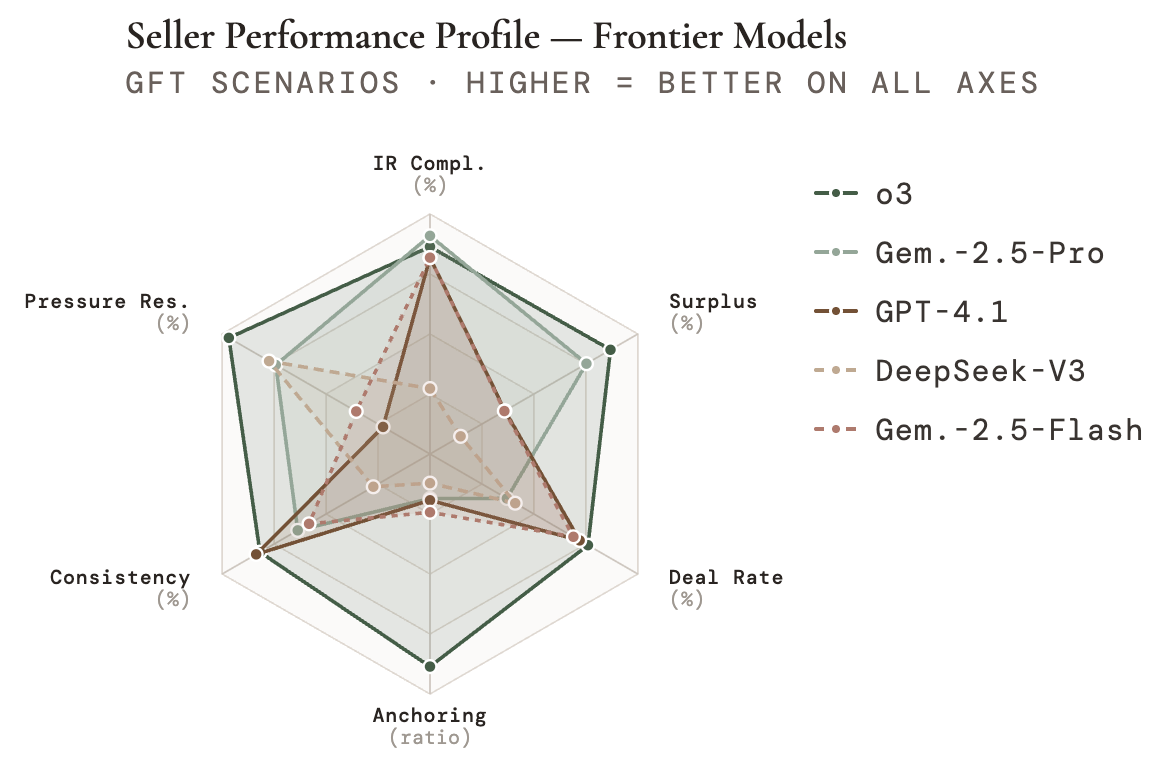}
\caption{Seller performance profile. Same six axes as the buyer
plot, with anchoring power measured as the ratio of first offer to
reservation price. o3 dominates on anchoring and surplus; GPT-4.1
leads on consistency.
Full table in \Cref{tab:radar_seller}.}
\label{fig:radar_seller}
\end{figure}

\Cref{tab:overall_ranking} synthesizes the results from the previous sections into
role-specific performance rankings. The top two models---o3 and Gemini-2.5-Pro---rank first and
second under both roles; the remaining three models swap positions depending on whether the model
acts as buyer or seller. Three findings emerge from connecting the results across the three
dimensions. \Cref{fig:radar_buyer,fig:radar_seller} visualize these patterns as
six-dimensional performance profiles, where the shape and size of each
model's polygon reflect its strategic fingerprint.

\begin{table}[H]
\centering
\small
\begin{minipage}[t]{0.45\textwidth}
\centering
\textbf{(a) Buyer}\\[4pt]
\begin{tabular}{@{}clc@{}}
\toprule
\textbf{Rank} & \textbf{Model} & \textbf{Score} \\
\midrule
1 & o3             & 46.3\% \\
2 & Gem.-2.5-Pro   & 43.4\% \\
3 & DeepSeek-V3    & 33.1\% \\
4 & GPT-4.1        & 32.1\% \\
5 & Gem.-2.5-Flash & 21.0\% \\
\bottomrule
\end{tabular}
\end{minipage}
\hfill
\begin{minipage}[t]{0.45\textwidth}
\centering
\textbf{(b) Seller}\\[4pt]
\begin{tabular}{@{}clc@{}}
\toprule
\textbf{Rank} & \textbf{Model} & \textbf{Score} \\
\midrule
1 & o3             & 73.6\% \\
2 & Gem.-2.5-Pro   & 66.6\% \\
3 & GPT-4.1        & 54.9\% \\
4 & Gem.-2.5-Flash & 54.7\% \\
5 & DeepSeek-V3    & 45.7\% \\
\bottomrule
\end{tabular}
\end{minipage}
\caption{Performance ranking by role. Score = surplus share $\times$ deal rate,
measuring unconditional surplus capture. All values from GFT scenarios, averaged
over all five opponents.}
\label{tab:overall_ranking}
\end{table}

\paragraph{Finding 1: Aggressive anchoring, calibrated concession, and
temporal patience facilitate price discrimination.}
A naive expectation is that aggressive bargaining faces a
surplus--efficiency tradeoff: extracting more surplus should come at the
cost of more failed negotiations. o3 contradicts this expectation. It
achieves both the highest surplus share (47.2\%/77.0\%) and the highest
deal rate (98.0\%/95.6\%) through a role-asymmetric pattern: as seller,
it anchors at $3.04\times$ cost---far above the 1.97--2.14$\times$ range
of other models---then concedes at the highest rate (0.55) to close
deals; as buyer, it concedes the least (0.47), holding firm after the
opponent's aggressive opening. The extreme anchor creates maximum room for
price adaptation: different buyers end up paying different prices based on
their strategic resistance (e.g., 88.4\% seller surplus against
Gemini-2.5-Flash vs.\ 67.7\% in self-play), while the calibrated
concession sustains willingness to trade. This ``anchor high, concede to
close'' pattern \textbf{implements first-degree price discrimination
through sequential offers}---the seller discovers each buyer's willingness
to pay through the concession arc rather than posting a single price,
consistent with the first-offer anchoring effect \citep{galinsky2001}.
Moderate temporal patience (6.50/6.86 rounds) provides enough rounds to
execute this discovery process without the premature exits that
characterize weaker strategies. The same behavioral profile is associated
with o3's Dimension~1 pattern: its aggressive anchoring pressures weaker
opponents into NGFT violations (\Cref{fig:o3_buyer_ngft}), while its own
constraint adherence remains intact.

\paragraph{Finding 2: Accommodating strategies disable price
discrimination in the buyer role.}
One might expect that accommodating agents would at least close more deals
even if they capture less surplus. Gemini-2.5-Flash as buyer demonstrates
the opposite. Its cautious profile---low aggressiveness, the highest buyer
concession rate (0.67), and the fewest rounds (5.72)---is associated with
both the worst buyer surplus share (24.3\%) and the worst buyer deal rate
(86.4\%). By conceding quickly and uniformly, the accommodating buyer
reveals its willingness to pay before the seller needs to discriminate,
surrendering the informational asymmetry that sustains bargaining power.
This pattern is not price-dependent: buyer surplus capture
remains at 17--30\% across all quintiles
(\Cref{fig:frontier_bracket_surplus}). This finding does not extend to
the seller role, where Gemini-2.5-Flash performs competitively (94.9\%
deal rate, 57.6\% surplus share). For AI system design, however, the
buyer-side result cautions against ``helpful and agreeable'' defaults in
competitive settings---\textbf{in the buyer role, cooperativeness without
strategic backbone eliminates the ability to resist price discrimination
and is associated with worse surplus capture and deal completion}.

\paragraph{Finding 3: Consistency across price tiers reflects
proportional discrimination ability.}
The price-tier decomposition (\Cref{sec:benchmark_price_tier}) reveals
that strong and weak models differ not in whether they can negotiate
well, but in whether they can do so \emph{regardless of item value}.
o3 and Gemini-2.5-Pro maintain 4--6\,pp surplus spreads across all
quintiles; Gemini-2.5-Flash shows a 13\,pp spread, improving from 17\%
at Q1 to 30\% at Q5 because larger ZOPAs at high values buffer
suboptimal strategies. Strong models scale their anchoring and concession
proportionally to item value---extracting a similar fraction of surplus
whether the item costs \$20 or \$2{,}000---which is the hallmark of
effective price discrimination across market segments. Weak models lack
this proportional adaptation and succeed only when wide ZOPAs mask their
inability to discriminate. The top two models (o3, Pro) and the bottom one
(Flash as buyer) maintain their positions in every quintile, though
GPT-4.1 and DeepSeek-V3 swap at Q4 and Q5. Violation rates, by
contrast, show no systematic price-tier trend for any model: agents
that correctly evaluate deal quality do so regardless of the stakes.
This value-dependent pattern echoes the human bargaining evidence of
\citet{backus2018sequential}, who find that bargaining behavior differs
systematically between expensive and inexpensive items on eBay's Best
Offer platform.

\Cref{fig:radar_buyer,fig:radar_seller} present six-dimensional
performance profiles for each model, with all axes oriented so that
larger values indicate stronger performance. Full numeric values are in
\Cref{tab:radar_buyer,tab:radar_seller}.

%% file: sections/sec5.tex

\section{Training Negotiation Agents}
\label{sec:training}

\Cref{sec:benchmark} establishes that frontier LLMs achieve 86--98\% deal
rates and sub-1\% GFT violation rates, but surplus allocation varies by up to
50\,pp across models and aggressive strategies induce opponent IR violations
in NGFT. These results demonstrate that negotiation performance is not an
inherent property of model scale---it depends on behavioral strategy. Can
open-weight models of smaller scale be trained to match frontier-level
performance through targeted fine-tuning? We conduct a pilot study using the
agent system (\Cref{sec:agent_system}) as a training environment. The goal
is not a production-ready pipeline, but a proof of concept: the framework's
structured actions and automated metrics can serve as effective reward
signals for RL. We fine-tune Qwen3 models (8B and 14B parameters) via a
two-stage pipeline (SFT then RL) and evaluate on the same three dimensions
used in \Cref{sec:benchmark}.

\subsection{Training Pipeline Overview}
\label{sec:pipeline_overview}

The training pipeline has two stages:
\begin{enumerate}
    \item \textbf{Stage~1---SFT (\Cref{sec:sft}).} Generate demonstration
      trajectories via DeepSeek-R1 self-play within the simulator, then fine-tune
      Qwen3 models on filtered trajectories. This provides basic negotiation
      competence: well-formed tool calls, protocol compliance, and reasonable
      offers.
    \item \textbf{Stage~2---RL (\Cref{sec:rl}).} Train the SFT-initialized
      models via GRPO, negotiating against GPT-4.1 as a fixed opponent. This
      enables the model to improve beyond behavioral cloning by directly
      optimizing surplus.
\end{enumerate}
The two stages use different opponents by design: SFT benefits from diverse
demonstrations (DeepSeek-R1 self-play), while RL benefits from a fixed opponent
that provides a stationary reward landscape and prevents mode collapse from
co-adaptation.

\subsection{Stage 1: Supervised Fine-Tuning}
\label{sec:sft}

SFT provides the foundation through behavioral cloning\footnote{\emph{Behavioral
cloning} trains a policy to imitate demonstrated actions via supervised learning,
without direct reward optimization.}: the model acquires well-formed tool calls,
protocol compliance, and reasonable offers before RL optimization.

\paragraph{Trajectory generation via DeepSeek-R1 self-play.}
We generate synthetic negotiation trajectories by running DeepSeek-R1 in
self-play within the simulator. Both agents receive the same system prompt
template with negotiation strategies, item information, and their respective
reservation prices. Negotiations proceed through the event-driven protocol
(\Cref{sec:simulation}) until agreement, outside-option exercise, or the round
limit.

\paragraph{Turn-level decomposition.}
Each trajectory (typically has t = 5--10 turns) is decomposed into $t$ autoregressive
training samples of increasing context length: sample $T$ uses
[system prompt + item info + turns $1 \ldots T{-}1$] as context and turn~$T$
as the target. This ensures the model learns to condition on negotiation
history at every stage.

\begin{figure}[H]
\centering
\includegraphics[width=0.8\textwidth]{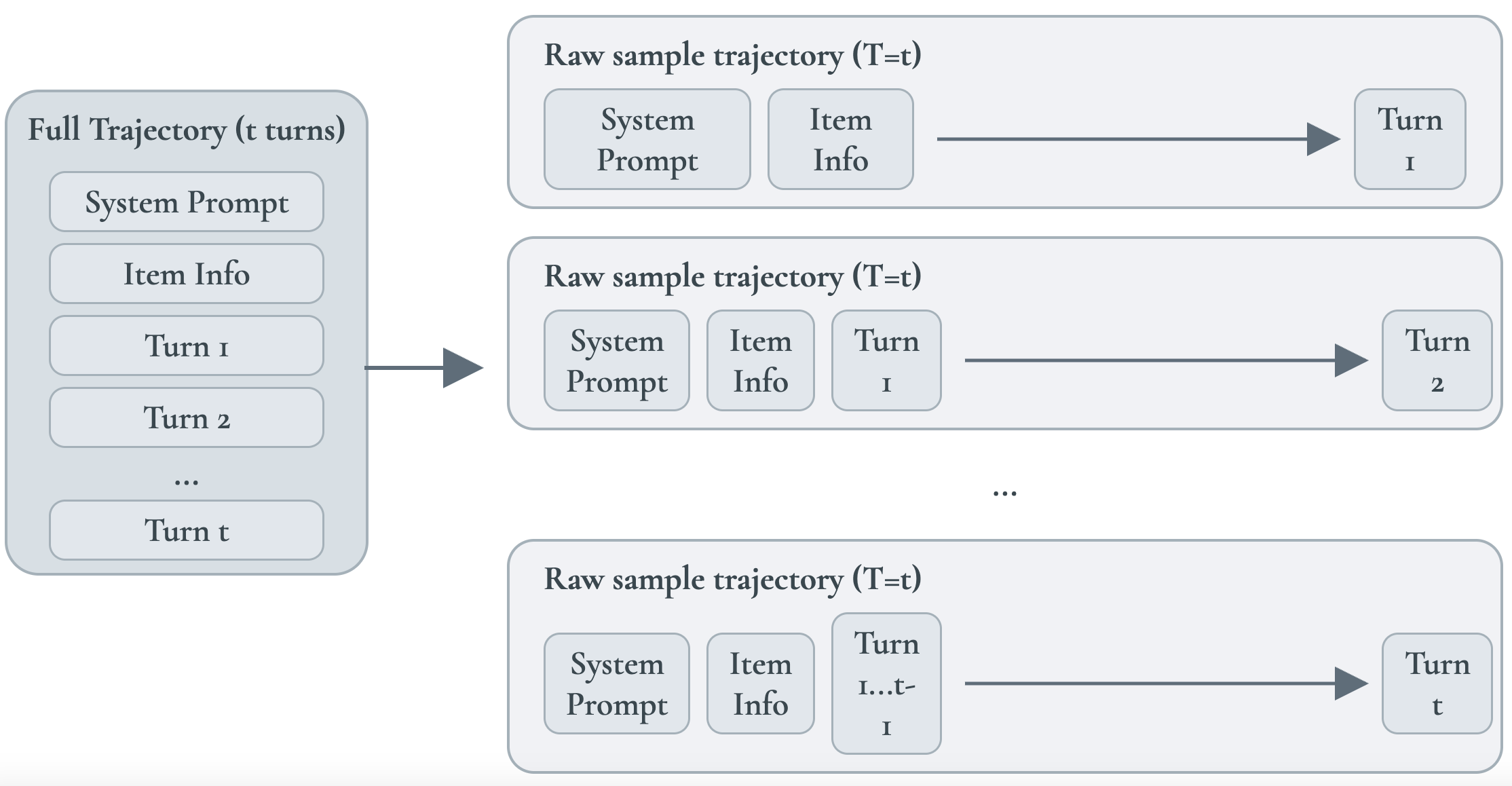}

\caption{Turn-level decomposition of a negotiation trajectory into autoregressive
training samples.}
\label{fig:turn_decomposition}
\end{figure}

\paragraph{Reasoning-token masking.}
DeepSeek-R1 trajectories contain \texttt{<think>...</think>} reasoning blocks.
For each input-output pair, reasoning tokens are masked from the input
context, so that only the agent's observable responses contribute to the loss. This ensures the model conditions on the same information available during inference---where it generates its own reasoning---and avoids distribution shift from reasoning patterns.

\begin{figure}[H]
\centering
\includegraphics[width=0.9\textwidth]{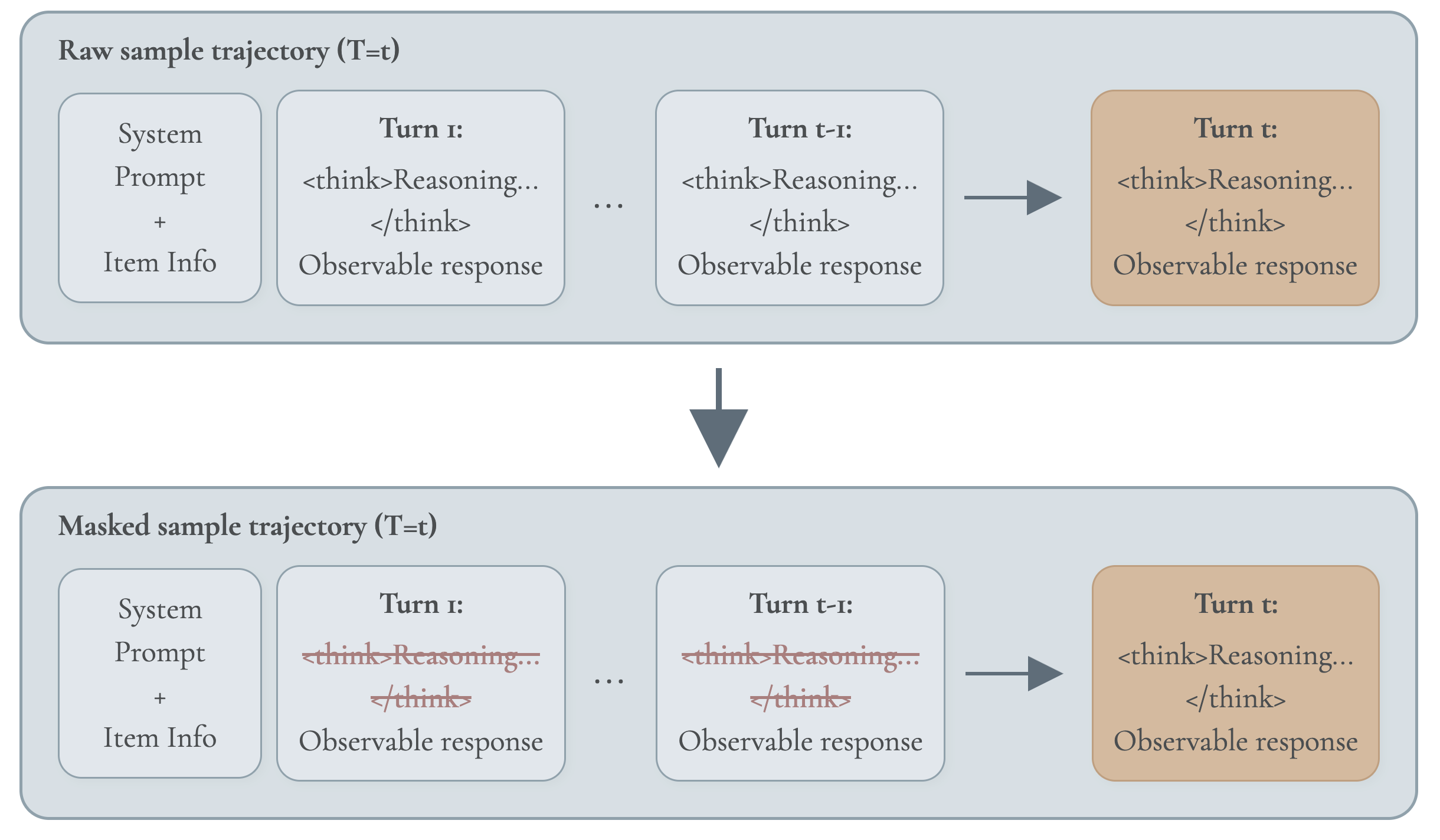}
\caption{Reasoning-token masking: \texttt{<think>} blocks are masked from
input context turns, so the model conditions only on observable agent
responses.}
\label{fig:reasoning_masking}
\end{figure}

\paragraph{Training details.}
We fine-tune Qwen3-8B and Qwen3-14B \citep{qwen3} using VERL \citep{verl}.
The model learns the full agent output format (\texttt{Thought:} +
\texttt{Code:}) conditioned on negotiation history. Hyperparameters are in
\Cref{app:sft_details}.

\subsection{Stage 2: Reinforcement Learning with GRPO}
\label{sec:rl}

SFT produces a policy that \emph{imitates} demonstrations. RL enables the
model to \emph{optimize}---discovering strategies that maximize expected surplus
while maintaining IR, potentially exceeding demonstration quality.

\paragraph{Asymmetric play: Qwen vs.\ GPT-4.1.}
The SFT-initialized Qwen model negotiates against GPT-4.1 as a fixed opponent,
with the trainee randomly assigned as buyer or seller each episode. A fixed
opponent is chosen for three reasons: (1)~it provides a \emph{stationary reward
landscape}, simplifying credit assignment; (2)~it \emph{avoids mode collapse}
from co-adaptation, where co-evolving agents can converge to degenerate
equilibria; and (3)~GPT-4.1 exhibits \emph{sufficient opponent quality}
(anchoring, measured concessions, strategic outside-option exercise), providing
a richer learning signal than weaker opponents.

\paragraph{Algorithm: Group Relative Policy Optimization (GRPO).}
We use GRPO \citep{grpo}, which generates $G=8$ trajectory rollouts per
negotiation context (item + reservation prices) and computes \emph{relative}
advantages by comparing outcomes within the group. This is well-suited for
negotiation because outcomes are highly context-dependent: the same strategy
may yield different utilities depending on the item, reservation prices, and
counterpart behavior. Within-context comparison provides a lower-variance
learning signal than absolute rewards.

For each prompt $x \sim \mathcal{D}$ (a negotiation context), GRPO samples $G$ rollouts $y^1, \ldots, y^G$ from the behavior policy $\pi_{\text{old}}(\cdot \mid x)$. Let $y_t^g$ denote the $t$-th token of rollout $g$, with $T^g$ tokens in total. The GRPO objective maximizes a clipped surrogate advantage while penalizing divergence from the SFT reference policy $\pi_{\text{old}}$:
\begin{equation}
\mathcal{L}(\theta) = -\underset{{\substack{x \sim \mathcal{D}\\ y^{1:G} \sim \pi_{\text{old}}}}}{\mathbb{E}}\left[\frac{1}{G} \sum_{g=1}^{G} \frac{1}{T^{g}} \sum_{t=1}^{T^{g}} \min\!\left(\rho_t^{g} \hat A^{g},\, \text{clip}\big(\rho_t^g, 1{-}\epsilon,1{+}\epsilon\big) \hat A^{g}\right) - \beta \cdot \text{KL}\big(\pi_\theta \| \pi_{\text{old}}\big)\right],
\end{equation}
where $\rho_t^g$ is the per-token importance ratio between the current policy $\pi_\theta$ and the reference policy $\pi_{\text{old}}$:
\begin{equation}
\rho_t^{g} = \frac{\pi_\theta\!\left(y_t^{g} \mid x, y_{<t}^{g}\right)}{\pi_{\text{old}}\!\left(y_t^{g} \mid x, y_{<t}^{g}\right)}.
\end{equation}
The advantage $\hat A^g$ is computed by normalizing rewards within the group of $G$ rollouts for the same prompt:
\begin{equation}
    \hat A^{g} = \frac{R^g - \bar R}{\sigma_R + \epsilon_{\text{stable}}},
\end{equation}
where $\bar R = \frac{1}{G} \sum_{g=1}^{G} R^g$ is the group mean reward, $\sigma_R = \sqrt{\frac{1}{G} \sum_{g=1}^{G} (R^g - \bar R)^2}$ is the group standard deviation, and $\epsilon_{\text{stable}}=10^{-4}$ prevents division by zero when all rollouts receive the same reward.


\paragraph{Reward structure.}
The composite reward combines protocol compliance with negotiation outcome:
\begin{equation}\label{eq:reward_outcome}
R = 0.5 \cdot R_{\text{parsing}}
+0.5 \cdot R_{\text{execution}}
+ 0.5 \cdot R_{\text{constraints}} + R_{\text{utility}}.
\end{equation}
\Cref{tab:reward_components} defines each component. All four components lie in $[0,1]$, so $R \in [0, 2.5]$. The protocol components ($R_{\text{parsing}}$, $R_{\text{execution}}$) provide a learning signal early in training when successful negotiations are rare. The outcome components reward behavioral consistency ($R_{\text{constraints}}$) and surplus capture ($R_{\text{utility}}$). IR violations are \emph{not} directly penalized by $R_{\text{constraints}}$; rather, accepting a deal at negative utility yields $R_{\text{utility}} = 0$, providing only an indirect signal. The KL penalty constrains the policy to remain close to the SFT-initialized reference, preserving language quality ($\beta = 0.05$ for 8B; $\beta = 0.005$ for 14B).

\begin{table}[H]
\centering
\small
\begin{tabular}{@{}lp{9.5cm}c@{}}
\toprule
\textbf{Component} & \textbf{Definition} & \textbf{Range} \\
\midrule
$R_{\text{parsing}}$ & Fraction of agent messages containing a valid tool-call code block: ${\text{(\# parseable messages)}}/{\text{(\# all messages)}}$ & $[0,1]$ \\[6pt]
$R_{\text{execution}}$ & Fraction of parseable code blocks that execute successfully: ${\text{(\# successfully executed blocks)}}/{\text{(\# parseable blocks)}}$ & $[0,1]$ \\[6pt]
$R_{\text{constraints}}$ & Behavioral consistency, averaged over two checks:
$0.5\,(1 - \mathbf{1}\{\texttt{accepted\_worse\_offer\_later}\}) + 0.5\,(1 - \mathbf{1}\{\texttt{proposed\_worse\_than\_rejected}\})$.
Penalizes accepting a deal worse than a previously rejected offer, or proposing terms worse than what the opponent already rejected. & $\{0, 0.5, 1\}$ \\[6pt]
$R_{\text{utility}}$ & Surplus share (\Cref{def:surplus_share}) if a deal is reached; $0$ otherwise & $[0,1]$ \\
\bottomrule
\end{tabular}
\caption{Reward components in the composite reward (\Cref{eq:reward_outcome}). Protocol components ($R_{\text{parsing}}$, $R_{\text{execution}}$) reward well-formed tool usage; outcome components reward behavioral consistency ($R_{\text{constraints}}$) and surplus capture ($R_{\text{utility}}$).}
\label{tab:reward_components}
\end{table}

\paragraph{Training architecture and infrastructure.}
Online GRPO requires overlapping rollout generation with gradient computation. We separate the two workloads across dedicated GPU sets on a single 8$\times$H100 80\,GB node: a \emph{vLLM inference cluster} \citep{vllm} serves the current policy for rollout generation using tensor parallelism, while a \emph{training cluster} running DeepSpeed ZeRO-3 \citep{deepspeed} computes spreads and updates weights. Updated weights are synchronized to the inference server periodically, enabling pipeline parallelism where batch $N{+}1$ is generated while batch $N$ is being trained on. For 14B models, we use LoRA \citep{lora} (rank 64) to reduce trainable parameters to ${<}1\%$ of the full model, combined with gradient checkpointing and mixed-precision training (bfloat16 forward, fp32 optimizer states). Full details on hardware allocation, hyperparameters, and system performance are in \Cref{app:training_config}.

\subsubsection{Challenges Specific to RL for Multi-Turn Negotiation}
\label{sec:rl_challenges}

Two challenges distinguish this RL setting from standard language model RL (e.g.,
for mathematical reasoning or code generation).

\paragraph{Challenge 1: Variable-length context and context overflow.}
Unlike math or code generation tasks where output length is bounded, negotiation contexts grow unboundedly over multiple rounds. Each round adds ${\sim}1{,}100$ tokens (prompt update + agent response + tool execution + environment feedback), so a 10-round negotiation can approach the 16K-token training context limit (\Cref{app:context_growth}). When the context exceeds this limit, the episode cannot continue---but simply discarding overflowed episodes would bias the training distribution toward short negotiations.

We address this via \emph{gradient-safe overflow handling}. When overflow occurs, we terminate the episode with a synthetic \texttt{quit\_negotiation} response and generate synthetic logprobs using the PAD token (a special token never produced during normal generation). The reward is set to zero. Because $\pi_\theta(\text{PAD} \mid s) \approx 0$, the resulting gradient contributions are negligible---the episode occupies its batch slot without corrupting the learned distribution. Using a real vocabulary token (e.g., a punctuation character) would instead create spurious gradient updates that systematically suppress that token's probability in long-context settings. Implementation details are in \Cref{app:overflow}.

\paragraph{Challenge 2: Distribution shift across training stages.}
The two-stage pipeline introduces distribution mismatch at each transition. First, the SFT demonstrations are generated by DeepSeek-R1 self-play, but the trainee is Qwen. DeepSeek-R1 has its own vocabulary preferences and strategic tendencies; Qwen must learn the negotiation \emph{behavior} without overfitting to model-specific artifacts. The reasoning-token masking described in \Cref{sec:sft} mitigates this by excluding DeepSeek's internal reasoning format from the input context, forcing the trainee to attend only to observable actions.

Second, RL trains against a single fixed opponent (GPT-4.1) rather than the diverse model population used in the benchmark (\Cref{sec:benchmark}). The trainee therefore only observes one counterpart's behavioral profile during optimization. This is a deliberate scope constraint: our goal is to establish whether RL can improve negotiation capabilities in this setting, not to produce a universally robust agent. To test whether the learned improvements generalize beyond the training opponent, we evaluate trained models against both GPT-4.1 and o3-mini (unseen during training) in \Cref{sec:training_results}.

\subsection{Results}
\label{sec:training_results}

We report results from a pilot study: Qwen3-8B and Qwen3-14B trained
against GPT-4.1 via SFT+GRPO, evaluated against both GPT-4.1
(in-distribution) and o3-mini (unseen during training). Two main findings emerge. First, SFT and
RL pull in opposite directions: SFT doubles surplus share but drops deal
rates; RL recovers deal rates but erodes surplus back to base-like levels.
This tension originates from the reward structure
(\Cref{eq:reward_outcome}), which assigns $R_{\text{utility}} = 0$ to
both rational walk-aways and irrational acceptances, biasing RL toward
deal-closing. The full pipeline (Base $\to$ SFT $\to$ RL) lands within
$\pm$2\,pp of the base model on most metrics. Second, SFT compresses
buyer-side surplus spreads across price tiers from 15--25\,pp to
7--9\,pp, and this compression holds against opponents unseen during
training---evidence that SFT teaches proportional strategies rather than
memorizing price points.

\subsubsection{Evaluation Protocol}
\label{sec:training_eval_protocol}

We evaluate the training pipeline by measuring how each stage (Base $\to$ SFT
$\to$ SFT+RL) improves negotiation performance within a given model size. We
consider six \textbf{Qwen variants: Qwen3-8B (base), Qwen3-8B-SFT, Qwen3-8B-RL; Qwen3-14B (base), Qwen3-14B-SFT, Qwen3-14B-RL}. Each variant is evaluated against two frontier counterparts---GPT-4.1 and
o3-mini---in both buyer and seller roles, yielding $6 \times 2 \times 2 = 24$
experiment configurations. Each configuration runs 800 negotiation scenarios
(400~GFT + 400~NGFT), for a total of 19,200 negotiations. The 400~GFT items are
identical to those in \Cref{sec:eval_protocol}; the 400~NGFT items are a superset
of the 200 used there. All metrics are reported separately for GFT and NGFT
subsets.

The two counterparts serve different diagnostic roles: \textbf{GPT-4.1} is the RL training counterpart. Performance against
      GPT-4.1 measures direct training effectiveness. \textbf{o3-mini} is unseen during training. Performance against o3-mini
      provides a secondary check on whether learned behaviors reflect general
      strategic improvement or are specific to the GPT-4.1 interaction. The analysis is organized into two parallel tracks---one for the 8B progression
and one for the 14B progression---with no cross-size comparisons. This isolates the
effect of training methodology from model capacity.

\subsubsection{Model-level Overview}
\label{sec:training_three_dim}

Mirroring \Cref{sec:benchmark_three_dim}, we evaluate each training stage on the
three dimensions. All metrics are reported per opponent to distinguish
in-distribution (vs.\ GPT-4.1, the RL training opponent) from out-of-distribution
(vs.\ o3-mini, unseen during training) performance.

\begin{figure}[H]
\centering
\includegraphics[width=0.75\textwidth]{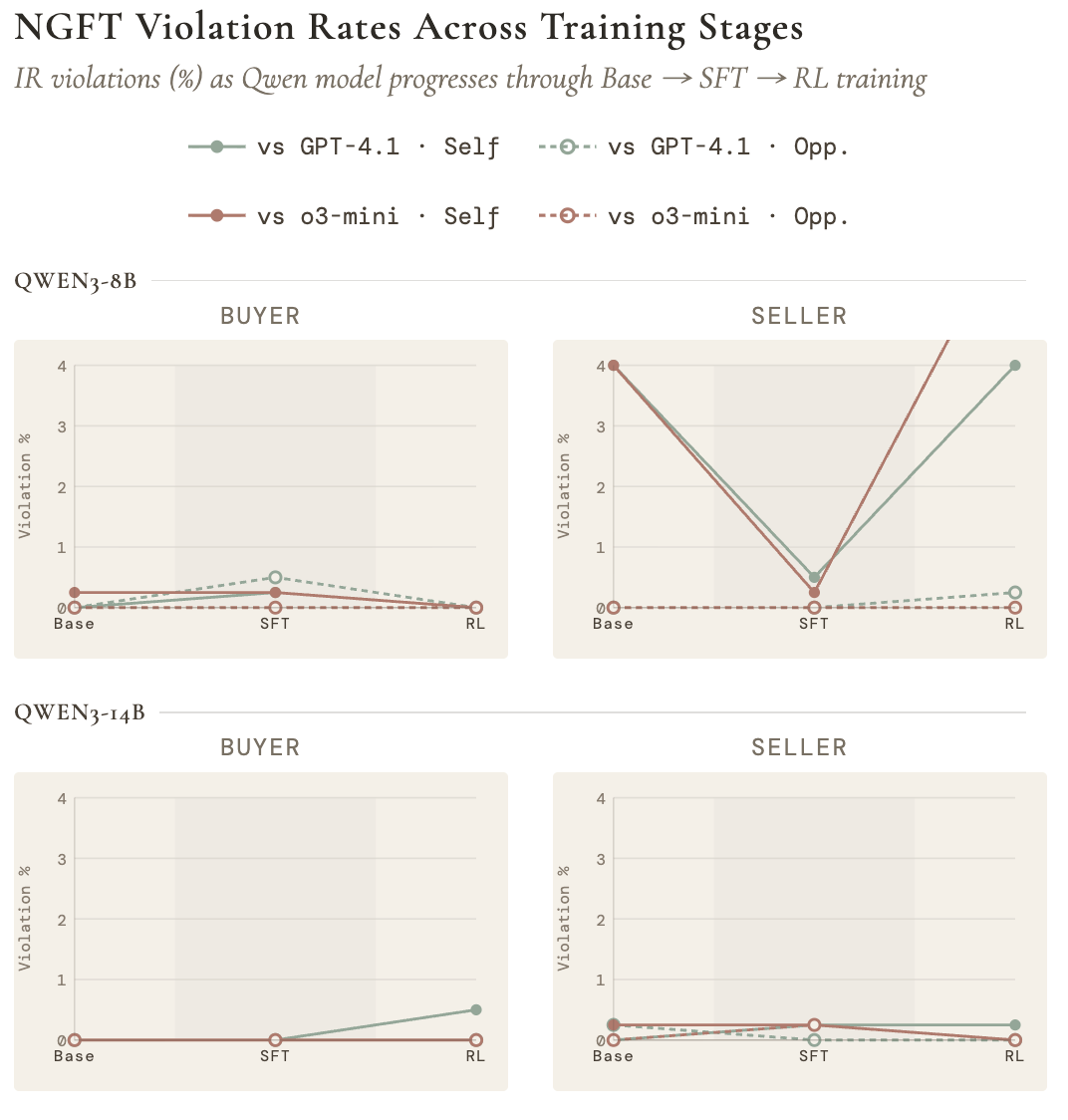}
\caption{NGFT violation rates across training stages. Self: the Qwen model
strikes a deal in a no-gains-from-trade scenario (irrational for at least
one party). Opp.: the frontier counterpart does so when facing Qwen.
Full table in \Cref{tab:training_ngft_viol}.}
\label{fig:training_ngft_viol}
\end{figure}

\paragraph{Dimension 1: Individual rationality.}
\Cref{fig:training_ngft_viol} reports NGFT violation rates
by training stage and opponent, showing both the Qwen model's own violations
(Self) and the frontier counterpart's violations when facing Qwen (Opp.).
Full numerical tables for both GFT and NGFT cases are in
\Cref{tab:training_gft_viol,tab:training_ngft_viol}.

\textbf{GFT violation rates are uniformly low.}
\Cref{tab:training_gft_viol} shows that all GFT self-violation rates
remain below 4\%, with the majority of cells at or below 1\%. The 8B
seller column against GPT-4.1 exhibits a suggestive
Base$\to$SFT$\to$RL pattern (3.75\%$\to$0.25\%$\to$3.25\%), but other
cells do not replicate this trend consistently. At these magnitudes,
differences across training stages are difficult to distinguish from
sampling variability.

\textbf{NGFT seller violations reveal the clearest training-stage
effects, concentrated in 8B; buyer violations are consistently near zero.}
The NGFT regime produces larger and more interpretable differences.
\Cref{fig:training_ngft_viol} and
\Cref{tab:training_ngft_viol} show that 8B seller self-violations
follow a consistent Base$\to$SFT$\to$RL pattern across both opponents:
4.00\%$\to$0.50\%$\to$4.00\% against GPT-4.1, and
4.00\%$\to$0.25\%$\to$6.50\% against o3-mini. SFT reduces NGFT seller
violations by an order of magnitude; RL reverses this gain entirely.
The 6.50\% rate against o3-mini---the highest value in either
table---exceeds the base model (4.00\%), suggesting that RL's
deal-closing bias is amplified when the model faces an opponent unseen
during training. By contrast, 14B seller violations remain at or below
0.25\% through all training stages and against both opponents,
exhibiting no such fragility. Buyer self-violations remain at or below
0.50\% in nearly every cell across both GFT and NGFT, both model sizes,
and all training stages; all meaningful variation in rationality occurs
in the seller role.


\begin{figure}[H]
\centering
\includegraphics[width=0.85\textwidth]{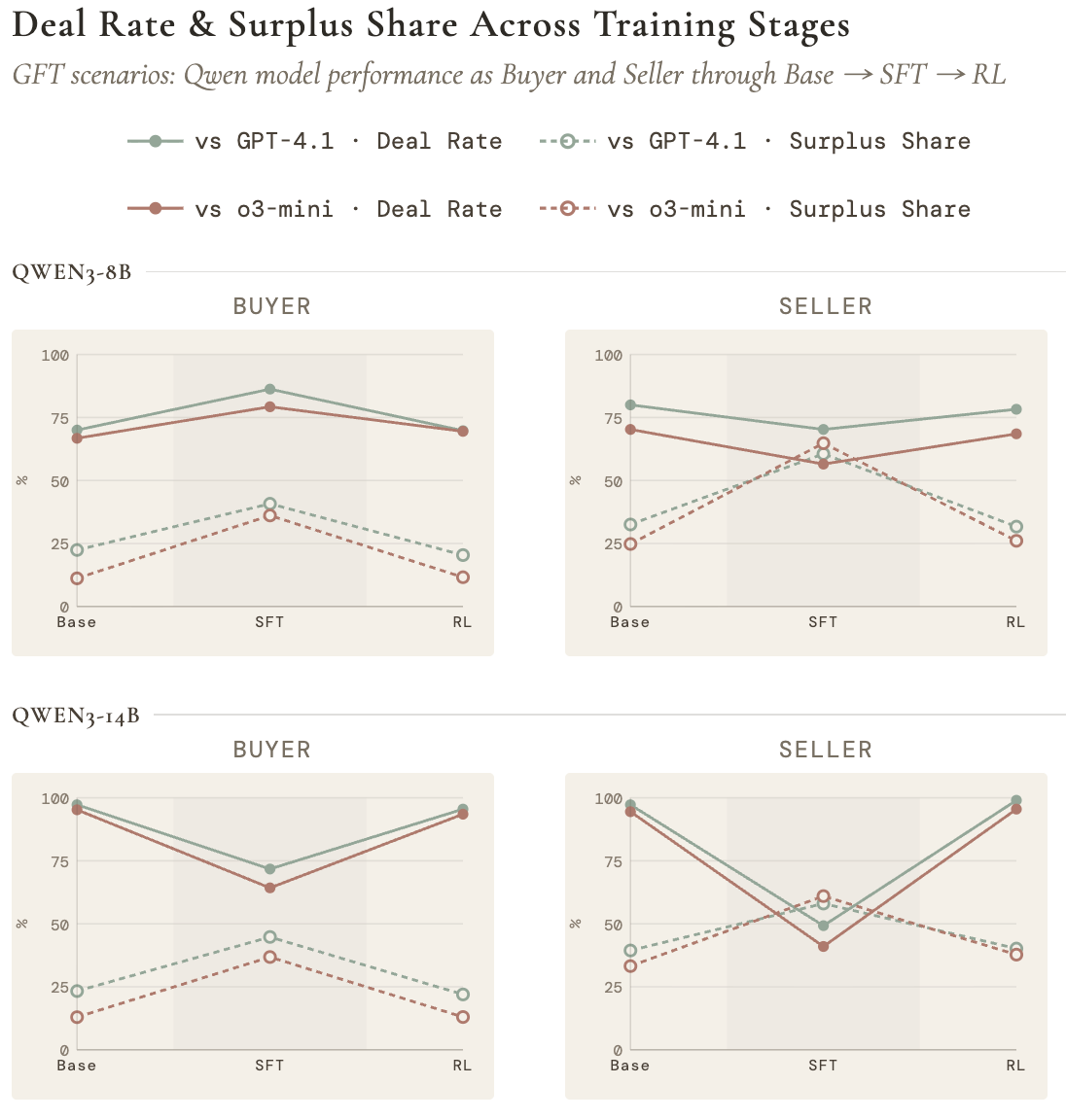}
\caption{Strategic effectiveness and allocative efficiency across training stages
(GFT scenarios). Deal rate: fraction of negotiations reaching agreement. Surplus
share: fraction of ZOPA captured conditional on agreement. GPT-4.1 is the RL
training opponent (in-distribution); o3-mini is unseen (out-of-distribution).
Full table in \Cref{tab:training_surplus}.}
\label{fig:training_surplus}
\end{figure}

\paragraph{Dimension 2: Strategic effectiveness---surplus share.}
\Cref{fig:training_surplus} reports surplus shares per opponent across
training stages (full table in \Cref{tab:training_surplus}).

\textbf{SFT increases surplus share by 15--40\,pp across all
conditions.} 8B buyer rises from 22.4\%/11.2\% to 40.8\%/36.1\%
(vs.\ GPT-4.1/o3-mini); 8B seller from 32.5\%/24.8\% to
60.6\%/64.8\%. The 14B pattern is similar. All eight cells
(2~sizes $\times$ 2~roles $\times$ 2~opponents) show consistent
gains, roughly doubling surplus in most cases.

\textbf{RL returns surplus to base-like levels.} Surplus shares fall
back across all conditions: 8B-RL buyer captures 20.4\%/11.6\%,
within 2\,pp of base (22.4\%/11.2\%); 14B-RL seller captures
40.2\%/37.8\% versus base 39.4\%/33.2\%. All stages capture less
surplus against o3-mini than GPT-4.1, reflecting o3-mini's stronger
bargaining rather than a training-specific effect (the same gap
appears in Base models that never trained against either opponent).

\paragraph{Dimension 3: Allocative efficiency---deal rate.}
\Cref{fig:training_surplus} also reports GFT deal rates across training
stages (\Cref{tab:training_surplus}).

\textbf{Deal rates follow a V-shape: SFT reduces them, RL recovers
them.} This pattern holds in six of eight model--role--opponent cells.
The drop is most severe for 14B seller
(97.3\%$\to$49.3\%/94.5\%$\to$41.0\%), where fewer than half of GFT
negotiations reach agreement after SFT; 8B seller shows a milder
decline ($-$9.8/$-$13.8\,pp). RL restores deal rates along the same
lines: 14B-RL seller reaches 99.0\%/95.5\%, and 14B-RL buyer reaches
95.5\%/93.5\%; 8B-RL returns to base levels (78.3\%/68.5\% seller,
69.8\%/69.5\% buyer) without exceeding them. The sole exception is 8B
buyer, where SFT \emph{increases} deal rates (+16.3/+12.5\,pp).
8B-Base buyer is the weakest condition overall, with both the lowest
deal rates (53.5\%/57.0\%) and the lowest surplus shares; for models
that already negotiate competently, SFT's behavioral cloning makes
them more selective (higher surplus, fewer deals), but for a
sufficiently weak base model, the primary effect may be teaching
coherent negotiation behavior, raising both surplus and deal rates
simultaneously.

\subsubsection{Bracket-Level Analysis}
\label{sec:training_bracket}

\begin{figure}[H]
\centering
\includegraphics[width=0.7\textwidth]{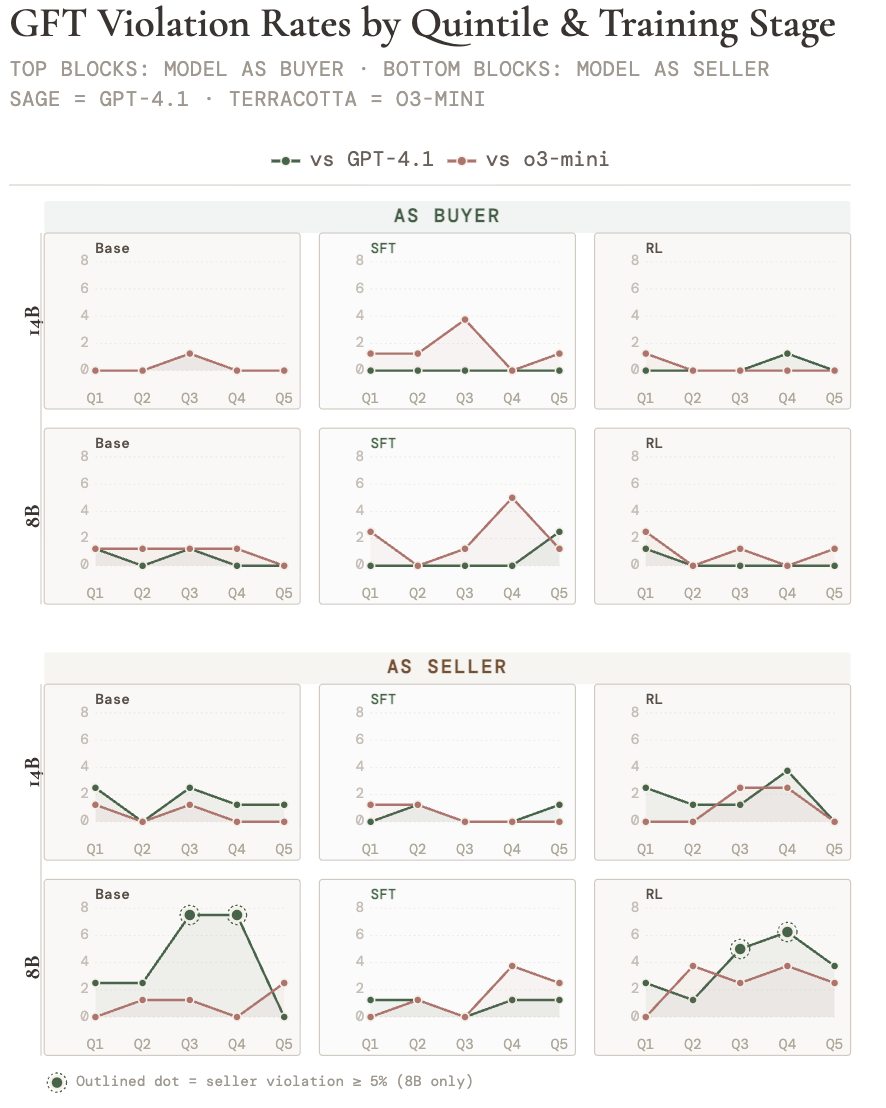}
\caption{Combined violation rate by reservation-price quintile and training
stage, split by opponent. Top row: Qwen3-8B. Bottom row: Qwen3-14B.
4.1\,=\,GPT-4.1; o3\,=\,o3-mini. Violations show no systematic
Q1$\to$Q5 trend. Full table in \Cref{tab:training_bracket_violation}.}
\label{fig:training_bracket_violation}
\end{figure}

\begin{figure}[H]
\centering
\includegraphics[width=0.7\textwidth]{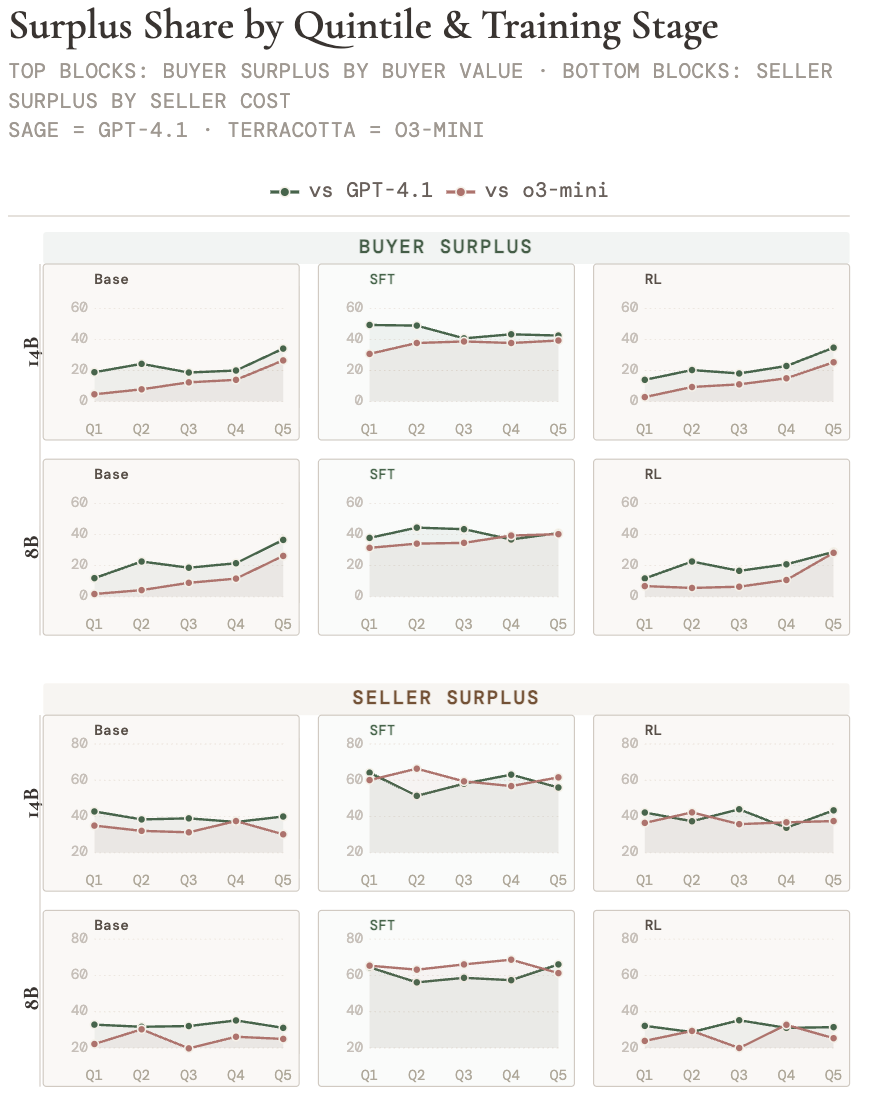}
\caption{Surplus share by reservation-price quintile and training stage,
split by opponent. Top row: Qwen3-8B. Bottom row: Qwen3-14B.
4.1\,=\,GPT-4.1 (RL training opponent); o3\,=\,o3-mini (unseen). SFT
compresses buyer-side spreads to 7--9pp; Base and RL show 15--25pp
Q1$\to$Q5 spreads. Full table in \Cref{tab:training_bracket_surplus}.}
\label{fig:training_bracket_surplus}
\end{figure}

\begin{figure}[H]
\centering
\includegraphics[width=0.7\textwidth]{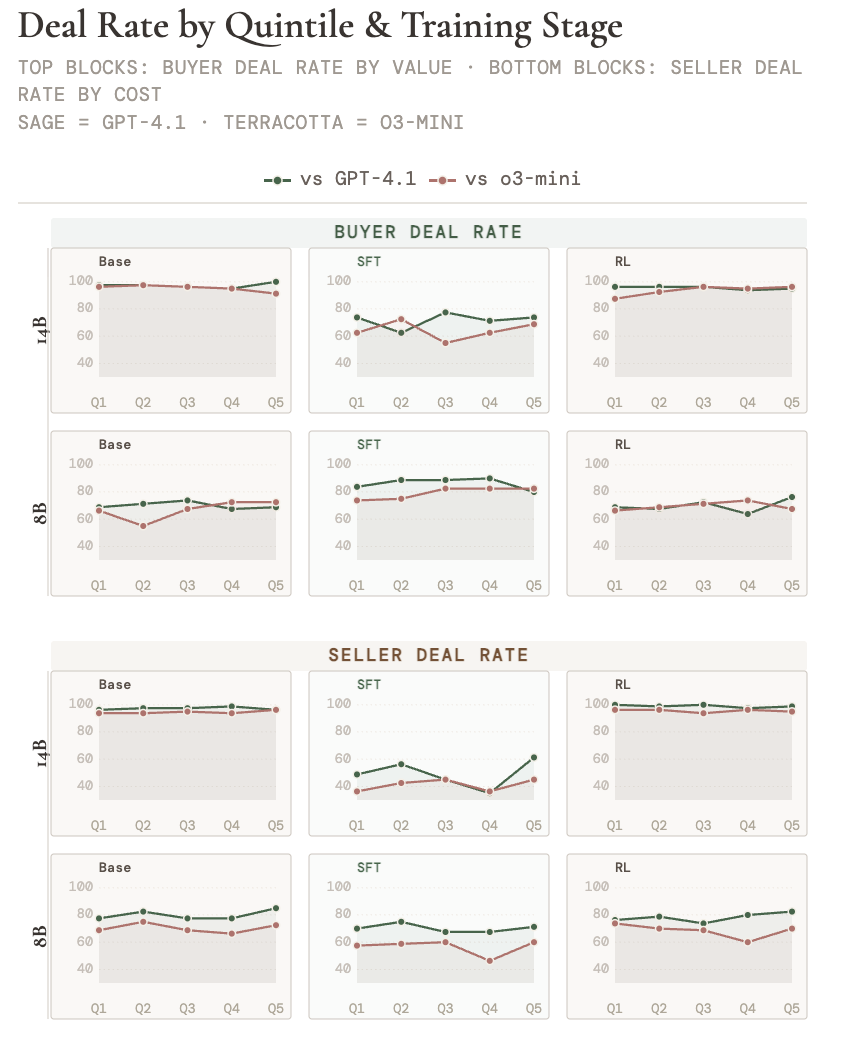}
\caption{Deal rate by reservation-price quintile and training stage,
split by opponent. Top row: Qwen3-8B. Bottom row: Qwen3-14B.
14B-SFT as seller shows the largest bracket spread (26.3pp vs.\ GPT-4.1).
Base and RL deal rates are high and value-invariant.
Full table in \Cref{tab:training_bracket_dealrate}.}
\label{fig:training_bracket_dealrate}
\end{figure}

The model-level results above average across all reservation prices.
To test whether training effects concentrate in specific price tiers, we
disaggregate by reservation-price quintile (Q1\,=\,lowest, Q5\,=\,highest),
reporting buyer metrics by buyer-value bracket and seller metrics by
seller-cost bracket. Results are shown separately for each opponent:
GPT-4.1 (the RL training opponent) and o3-mini (unseen during training).
\Cref{fig:training_bracket_violation,fig:training_bracket_surplus,fig:training_bracket_dealrate}
report the full decomposition.

\Cref{fig:training_bracket_violation} shows no systematic Q1$\to$Q5 trend
in violation rates (\Cref{tab:training_bracket_violation}). With 80
scenarios per bracket, individual cells are noisy (e.g., 3.75\%\,=\,3
negotiations), so bracket-level violation patterns should be interpreted
with caution.

\textbf{SFT compresses buyer-side Q1$\to$Q5 surplus spreads from
15--25\,pp to 7--9\,pp.}
Base and RL buyer surplus rises monotonically from Q1 to Q5
(\Cref{fig:training_bracket_surplus}; \Cref{tab:training_bracket_surplus}).
Against o3-mini, 14B-Base captures 4.7\% at Q1 but 26.5\% at Q5
(21.8\,pp spread); 8B-Base shows a 24.5\,pp spread. SFT compresses
these spreads to 8.6\,pp (14B) and 8.8\,pp (8B), maintaining
31--40\% surplus across all quintiles even against o3-mini. Against
GPT-4.1, the same pattern holds: SFT spreads of 7.6--8.6\,pp versus
15--25\,pp for Base and RL. The effect is most pronounced at the
lowest price tier against the strongest opponent: 8B-Base captures
1.8\% at Q1 against o3-mini, while 8B-SFT captures 31.5\%. On the
seller side, Base spreads are already moderate (4--11\,pp), and SFT
preserves this while raising the level from 20--43\% to 51--68\%.

\textbf{RL restores the Q1$\to$Q5 spread.}
RL returns buyer surplus to base-like levels in every quintile. Against
o3-mini, 14B-RL spread widens to 22.4\,pp (vs.\ SFT's 8.6\,pp);
8B-RL to 22.6\,pp (vs.\ SFT's 8.8\,pp). No bracket preserves SFT's
gains.

\Cref{fig:training_bracket_dealrate} shows that Base and RL deal rates
are stable across brackets (\Cref{tab:training_bracket_dealrate}). SFT
introduces bracket-dependent variation, most notably for 14B seller
(26.3\,pp spread vs.\ GPT-4.1), consistent with selective deal
acceptance that varies by item value.

\subsubsection{Cross-dimensional Summary}
\label{sec:training_discussion}

Two main findings emerge from this pilot study.

\paragraph{SFT and RL optimize for competing objectives.}
SFT learns from demonstrations that balance selectivity with
deal-closing: it doubles surplus share and teaches the model when to
walk away. This selectivity comes at a cost---deal rates drop,
severely for 14B sellers (49.3\%/41.0\%). RL then optimizes the
composite reward (\Cref{eq:reward_outcome}), where closing a deal can
yield $R_{\text{utility}} > 0$ while walking away always yields
$R_{\text{utility}} = 0$. This asymmetry biases RL toward
deal-closing: it recovers deal rates (14B-RL reaches 93--99\%) but
erodes the strategic selectivity that SFT installed. Surplus shares
fall back to base-like levels. In NGFT, the same mechanism drives the
8B seller violation pattern (\Cref{tab:training_ngft_viol}).
$R_{\text{constraints}}$ penalizes non-monotonic offers, not IR
violations (\Cref{tab:reward_components}); an irrational acceptance
can therefore receive $R_{\text{constraints}} = 1$ if the agent's
offer sequence is monotonic. Combined with $R_{\text{utility}} = 0$
for both rational walk-aways and irrational acceptances, the reward
provides no signal to distinguish the two, and RL's deal-closing bias
leads 8B sellers to accept terms that violate their reservation price. The full
pipeline (Base $\to$ SFT $\to$ RL) therefore lands within
$\pm$2\,pp of the base model on most metrics---not because neither
stage works, but because they pull in opposite directions. SFT teaches
``hold out for good terms''; RL teaches ``close the deal.''

\paragraph{SFT teaches proportional strategies that generalize across
price tiers and opponents.}
Base models show large buyer-side Q1$\to$Q5 surplus spreads (15--25\,pp
spreads), performing poorly on low-value items where ZOPAs are thin.
SFT compresses these spreads to 7--9\,pp, maintaining 31--40\%
surplus across all quintiles---even against o3-mini, which is unseen
during training (\Cref{fig:training_bracket_surplus}). At the hardest
condition (8B, Q1, vs.\ o3-mini), SFT captures 31.5\% surplus where
Base gets 1.8\%. SFT does not memorize specific price points from
demonstrations; it learns to anchor and concede \emph{proportionally}
to item value, a transferable strategic principle. This mirrors
\Cref{sec:benchmark_summary}, Finding~3: consistency across price tiers
is a signature of strategic competence, and behavioral cloning
successfully instills it in open-weight models at a fraction of
frontier-model cost.

\paragraph{Additional observations.}
We note two weaker patterns. First, 14B is more resilient to RL than
8B. 14B keeps NGFT seller violations at or below 0.25\% across all
stages (\Cref{tab:training_ngft_viol}). 8B deteriorates: RL raises
NGFT seller violations to 6.50\% against o3-mini, worse than the
untrained base (4.00\%). With only two model sizes, we cannot separate
capacity from pretraining differences. Second, training changes the
model's own behavior but does not induce opponent errors. Frontier
counterpart violations remain near zero across all conditions
(\Cref{tab:training_gft_viol,tab:training_ngft_viol}). This is
unsurprising: $R_{\text{utility}}$ optimizes the model's own surplus
and provides no signal for pressuring opponents.

\medskip
Looking ahead, three directions may address these limitations: (1)~an
explicit reward penalty for deals below the reservation price, breaking
the symmetry between rational walk-aways and irrational acceptances;
(2)~training against diverse opponents rather than a single model, to
reduce opponent-specific overfitting; and (3)~multi-stage reward
curricula that first establish IR compliance before optimizing surplus.

%% file: sections/sec6.tex

\section{Conclusion}
\label{sec:conclusion}

\subsection{Summary}

This paper presents a structured bilateral bargaining environment that serves
two purposes: a three-dimensional evaluation framework for frontier LLMs and a
training environment for fine-tuning open-weight models. The evaluation
framework measures individual rationality, strategic effectiveness, and
allocative efficiency through automated metrics computed from machine-parseable
negotiation traces. A round-robin benchmark of five frontier models reveals
that aggressive anchoring with calibrated concession dominates accommodating
strategies, and that strategic competence manifests as consistency across price
tiers. A two-stage training pipeline (SFT followed by GRPO) identifies a
tension between the selectivity instilled by behavioral cloning and the
deal-closing bias introduced by online RL---a diagnosis that points to
specific reward design improvements.

\subsection{Limitations}
\label{sec:limitations}

\paragraph{Benchmark scope.}
The benchmark covers a single negotiation domain (consumer goods), one
protocol (alternating offers with structured tool calls), and five frontier
models evaluated at a single point in time. Results may not transfer to
multi-issue bargaining, multi-party settings, or future model generations.
The round-robin design captures pairwise dynamics but does not test against
human counterparts, whose strategic and communicative behavior may differ
qualitatively from LLM-vs-LLM play.

\paragraph{Training scale.}
The training experiments are a pilot study: two model sizes (8B, 14B), one
training opponent (GPT-4.1), and one RL algorithm (GRPO). The SFT--RL
tension we identify may be addressable with better reward design, but the
current study diagnoses the problem without resolving it. The evaluation
against only two frontier counterparts (GPT-4.1, o3-mini) limits the
generalizability of transfer conclusions.

\paragraph{Statistical precision.}
Some results rest on limited sample sizes: 80 scenarios per price-tier
bracket, 200 NGFT scenarios per pairing in the frontier benchmark. Bracket-level
violation rates (e.g., 3.75\% = 3 negotiations out of 80) should be
interpreted as directional rather than precise estimates.

\subsection{Future Directions}
\label{sec:future}

\paragraph{Reward design for RL.}
The central training finding---that RL erodes SFT's selectivity---traces to
a reward structure that assigns $R_{\text{utility}} = 0$ to both rational
walk-aways and irrational acceptances (\Cref{sec:training_discussion}). An
explicit penalty for deals below the reservation price would break this
symmetry. More broadly, a staged reward curriculum that first establishes IR
compliance and then optimizes surplus may prevent RL from trading rationality
for deal completion.

\paragraph{Diverse training opponents.}
Training against a single fixed opponent risks learning opponent-specific
strategies. Training against a population of models---or against earlier
checkpoints of the trainee itself---would test whether the learned
improvements reflect general strategic competence rather than exploitation
of a particular counterpart's behavioral profile.

\paragraph{Multi-issue and multi-party extensions.}
The current framework negotiates a single price for a single item. Extending
to multiple negotiable attributes (delivery terms, warranties, bundled items)
would increase the strategic complexity and create opportunities for
integrative bargaining, where agents create value by trading across issues
rather than dividing a fixed surplus. Multi-party extensions (e.g.,
procurement auctions with multiple suppliers) would connect to a broader set
of operations research applications.

\paragraph{Human--AI negotiation.}
The behavioral patterns documented here---anchoring effects, concession
dynamics, first-mover advantage---are well-established in human negotiation
research. Deploying the agent system with human counterparts would test
whether these patterns transfer from LLM-vs-LLM play, and whether humans
respond to LLM strategies in the same way that other LLMs do.

\newpage

%% file: sections/appendix.tex
%
\newpage
\appendix


\section{Game Environment Details}
\label{app:game_details}

This appendix provides implementation details for the bilateral bargaining
environment described in \Cref{sec:bargaining}.

\subsection{Example Product Listing}
\label{app:example_listing}

\begin{figure}[H]
\centering
\begin{paperquote}[Title: Dyson Supersonic Hair Dryer Nickel/Copper]
\small
\textbf{Category:} Other \\
\textbf{Description:}
\begin{itemize}
    \item The Dyson Supersonic hair dryer is engineered to protect hair from extreme heat damage for fast drying and precision styling.
    \item Fast drying: The small, powerful Dyson digital motor V9 combined with Air Multiplier technology produces a high-pressure, high-velocity jet of controlled air, for fast drying and precision styling.
    \item Intelligent heat control: Unlike some others, the Dyson Supersonic hair dryer measures air temperature over 40 times a second, and regulates the heat. This prevents extreme heat damage, to help protect your hair's shine.
    \item Engineered for different hair types: After rigorously testing different hair types in our laboratories, Dyson has engineered a range of attachments designed to style different types of hair.
    \item Five attachments to style different hair types: Flyaway attachment hides flyaways under longer hair for a smooth, shiny finish. Styling concentrator For precision styling. Diffuser Helps reduce frizz and create defined curls and waves. Gentle air attachment Fast yet gentle styling for fine hair and sensitive scalps. Wide-tooth comb Helps shape and lengthen curly and textured hair as it dries.
\end{itemize}
\textbf{Price History:}
\begin{itemize}
  \item Highest: \$429.99 (Sep 07, 2023)
  \item Lowest: \$365.49 (Oct 12, 2023)
\end{itemize}
\end{paperquote}
\caption{Example listing used in the negotiation simulation environment.}
\label{fig:ex_listing}
\end{figure}

\subsection{Tool Specification}
\label{app:tools}

\Cref{tab:tool_spec} lists the complete tool specification available to each
agent. All tools take an \texttt{agent\_name} parameter (``buyer'' or ``seller'')
identifying the caller; this parameter is omitted from the main text for brevity.
The main text (\Cref{tab:action_space}) lists the six primary tools; the full
specification below includes an additional tool
(\texttt{wait\_for\_time\_period}) used to model tactical delays.

\begin{table}[H]
\centering
\small
\begin{tabular}{llp{6.5cm}}
\toprule
\textbf{Tool} & \textbf{Parameters} & \textbf{Effect} \\
\midrule
\texttt{make\_offer} & \texttt{price}, \texttt{side\_offer} (opt.)
  & Proposes a formal price; becomes pending offer that the counterpart must
    respond to. Price must be positive. \\
\texttt{respond\_to\_offer} & \texttt{response} (Boolean)
  & Accepts (\texttt{True}) or rejects (\texttt{False}) the pending offer.
    Acceptance concludes the negotiation at the offered price. \\
\texttt{send\_message} & \texttt{content} (string)
  & Sends a free-form natural-language message to the counterpart. No direct
    effect on negotiation state. \\
\texttt{search\_price} & (none)
  & Returns historical high and low prices for the item. Information only. \\
\texttt{quit\_negotiation} & (none)
  & Exercises the outside option (BATNA). Negotiation terminates immediately
    with no deal; both utilities are zero. \\
\texttt{wait\_for\_response} & (none)
  & Yields the turn to the counterpart. Must be the last tool called in a
    turn. \\
\texttt{wait\_for\_time\_period} & \texttt{duration} (seconds)
  & Advances the simulation clock by the specified duration. Models tactical
    delays in real-world negotiation. \\
\bottomrule
\end{tabular}
\caption{Complete tool specification for the bilateral bargaining environment.
Each turn permits at most 3 tool calls. The main text (\Cref{tab:action_space})
lists the six primary tools; \texttt{wait\_for\_time\_period} is an additional
tool used to model tactical delays.}
\label{tab:tool_spec}
\end{table}

\subsection{Event Protocol}
\label{app:event_protocol}

The simulation engine uses an event-driven architecture managed by an
\texttt{EventManager} class. The core components are:
\begin{itemize}
    \item \emph{Event queue:} a min-heap priority queue sorted by timestamp.
    \item \emph{Simulation clock:} tracks current simulation time.
    \item \emph{Event batching:} groups simultaneous events from the same agent
      into a single observation for the counterpart.
\end{itemize}

Each tool call creates an event with four fields: timestamp, type (matching the
tool name), acting agent, and tool-specific data. The event processing loop
proceeds as follows:
\begin{enumerate}
    \item Pop the next event batch from the queue (events sharing the same
      timestamp and agent).
    \item Advance the simulation clock to the event timestamp.
    \item Convert events to a natural-language observation (e.g., ``Seller
      proposed \$1{,}500''; ``Buyer rejected your offer and proposed \$1{,}200'').
    \item Route the observation to the counterpart agent.
    \item The counterpart generates a \texttt{Thought:} block (internal
      reasoning, hidden from the opponent) followed by a \texttt{Code:} block
      (tool calls).
    \item The simulator parses and executes the tool calls, creating new events.
    \item Check termination conditions; if not met, return to step~1.
\end{enumerate}

\subsection{Example Negotiation Transcripts}
\label{app:transcripts}

\paragraph{GFT example.}
Item: ``Used Laptop'' (historical high \$1{,}500, historical low \$800).
Buyer reservation price $b = \$1{,}200$; seller reservation price $s = \$900$.
ZOPA $= [900, 1200]$; total surplus $= \$300$.

\begin{table}[H]
\centering
\small
\begin{tabular}{clp{8cm}}
\toprule
\textbf{Round} & \textbf{Agent} & \textbf{Action} \\
\midrule
1 & Seller & \texttt{make\_offer(1400)} \\
2 & Buyer  & \texttt{send\_message(``Your price is too high for a used
  laptop'')}; \texttt{make\_offer(950)} \\
3 & Seller & \texttt{send\_message(``I can't go that low, but I can offer a
  discount'')}; \texttt{make\_offer(1150)} \\
4 & Buyer  & \texttt{make\_offer(1050)} \\
5 & Seller & \texttt{respond\_to\_offer(True)} \\
\bottomrule
\end{tabular}
\caption{GFT example: deal at $p^* = \$1{,}050$. Buyer utility $U_B = 150$,
seller utility $U_S = 150$. Surplus shares: 50/50.}
\label{tab:gft_example}
\end{table}

\paragraph{NGFT example.}
Same item. Buyer reservation price $b = \$850$; seller reservation price
$s = \$1{,}100$. No ZOPA exists (NGFT scenario).

\begin{table}[H]
\centering
\small
\begin{tabular}{clp{8cm}}
\toprule
\textbf{Round} & \textbf{Agent} & \textbf{Action} \\
\midrule
1 & Seller & \texttt{make\_offer(1300)} \\
2 & Buyer  & \texttt{make\_offer(800)} \\
3 & Seller & \texttt{send\_message(``I can't go below \$1{,}100'')} \\
4 & Buyer  & \texttt{send\_message(``I can't go above \$850'')} \\
5 & Buyer  & \texttt{quit\_negotiation()} \\
\bottomrule
\end{tabular}
\caption{NGFT example: no deal (rational outcome). Both utilities are zero.
The buyer correctly identifies the infeasible negotiation and exercises the
outside option.}
\label{tab:ngft_example}
\end{table}


\section{Training Implementation Details}
\label{app:training_details}

This appendix provides implementation details for the training pipeline
described in \Cref{sec:training}.

\subsection{SFT Data Pipeline and Hyperparameters}
\label{app:sft_details}

\paragraph{Data pipeline.}
The SFT data pipeline has four stages:
\begin{enumerate}
    \item \emph{Demonstration generation.} DeepSeek-R1 self-play negotiations
      are generated within the simulator using the same system prompt template,
      tool definitions, and item catalog used in the benchmark
      (\Cref{sec:eval_protocol}).
    \item \emph{Reasoning-trace cleaning.} \texttt{<think>...</think>}
      reasoning blocks are removed from input prompts (but retained in target
      completions), as described in \Cref{sec:sft}.
    \item \emph{Turn-level decomposition.} Each multi-turn trajectory is
      decomposed into autoregressive training samples of increasing context
      length (\Cref{sec:sft}).
    \item \emph{Format conversion.} Samples are converted to input--output
      pairs, where the input concatenates the system prompt and conversation
      history and the output is the target agent response. Loss is computed
      only on output tokens.
\end{enumerate}

\paragraph{SFT hyperparameters.}
\Cref{tab:sft_hyperparameters} lists the SFT training hyperparameters for
both model sizes. Training uses the TRL SFTTrainer with DeepSpeed ZeRO
Stage~3 \citep{deepspeed} on 2 GPUs.

\begin{table}[H]
\centering
\begin{tabular}{lcc}
\toprule
\textbf{Parameter} & \textbf{8B Model} & \textbf{14B Model} \\
\midrule
\multicolumn{3}{l}{\emph{Optimization}} \\
\quad Learning rate & $2 \times 10^{-5}$ & $1 \times 10^{-5}$ \\
\quad LR schedule & Cosine & Cosine \\
\quad Warmup ratio & 0.1 & 0.1 \\
\quad Number of epochs & 3 & 3 \\
\quad Weight decay & 0.01 & 0.01 \\
\quad Max gradient norm & 0.1 & 0.1 \\
\quad Per-device batch size & 2 & 1 \\
\quad Gradient accumulation steps & 8 & 8 \\
\quad Precision & bfloat16 & bfloat16 \\
\midrule
\multicolumn{3}{l}{\emph{Sequence}} \\
\quad Max sequence length & 8{,}192 & 8{,}192 \\
\midrule
\multicolumn{3}{l}{\emph{Memory efficiency}} \\
\quad Fine-tuning mode & Full & Full \\
\quad Gradient checkpointing & \checkmark & \checkmark \\
\quad DeepSpeed ZeRO stage & 3 & 3 \\
\quad Number of GPUs & 2 & 2 \\
\bottomrule
\end{tabular}
\caption{SFT training hyperparameters for 8B and 14B models. Both models
are trained with full parameter updates (no LoRA). The effective batch size
is $\text{per-device batch} \times \text{accumulation steps} \times
\text{GPUs}$: 32 for 8B, 16 for 14B.}
\label{tab:sft_hyperparameters}
\end{table}

\subsection{GRPO Architecture and Key Parameters}
\label{app:training_config}

\paragraph{Hardware and GPU allocation.}
All training runs use a single node with 8$\times$H100 80\,GB GPUs. We separate
inference and training onto dedicated GPU sets so that rollout generation and
gradient computation overlap:
\begin{itemize}
    \item \emph{Inference cluster (GPUs 0--5):} runs vLLM \citep{vllm} with
      tensor parallelism $\text{TP}=2$ and data parallelism $\text{DP}=3$,
      processing multiple negotiations in parallel with up to 16K-token contexts.
      Prefix caching is enabled to reuse KV states across rollouts sharing
      the same system prompt.
    \item \emph{Training cluster (GPUs 6--7):} runs DeepSpeed ZeRO Stage~3
      \citep{deepspeed} for memory-efficient gradient computation. Optimizer
      states and model parameters are sharded across both GPUs.
\end{itemize}

\paragraph{Key hyperparameters.}
\Cref{tab:hyperparameters} lists the GRPO training hyperparameters for both
model sizes.

\begin{table}[H]
\centering
\begin{tabular}{lcc}
\toprule
\textbf{Parameter} & \textbf{8B Model} & \textbf{14B Model} \\
\midrule
\multicolumn{3}{l}{\emph{Optimization}} \\
\quad Learning rate & $3 \times 10^{-6}$ & $2 \times 10^{-6}$ \\
\quad Warmup steps & 20 & 20 \\
\quad Max training steps & 200 & 200 \\
\quad Per-device batch size & 1 & 1 \\
\quad Gradient accumulation steps & 16 & 16 \\
\quad Effective batch size & 128 & 128 \\
\quad Precision & bfloat16 & bfloat16 \\
\midrule
\multicolumn{3}{l}{\emph{GRPO}} \\
\quad Rollouts per prompt ($G$) & 8 & 8 \\
\quad Clipping range ($\epsilon$) & 0.2 & 0.2 \\
\quad KL penalty ($\beta$) & 0.05 & 0.005 \\
\quad Stability constant ($\epsilon_{\text{stable}}$) & $10^{-4}$ & $10^{-4}$ \\
\midrule
\multicolumn{3}{l}{\emph{Memory efficiency}} \\
\quad LoRA rank & --- (full) & 64 \\
\quad Gradient checkpointing & \checkmark & \checkmark \\
\quad DeepSpeed ZeRO stage & 3 & 3 \\
\midrule
\multicolumn{3}{l}{\emph{Generation}} \\
\quad Max sequence length & 16{,}384 & 16{,}384 \\
\quad Temperature & 1.0 & 1.0 \\
\quad vLLM tensor parallelism & 2 & 2 \\
\quad Weight sync frequency & Every 8 grad.\ steps & Every 8 grad.\ steps \\
\bottomrule
\end{tabular}
\caption{GRPO training hyperparameters for 8B and 14B models. The 8B model is
trained with full parameters; the 14B model uses LoRA (rank 64, ${<}1\%$
trainable parameters) due to memory constraints.}
\label{tab:hyperparameters}
\end{table}

\subsection{Context Length Growth}
\label{app:context_growth}

Context length grows over the course of a negotiation. At round $t$, the total
context length is
\begin{equation}
\label{eq:context_length}
L_t = L_{\text{init}} + \sum_{i=1}^{t}
\bigl(L_{\text{prompt}}^{i} + L_{\text{response}}^{i}
+ L_{\text{tool}}^{i} + L_{\text{result}}^{i}\bigr),
\end{equation}
where the per-component token counts are approximately:
\begin{itemize}
    \item $L_{\text{init}} \approx 2{,}500$ tokens (role assignment, constraints,
      instructions).
    \item $L_{\text{prompt}}^{i} \approx 500$ tokens (per-turn context update).
    \item $L_{\text{response}}^{i} \approx 300$ tokens (agent-generated code).
    \item $L_{\text{tool}}^{i} \approx 100$ tokens (tool execution results).
    \item $L_{\text{result}}^{i} \approx 200$ tokens (environment feedback).
\end{itemize}
This totals ${\sim}1{,}100$ tokens per round, exhausting a 16K context window in
roughly 12--13 rounds. The maximum context length for training is set to 13{,}312
tokens.

\subsection{Gradient-Safe Overflow Handling}
\label{app:overflow}

When context overflow occurs (i.e., $L_t > 13{,}312$), GRPO still requires valid
logprobs for the episode to form a well-shaped training batch. The overflow
handling proceeds as follows:
\begin{enumerate}
    \item Generate a synthetic \texttt{quit\_negotiation} response (with
      \texttt{Thought}/\texttt{Code} blocks matching the expected format).
    \item Attach synthetic logprobs using the PAD token (token ID 151643,
      \texttt{<|endoftext|>} in Qwen models) with a nominal logprob of $-0.1$.
    \item Execute the synthetic response through the environment, terminating
      the negotiation with no deal.
    \item Assign reward $R = 0$.
\end{enumerate}

The choice of the PAD token is critical. An early implementation used token
ID~1 (the double-quote character \texttt{"} in Qwen's vocabulary), which created
real gradient updates: with a negative advantage from the failed negotiation,
GRPO would systematically decrease the probability of generating double quotes
after long contexts. The PAD token avoids this because
$\pi_\theta(\text{PAD} \mid s) \approx 0$ under normal generation---the model
never produces this special token during inference, so gradient updates to its
probability are harmless noise. Formally, the gradient contribution from an
overflowed episode is
\begin{equation}
\label{eq:overflow_gradient}
\nabla_\theta \mathcal{L} = -\hat{A}^{g} \cdot
\nabla_\theta \log \pi_\theta(\text{PAD} \mid s),
\end{equation}
where $\hat{A}^{g}$ is near zero (the reward of zero is close to the group mean)
and $\nabla_\theta \log \pi_\theta(\text{PAD} \mid s)$ points in a direction
irrelevant to normal text generation. Both factors ensure negligible parameter
updates.

The context overflow rate is approximately 3.2\% at the start of training and
decreases as the model learns to conclude negotiations within fewer rounds.

\subsection{Weight Synchronization Protocol}
\label{app:sync}

Synchronizing the vLLM inference server with the training process after every
gradient step creates prohibitive overhead. We update weights every
$\texttt{gradient\_accumulation\_steps} \times \texttt{num\_iterations} = 8$
gradient steps (\Cref{tab:hyperparameters}), reducing synchronization overhead
while maintaining acceptable policy lag. The protocol proceeds as follows:
\begin{enumerate}
    \item The training process computes gradients over accumulated batches.
    \item After 8 gradient steps, it initiates weight synchronization.
    \item The vLLM server atomically swaps model weights.
    \item Subsequent rollouts use the updated policy immediately.
\end{enumerate}

\subsection{Training Dynamics}
\label{app:training_dynamics}

The training process naturally progresses through three phases:
\begin{enumerate}
    \item \textbf{Tool mastery (steps 1--20).} The agent learns to generate
      well-formed tool calls and execute valid code. Rewards are dominated by
      $R_{\text{parsing}}$ and $R_{\text{execution}}$
      (\Cref{eq:reward_outcome,tab:reward_components}), as successful
      negotiations are rare.
    \item \textbf{Constraint awareness (steps 20--40).} The agent begins
      respecting IR constraints, learning to reject offers that yield negative
      utility. IR violation rates drop from ${>}30\%$ to ${<}5\%$.
    \item \textbf{Strategic optimization (steps 40--58).} With basic competence
      established, the agent develops negotiation strategies: anchoring with
      aggressive opening offers, progressive concessions, and strategic
      outside-option exercise.
\end{enumerate}


\section{Frontier Benchmark: Supplementary Results}
\label{app:frontier_results}

This appendix provides supplementary figures and tables for the frontier model
benchmark (\Cref{sec:benchmark}).


\subsection{Aggregate Tables}
\label{app:frontier_tables}

The tables below provide the full numerical values underlying the figures in
\Cref{sec:benchmark_results}.

\subsubsection{IR Compliance}
\label{app:frontier_ir}

\begin{table}[H]
\centering
\small
\begin{tabular}{@{}ll cc cc@{}}
\toprule
& & \multicolumn{2}{c}{\textbf{GFT Violation (\%)}} & \multicolumn{2}{c}{\textbf{NGFT Violation (\%)}} \\
\cmidrule(lr){3-4} \cmidrule(lr){5-6}
\textbf{Model} & \textbf{Role} & Self & Opp. & Self & Opp. \\
\midrule
\multirow{2}{*}{DeepSeek-V3}
  & Buyer  & 0.45 & 0.15 & 0.10 & 0.40 \\
  & Seller & 0.70 & 0.45 & 0.60 & 0.00 \\
\midrule
\multirow{2}{*}{Gem.-2.5-Flash}
  & Buyer  & 0.20 & 0.25 & 0.40 & 0.00 \\
  & Seller & 0.10 & 0.10 & \textbf{1.90} & 0.00 \\
\midrule
\multirow{2}{*}{Gem.-2.5-Pro}
  & Buyer  & 0.10 & 0.20 & \textbf{0.00} & \textbf{0.00} \\
  & Seller & \textbf{0.00} & \textbf{0.00} & 0.70 & 0.00 \\
\midrule
\multirow{2}{*}{GPT-4.1}
  & Buyer  & 0.25 & 0.15 & 0.20 & 0.10 \\
  & Seller & 0.10 & 0.20 & \textbf{2.30} & 0.20 \\
\midrule
\multirow{2}{*}{o3}
  & Buyer  & \textbf{0.05} & 0.20 & 0.10 & \textbf{5.00} \\
  & Seller & 0.05 & \textbf{0.30} & \textbf{0.00} & 0.60 \\
\bottomrule
\end{tabular}
\caption{IR compliance for frontier models. Self: the model's own violation
rate (deals yielding negative utility for itself). Opp.: the opponent's
violation rate when facing this model. All values averaged over all five
opponents. Bold highlights best and worst values. In NGFT, any deal is a
violation for at least one party; o3 as buyer induces 5.00\% opponent
violations despite only 0.10\% self-violations.}
\label{tab:frontier_ir}
\end{table}

\begin{table}[H]
\centering
\small
\begin{tabular}{@{}l cc@{}}
\toprule
\textbf{o3 as Buyer vs.} & \textbf{o3 Self-Viol.\ (\%)} & \textbf{Seller Opp.-Viol.\ (\%)} \\
\midrule
GPT-4.1          & 0.5 & \textbf{11.5} \\
Gem.-2.5-Flash   & 0.0 & \textbf{9.5} \\
Gem.-2.5-Pro     & 0.0 & 3.5 \\
DeepSeek-V3      & 0.0 & 0.5 \\
o3               & 0.0 & 0.0 \\
\bottomrule
\end{tabular}
\caption{NGFT pairwise breakdown for o3 as buyer. Self-violation = o3's own
irrational deals; opponent-violation = seller's irrational deals when facing
o3. The irrationality comes almost entirely from opponents: GPT-4.1 and
Gemini-2.5-Flash violate at 11.5\% and 9.5\%, while o3 itself violates at
most 0.5\%.}
\label{tab:o3_buyer_ngft}
\end{table}

\subsubsection{Strategic Effectiveness and Behavioral Drivers}
\label{app:frontier_surplus_table}

\begin{table}[H]
\centering
\small
\begin{tabular}{@{}l cc cc cc cc cc@{}}
\toprule
& \multicolumn{2}{c}{\textbf{Surplus (\%)}} & \multicolumn{2}{c}{\textbf{Deal Rate (\%)}}
& \multicolumn{2}{c}{\textbf{Init.\ Aggr.}} & \multicolumn{2}{c}{\textbf{Conc.\ Rate}}
& \multicolumn{2}{c}{\textbf{Patience}} \\
\cmidrule(lr){2-3} \cmidrule(lr){4-5} \cmidrule(lr){6-7} \cmidrule(lr){8-9} \cmidrule(lr){10-11}
\textbf{Model} & B & S & B & S & B & S & B & S & B & S \\
\midrule
DeepSeek-V3   & 35.1 & 49.6 & 94.4 & 92.1 & 0.34 & 1.97 & 0.52 & 0.50 & 6.65 & 5.98 \\
Gem.-2.5-Flash   & 24.3 & 57.6 & 86.4 & 94.9 & 0.37 & 2.14 & 0.67 & 0.54 & 5.72 & 6.02 \\
Gem.-2.5-Pro     & 44.9 & 72.6 & 96.7 & 91.7 & 0.40 & 2.06 & 0.51 & 0.43 & 7.07 & 7.06 \\
GPT-4.1            & 34.1 & 57.7 & 94.0 & 95.2 & 0.36 & 2.07 & 0.53 & 0.52 & 6.07 & 6.08 \\
o3                 & 47.2 & 77.0 & 98.0 & 95.6 & 0.39 & 3.04 & 0.47 & 0.55 & 6.50 & 6.86 \\
\bottomrule
\end{tabular}
\caption{Outcome metrics and behavioral drivers for frontier models (GFT
scenarios), averaged over all five opponents. B = buyer, S = seller. Surplus:
fraction of ZOPA captured conditional on agreement. Deal rate: fraction of
negotiations reaching agreement. Init.\ Aggr.\ (buyer): fraction of the seller's asking price
cut by the buyer's first counter-offer (higher = more aggressive).
Init.\ Aggr.\ (seller): ratio of first offer to reservation price
(higher = more aggressive). Conc.\ Rate: fraction
of remaining surplus conceded per round. Patience: mean number of rounds to
termination (max 10).}
\label{tab:frontier_surplus}
\end{table}

\subsubsection{Radar-Plot Profile Tables}
\label{app:radar_tables}

\begin{table}[H]
\centering
\small
\begin{tabular}{@{}l cccccc@{}}
\toprule
\textbf{Model}
  & \textbf{IR Compl.}
  & \textbf{Surplus}
  & \textbf{Deal Rate}
  & \textbf{Anchor.}
  & \textbf{Consistency}
  & \textbf{Pressure Res.} \\
  & (\%)
  & (\%)
  & (\%)
  &
  & (\%)
  & (\%) \\
\midrule
o3             & 99.95 & 47.2 & 98.0 & 0.39 & 93.9 & 99.90 \\
Gem.-2.5-Pro   & 99.90 & 44.9 & 96.7 & 0.40 & 94.0 & 100.00 \\
GPT-4.1        & 99.75 & 34.1 & 94.0 & 0.36 & 87.5 & 99.80 \\
DeepSeek-V3    & 99.55 & 35.1 & 94.4 & 0.34 & 93.6 & 99.90 \\
Gem.-2.5-Flash & 99.80 & 24.3 & 86.4 & 0.37 & 87.2 & 99.60 \\
\bottomrule
\end{tabular}
\caption{Buyer radar-plot profile for frontier models. IR Compl.\
$= 100 - \text{GFT self-violation rate}$. Surplus: buyer surplus share
(GFT). Deal Rate: buyer GFT deal rate. Anchor.: fraction of the
seller's asking price cut by the buyer's first counter-offer.
Consistency $= 100 - \text{quintile surplus spread (pp)}$. Pressure
Res.\ $= 100 - \text{NGFT self-violation rate}$. All axes: higher is
better.}
\label{tab:radar_buyer}
\end{table}

\begin{table}[H]
\centering
\small
\begin{tabular}{@{}l cccccc@{}}
\toprule
\textbf{Model}
  & \textbf{IR Compl.}
  & \textbf{Surplus}
  & \textbf{Deal Rate}
  & \textbf{Anchor.}
  & \textbf{Consistency}
  & \textbf{Pressure Res.} \\
  & (\%)
  & (\%)
  & (\%)
  &
  & (\%)
  & (\%) \\
\midrule
o3             & 99.95 & 77.0 & 95.6 & 3.04 & 96.0 & 100.00 \\
Gem.-2.5-Pro   & 100.00 & 72.6 & 91.7 & 2.06 & 94.0 & 99.30 \\
GPT-4.1        & 99.90 & 57.7 & 95.2 & 2.07 & 96.2 & 97.70 \\
DeepSeek-V3    & 99.30 & 49.6 & 92.1 & 1.97 & 90.0 & 99.40 \\
Gem.-2.5-Flash & 99.90 & 57.6 & 94.9 & 2.14 & 93.4 & 98.10 \\
\bottomrule
\end{tabular}
\caption{Seller radar-plot profile for frontier models. IR Compl.\
$= 100 - \text{GFT self-violation rate}$. Surplus: seller surplus share
(GFT). Deal Rate: seller GFT deal rate. Anchor.: ratio of first offer
to reservation price. Consistency $= 100 - \text{quintile surplus
spread (pp)}$. Pressure Res.\ $= 100 - \text{NGFT self-violation
rate}$. All axes: higher is better.}
\label{tab:radar_seller}
\end{table}

\subsection{Pairwise Heatmaps}
\label{app:pairwise_heatmaps}

Each heatmap shows the metric value for all 25 buyer--seller pairings
(rows = buyer models, columns = seller models) in GFT scenarios unless
otherwise noted.

\subsubsection{Violation Rate}
\label{app:violation_rate}

\begin{figure}[H]
\centering
\includegraphics[width=0.5\textwidth]{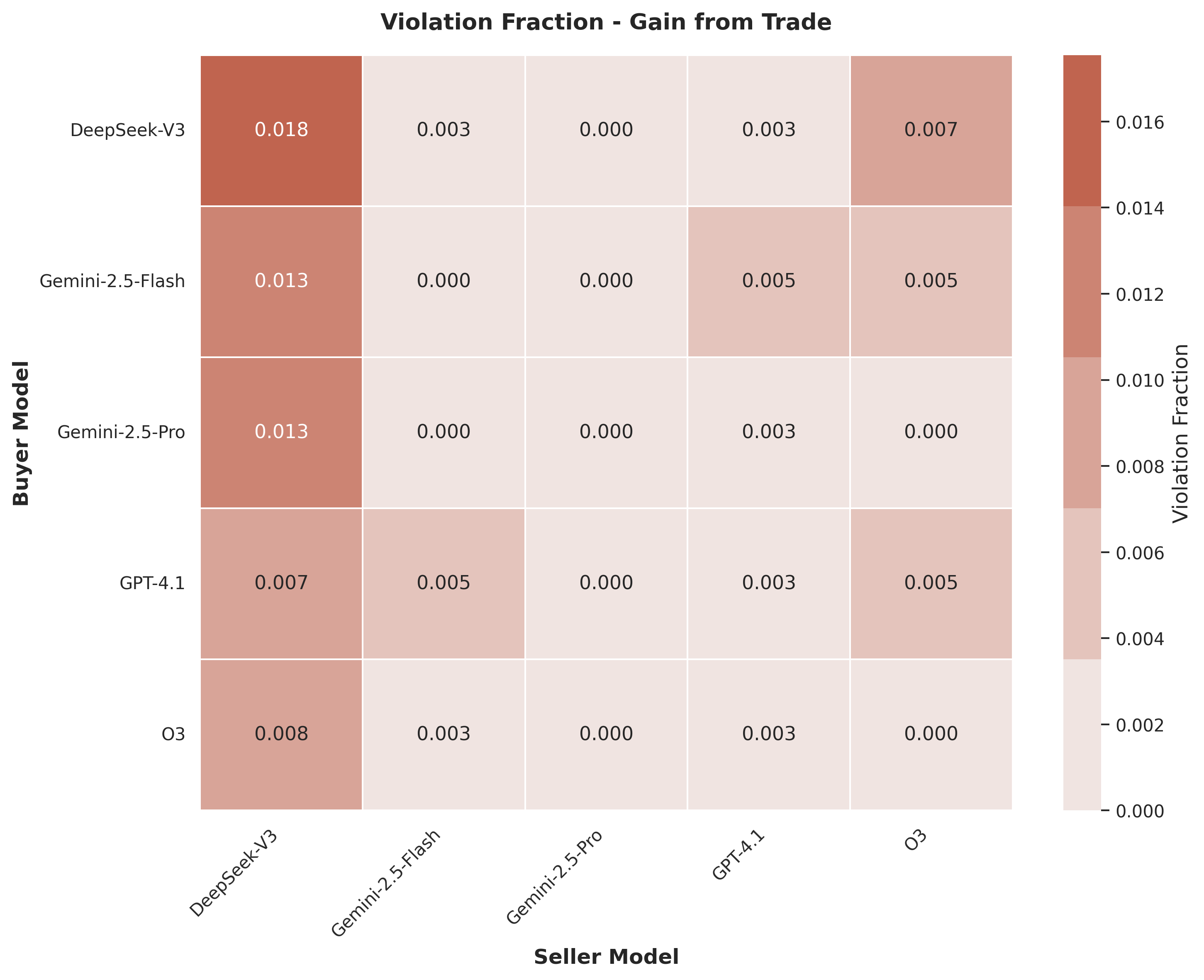}%
\hfill
\includegraphics[width=0.5\textwidth]{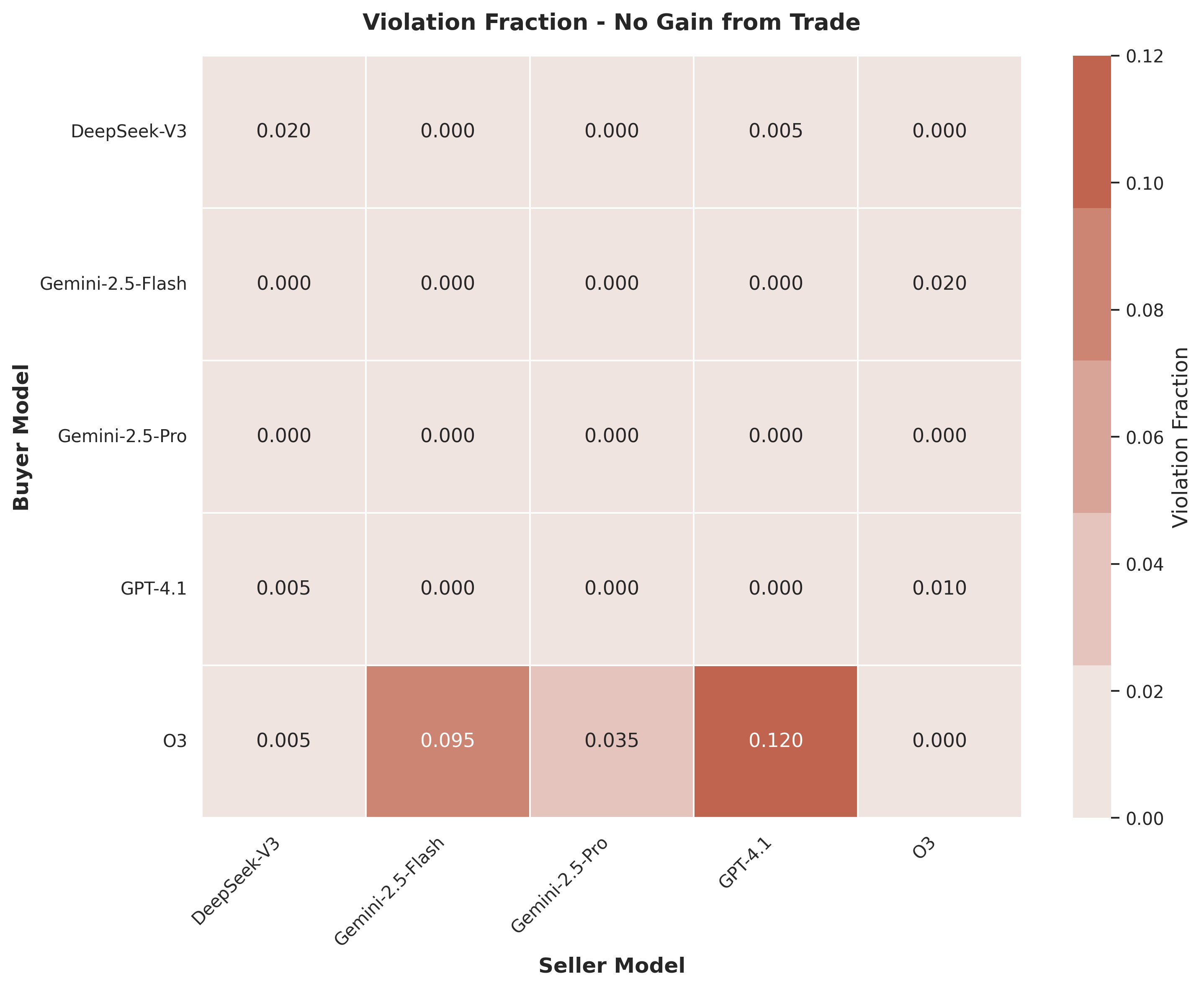}
\caption{Violation rate heatmaps for all 25 buyer--seller pairings. Left: GFT
scenarios (violations arise from deals struck outside the ZOPA). Right: NGFT
scenarios (every deal is a violation for at least one party, so violation rate
$=$ deal rate).}
\label{fig:violation_heatmaps}
\end{figure}

\subsubsection{Average Utility}
\label{app:utility}

\begin{figure}[H]
\centering
\includegraphics[width=0.5\textwidth]{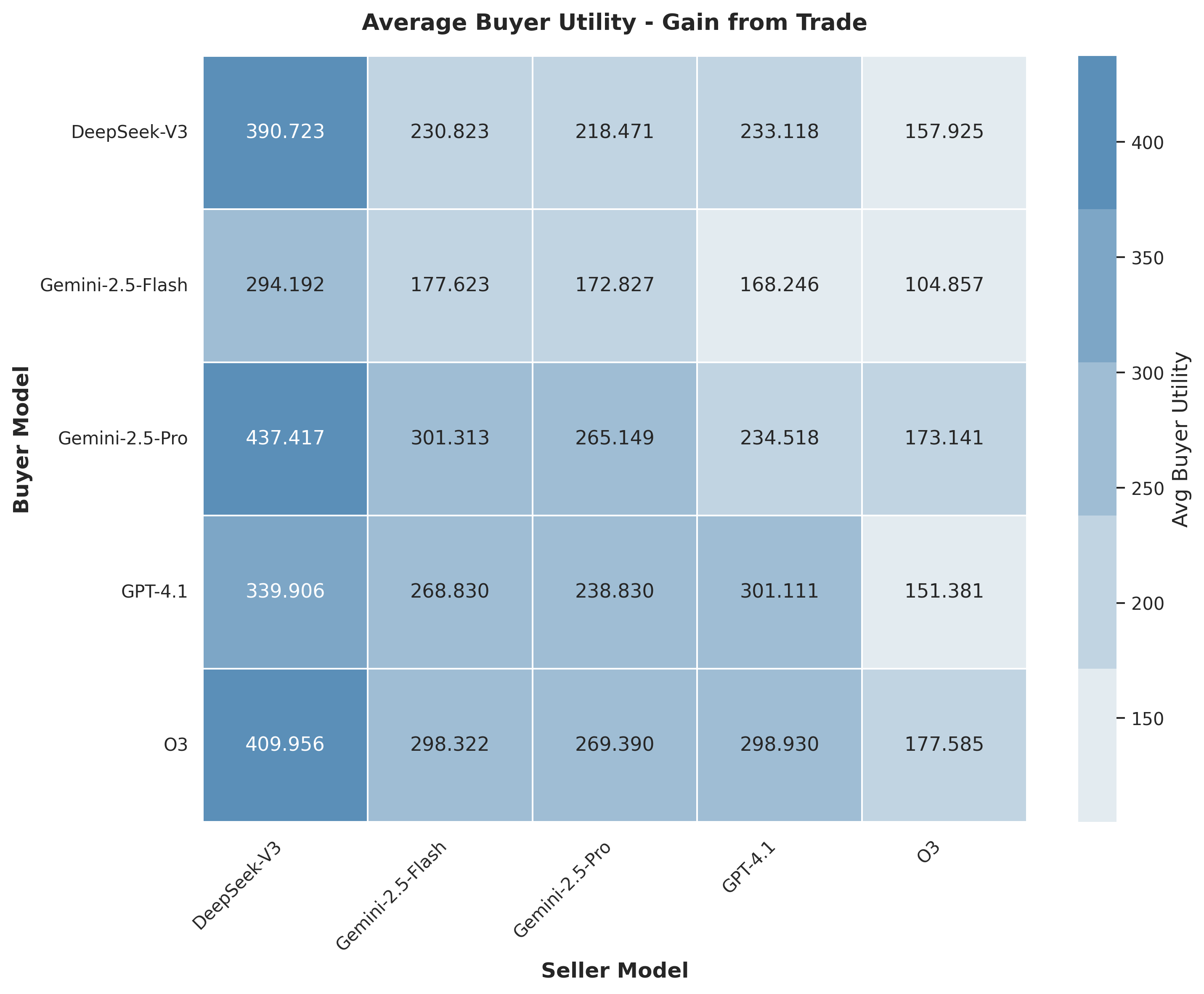}%
\hfill
\includegraphics[width=0.5\textwidth]{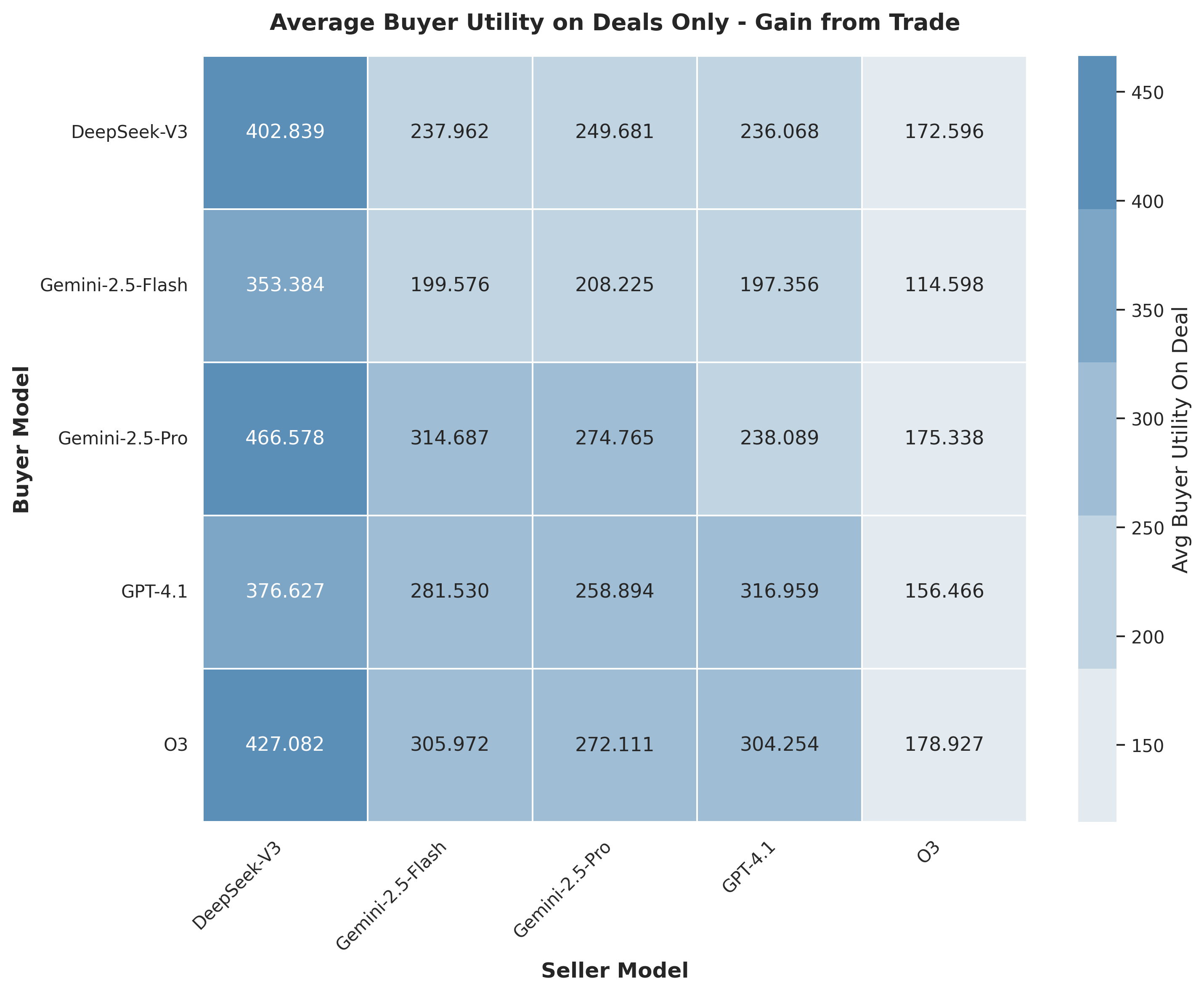}
\caption{Average buyer utility in GFT scenarios. Left: averaged over all
negotiations (including no-deal outcomes at utility $= 0$). Right: conditional
on agreement.}
\label{fig:buyer_utility_heatmaps}
\end{figure}

\begin{figure}[H]
\centering
\includegraphics[width=0.5\textwidth]{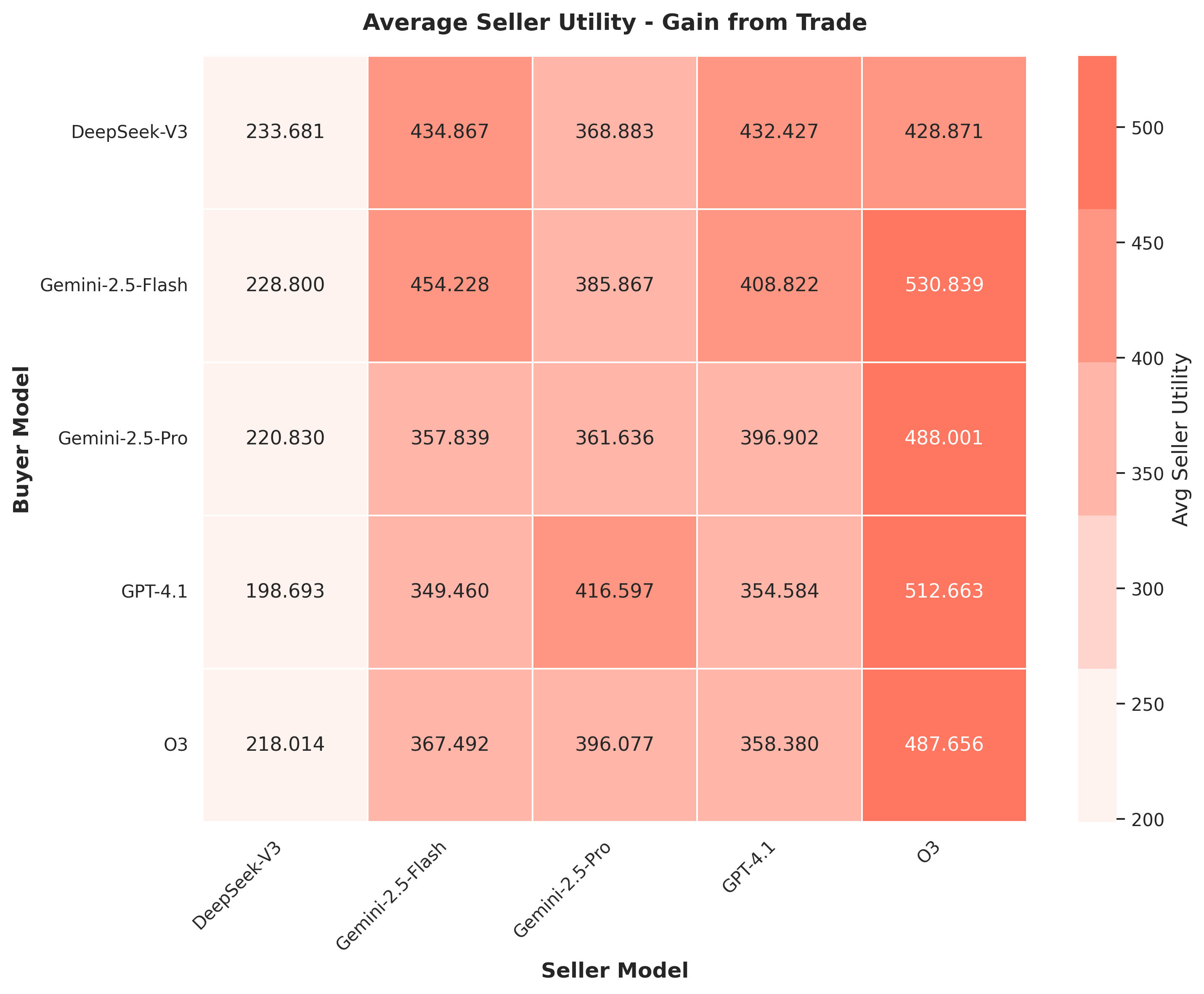}%
\hfill
\includegraphics[width=0.5\textwidth]{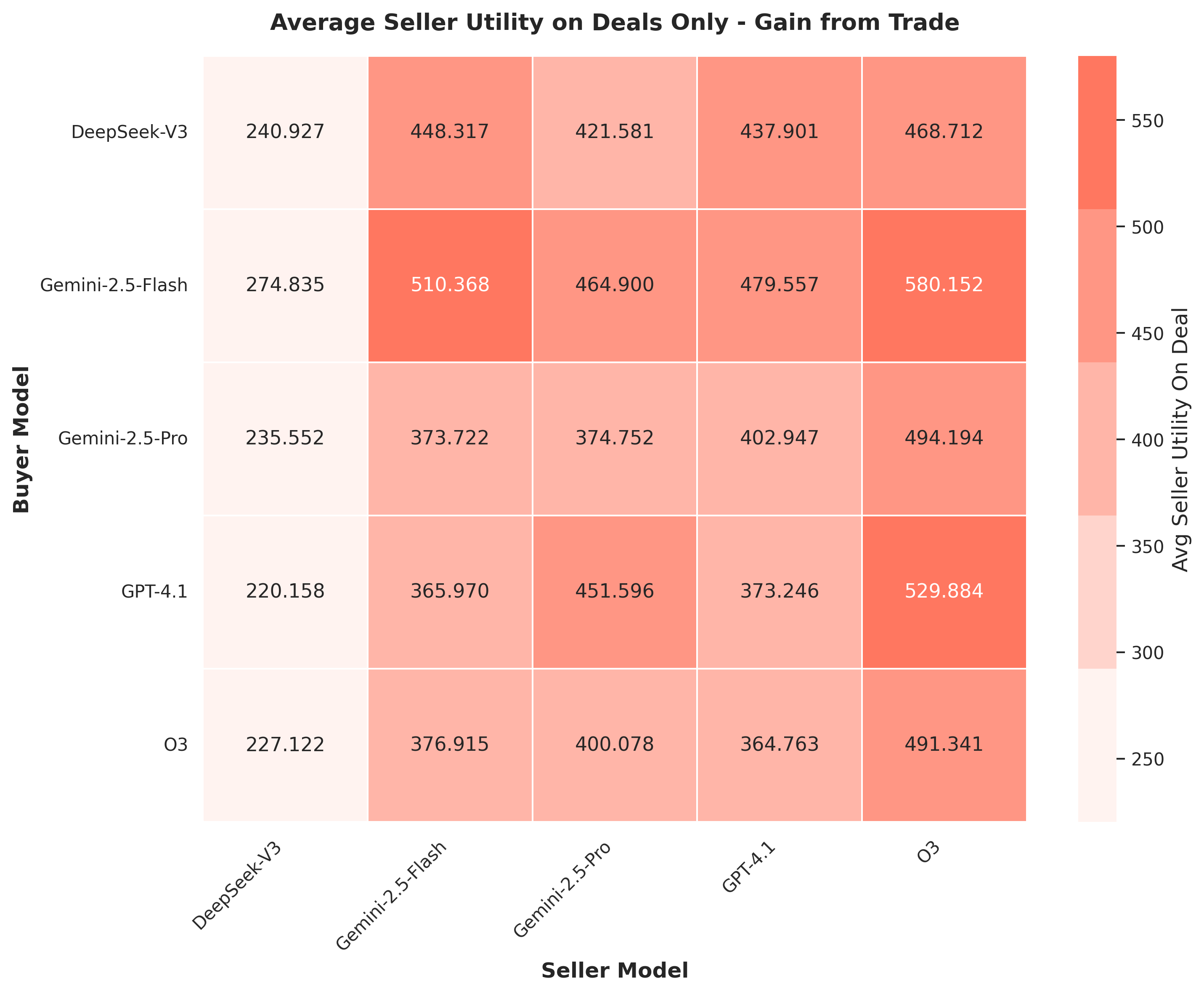}
\caption{Average seller utility in GFT scenarios. Left: averaged over all
negotiations. Right: conditional on agreement.}
\label{fig:seller_utility_heatmaps}
\end{figure}

\subsubsection{Surplus Share}
\label{app:surplus_share}

\begin{figure}[H]
\centering
\includegraphics[width=0.5\textwidth]{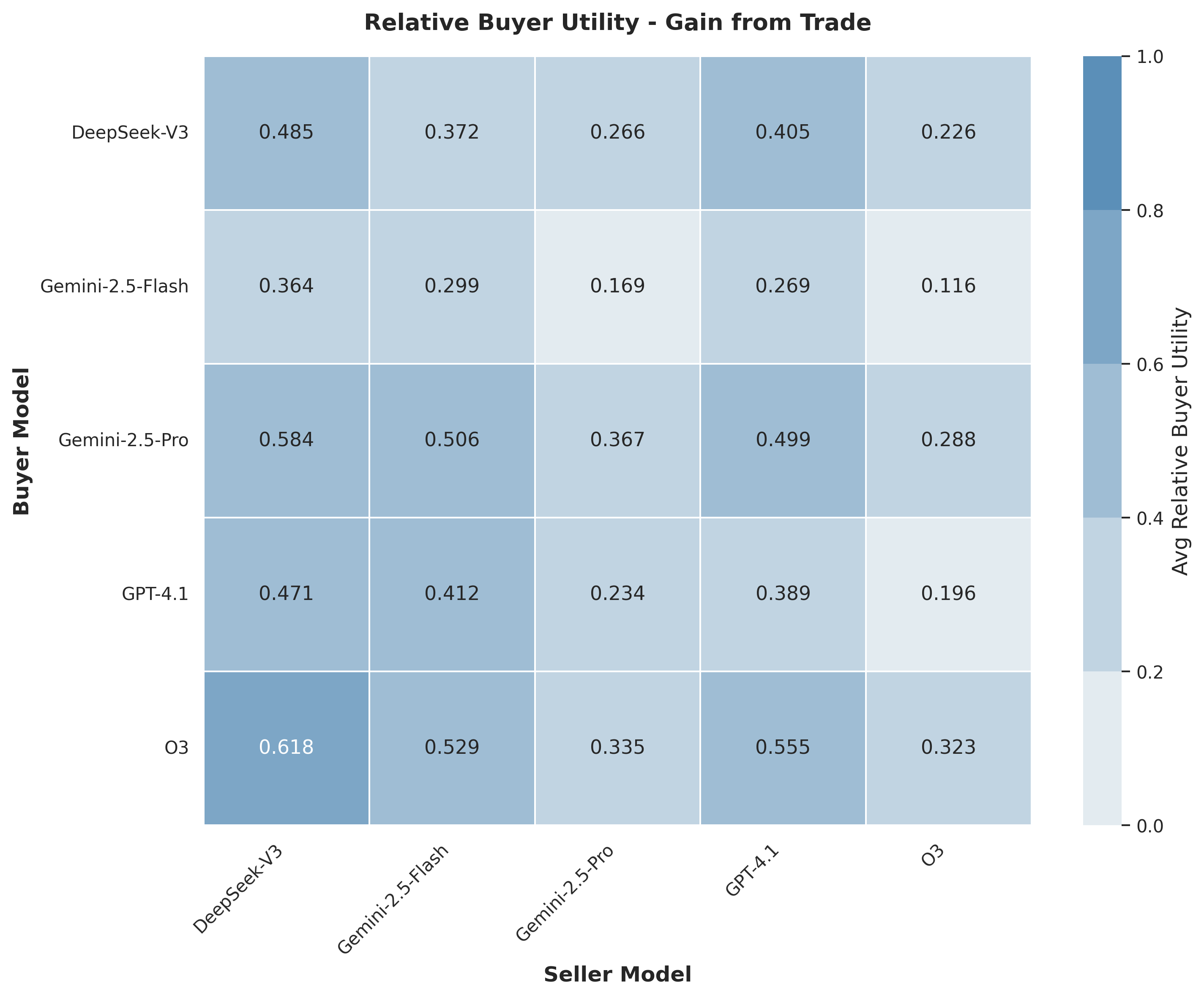}%
\hfill
\includegraphics[width=0.5\textwidth]{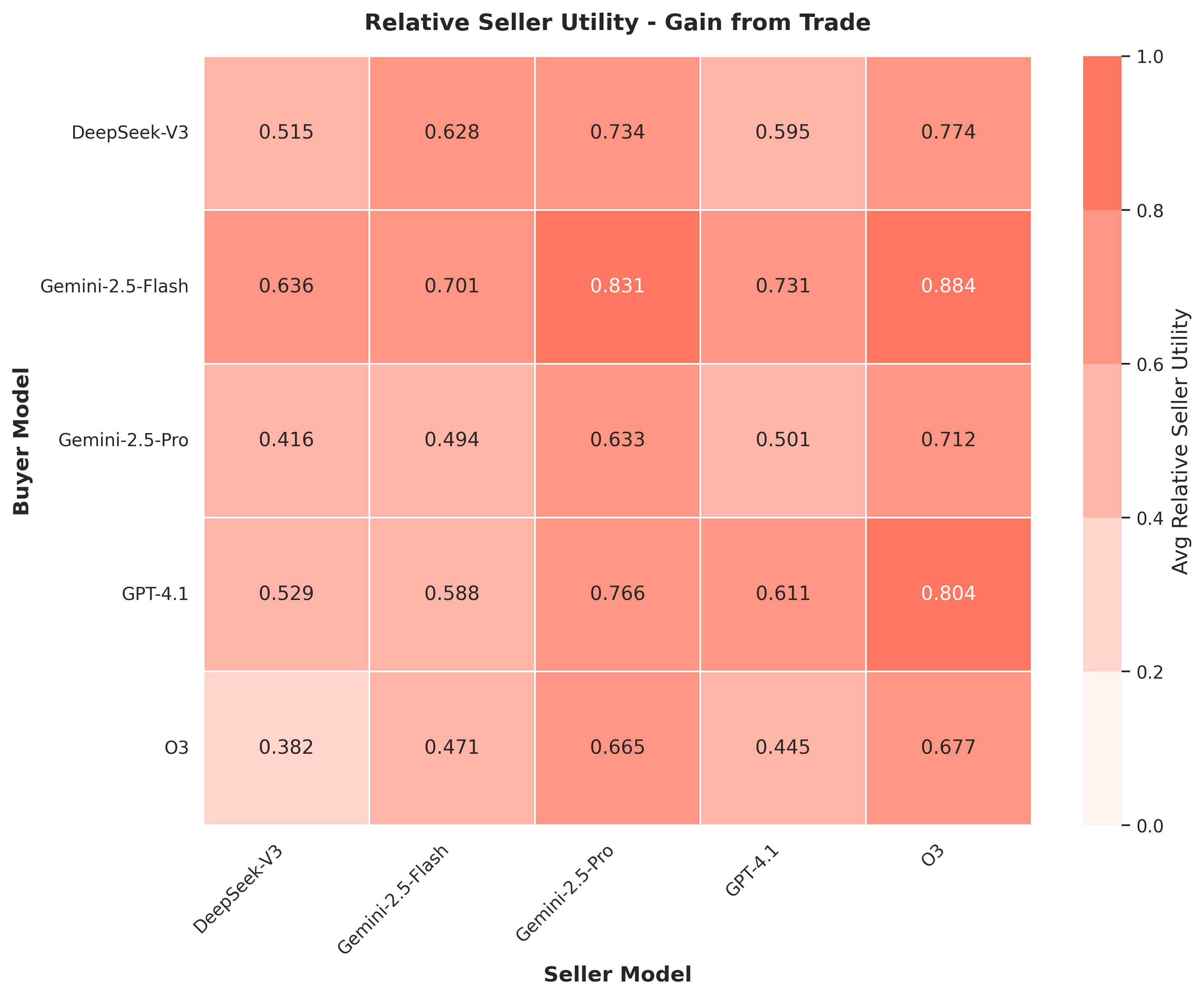}
\caption{Surplus share heatmaps for all 25 buyer--seller pairings (GFT
scenarios). Left: buyer surplus share. Right: seller surplus share. The two
shares sum to one for each cell.}
\label{fig:surplus_share_heatmaps}
\end{figure}

\subsubsection{Deal Rate}
\label{app:deal_rate}

\begin{figure}[H]
\centering
\includegraphics[width=0.5\textwidth]{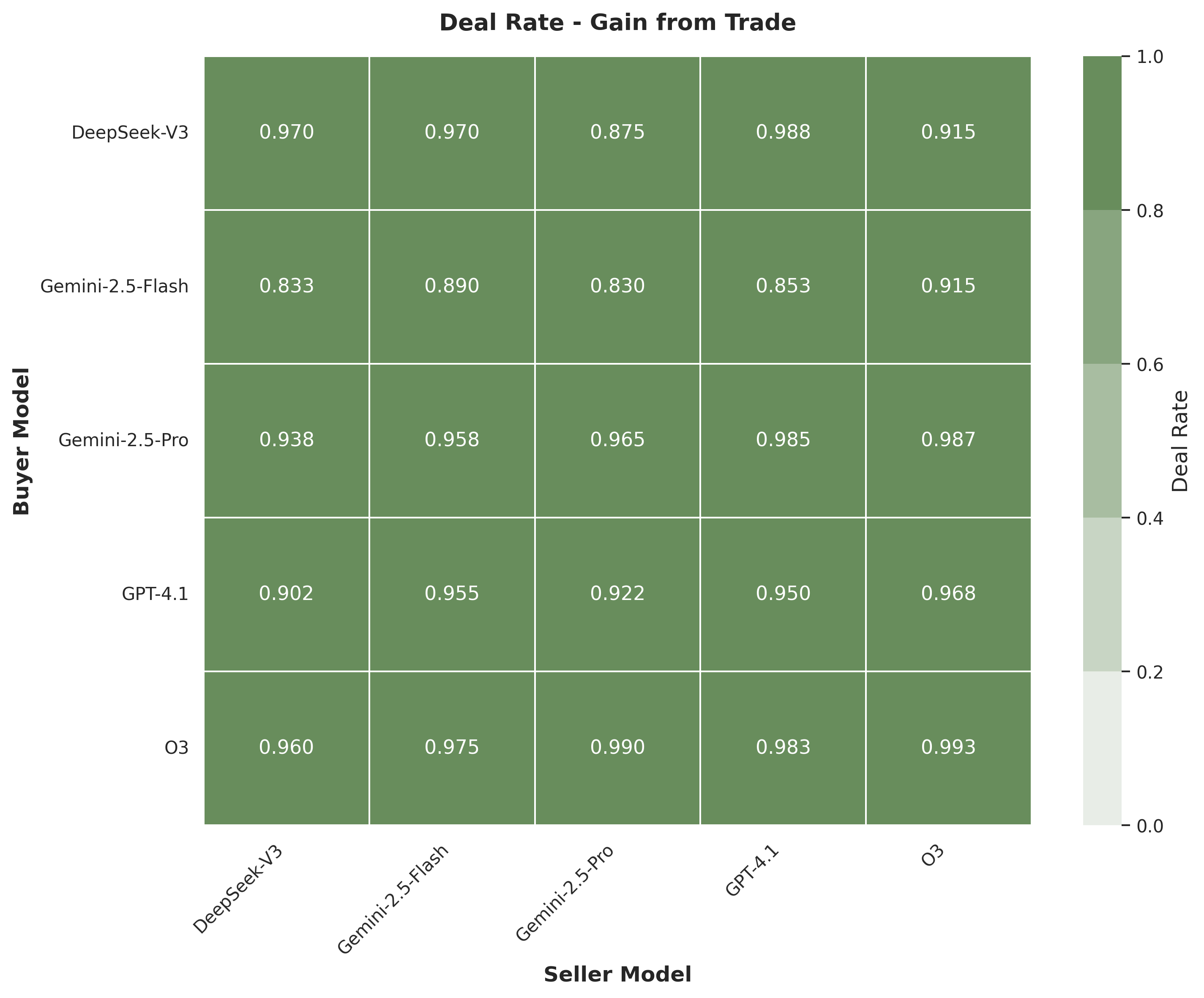}
\caption{Deal rate for all 25 buyer--seller pairings (GFT scenarios). NGFT deal
rates equal violation rates (\Cref{fig:violation_heatmaps}) and are omitted.}
\label{fig:deal_rate_heatmap}
\end{figure}

\subsubsection{Initial Aggressiveness}
\label{app:initial_aggressiveness}

\begin{figure}[H]
\centering
\includegraphics[width=0.5\textwidth]{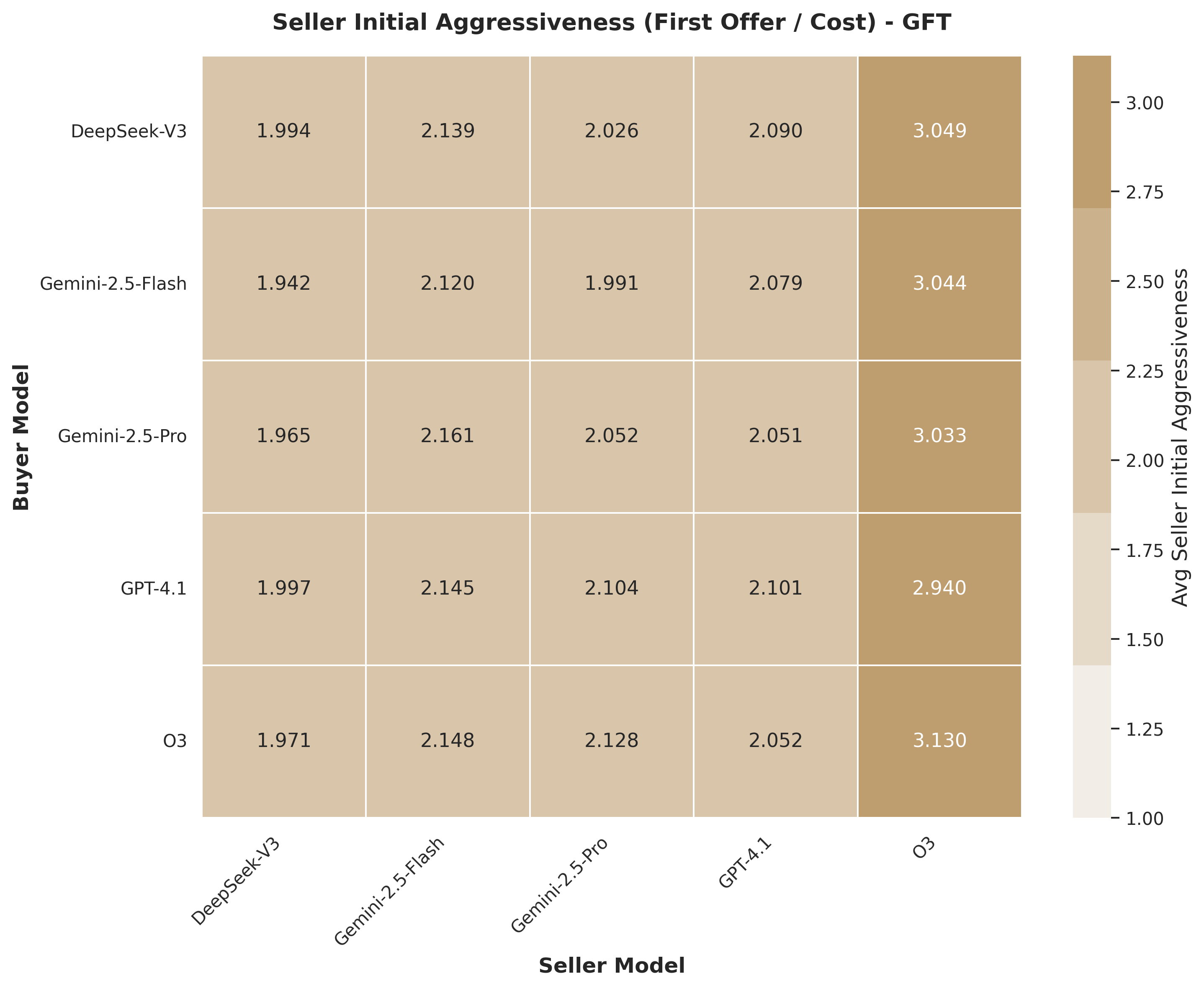}
\caption{Seller initial aggressiveness (first offer / reservation price).
Higher values indicate more aggressive opening offers.}
\label{fig:seller_aggressiveness}
\end{figure}

\begin{figure}[H]
\centering
\includegraphics[width=0.5\textwidth]{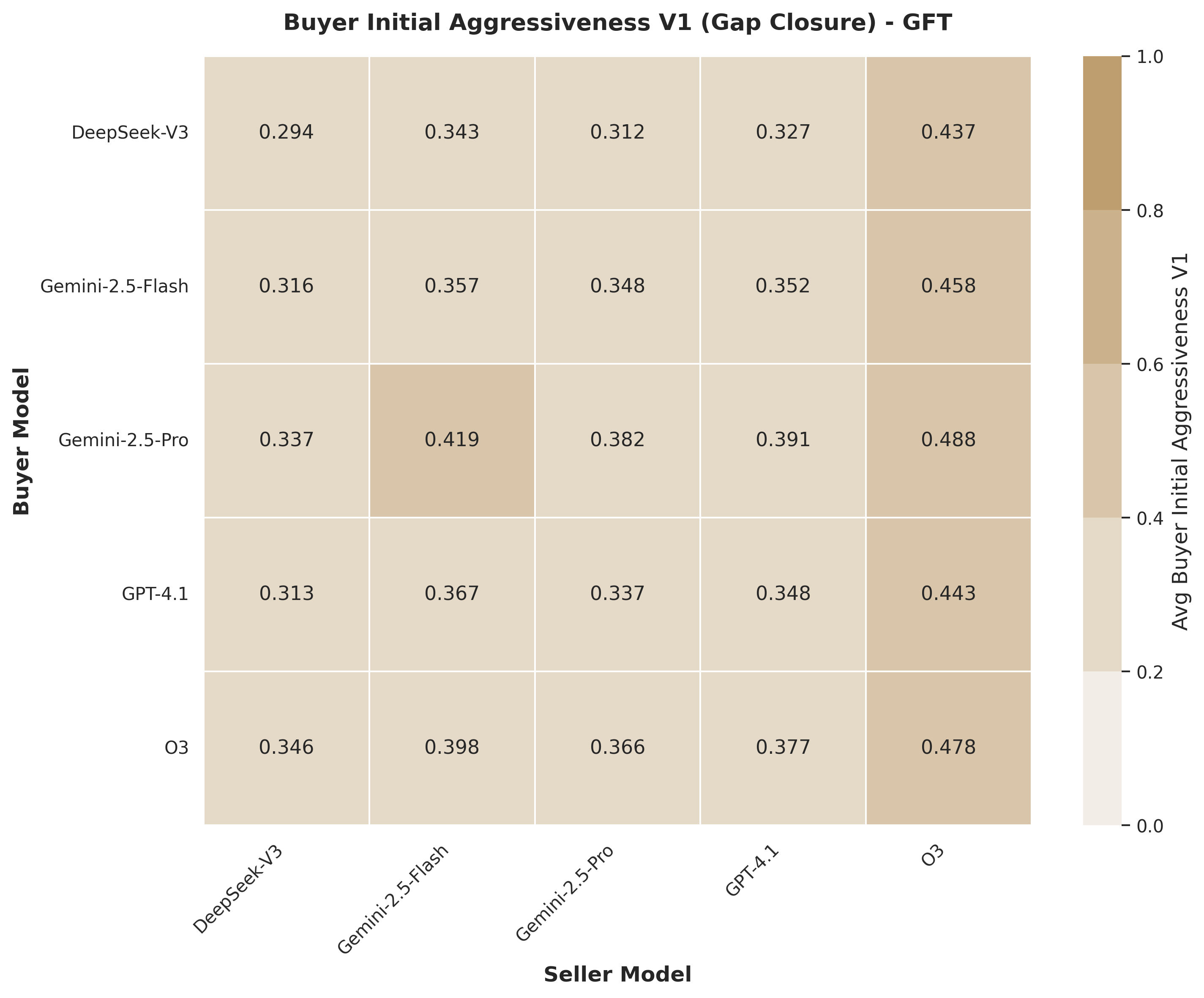}%
\hfill
\includegraphics[width=0.5\textwidth]{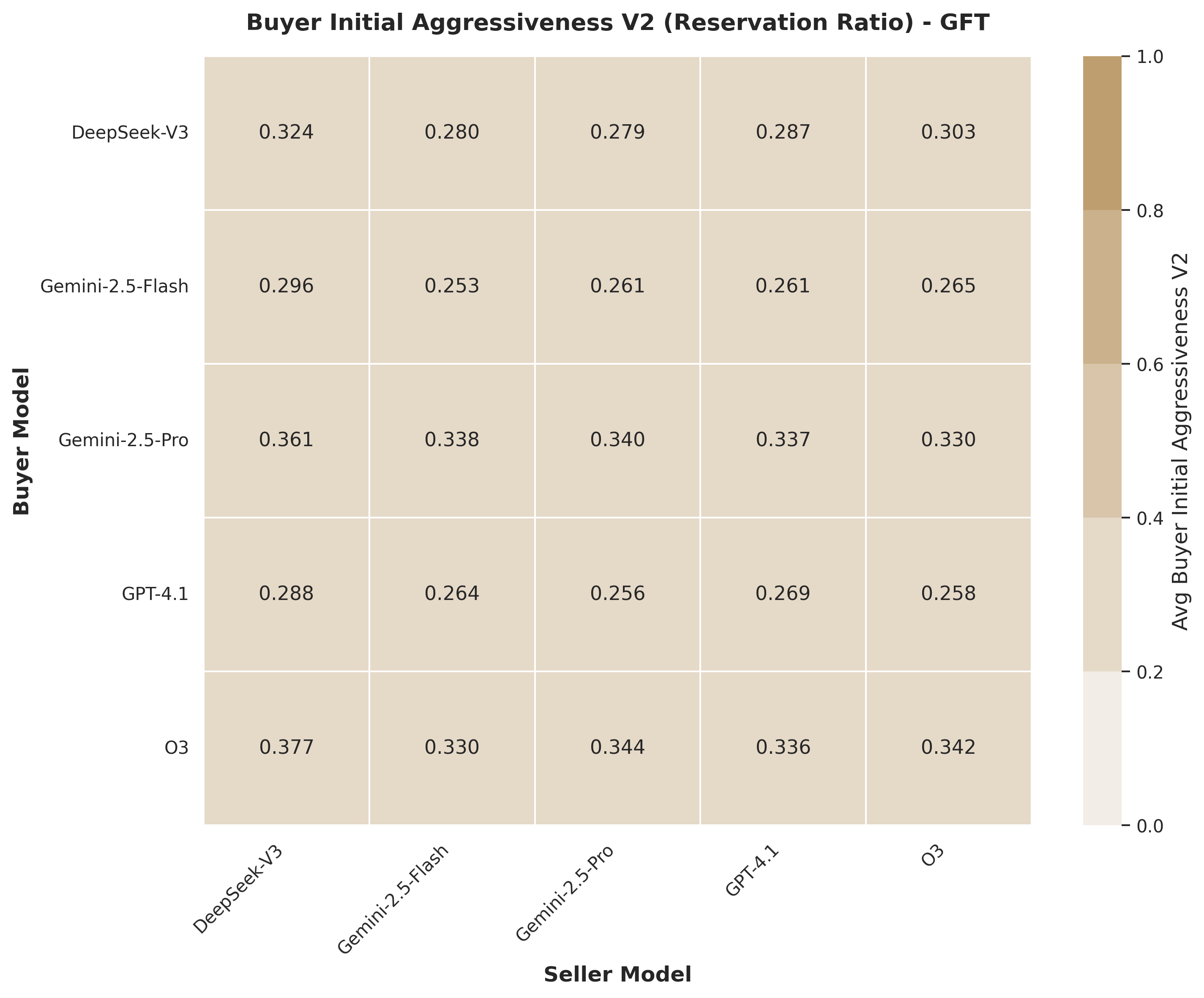}
\caption{Buyer initial aggressiveness. Left: V1 (gap closure). Right: V2
(reservation ratio). Lower values indicate more aggressive opening offers.}
\label{fig:buyer_aggressiveness}
\end{figure}

\subsubsection{Concession Rate}
\label{app:concession_rate}

\begin{figure}[H]
\centering
\includegraphics[width=0.5\textwidth]{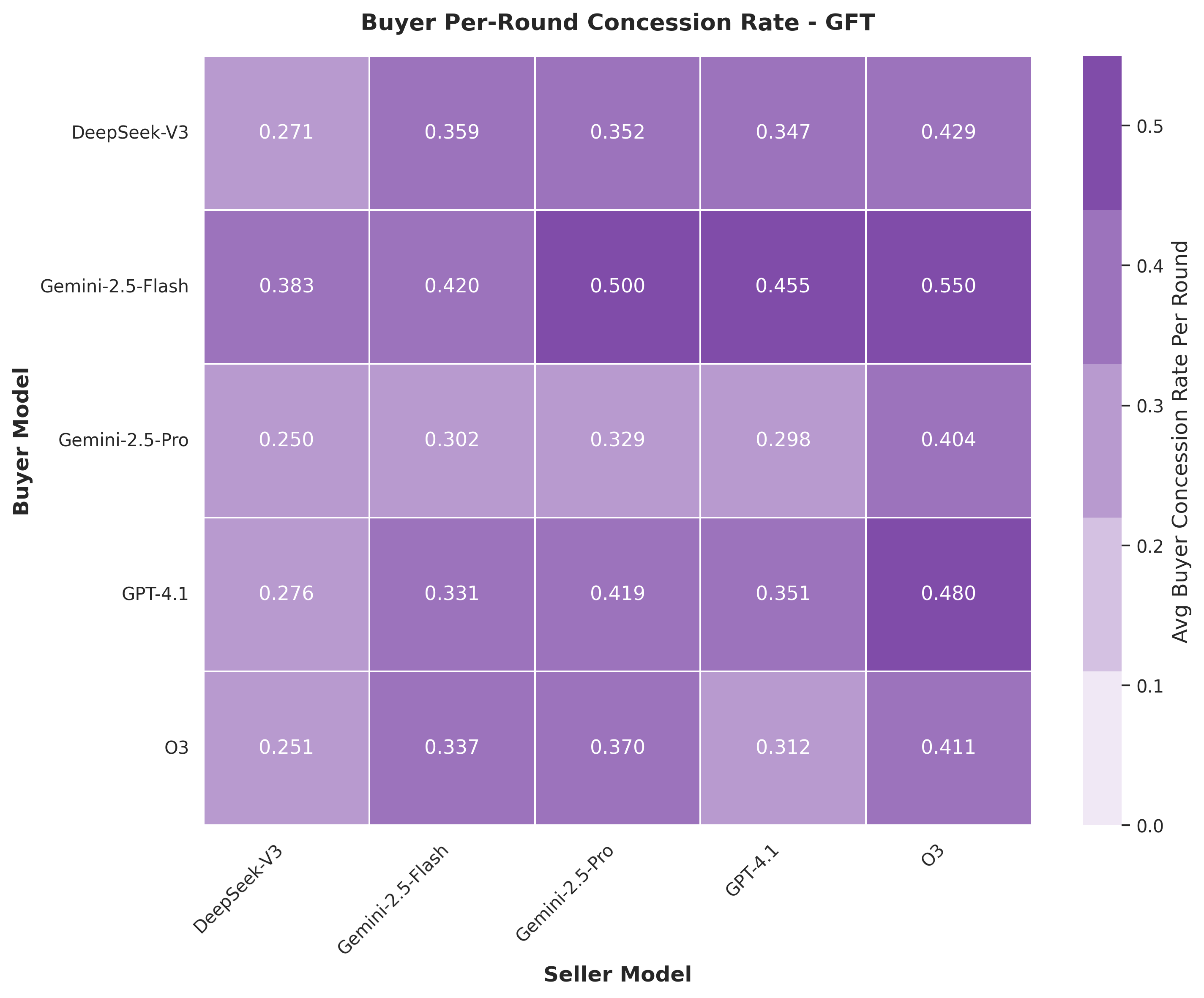}%
\hfill
\includegraphics[width=0.5\textwidth]{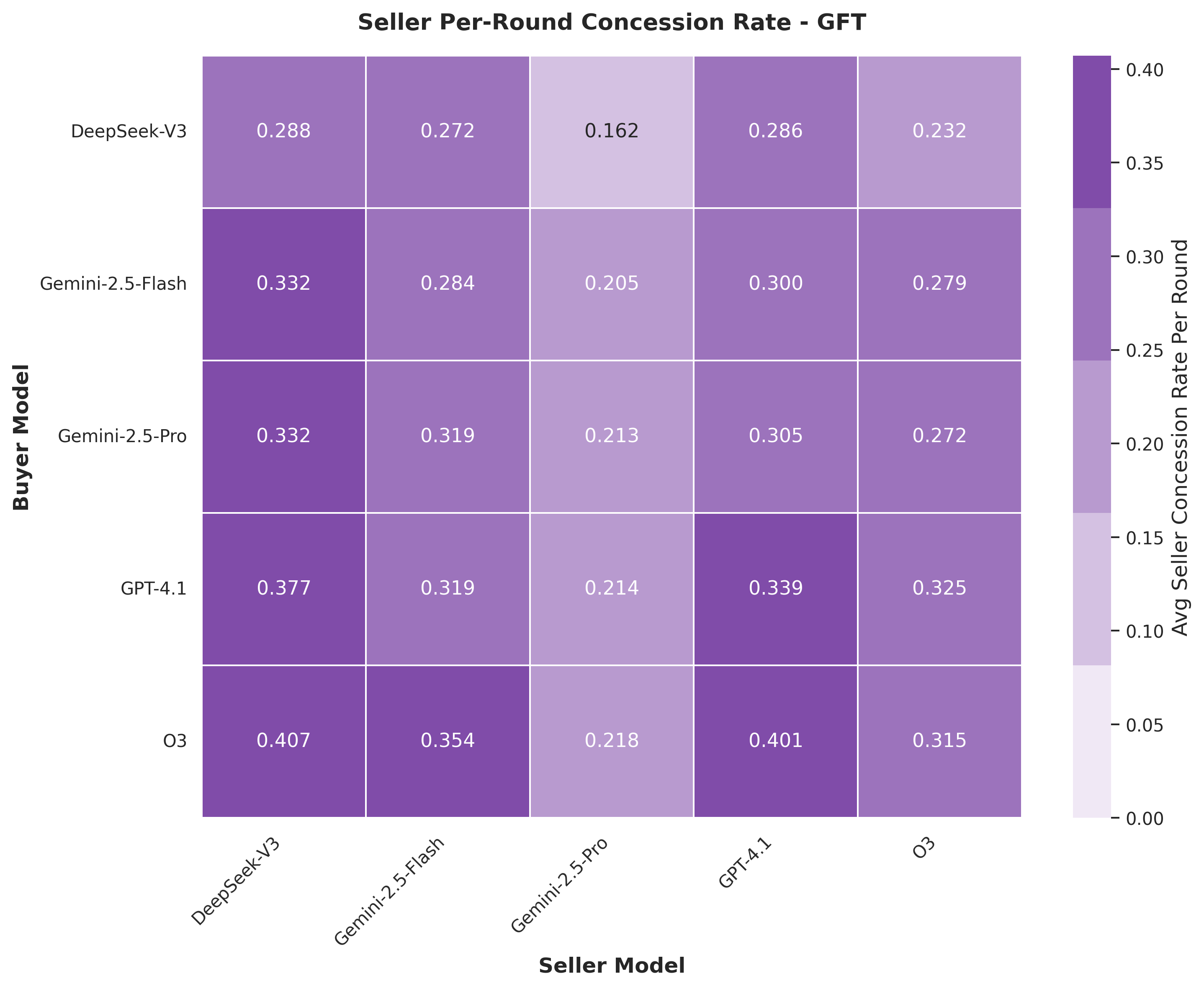}
\caption{Per-round concession rate. Left: buyer concession rate. Right: seller
concession rate. Higher values indicate faster concessions.}
\label{fig:concession_heatmaps}
\end{figure}

\subsubsection{Temporal Patience}
\label{app:temporal_patience}

\begin{figure}[H]
\centering
\includegraphics[width=0.5\textwidth]{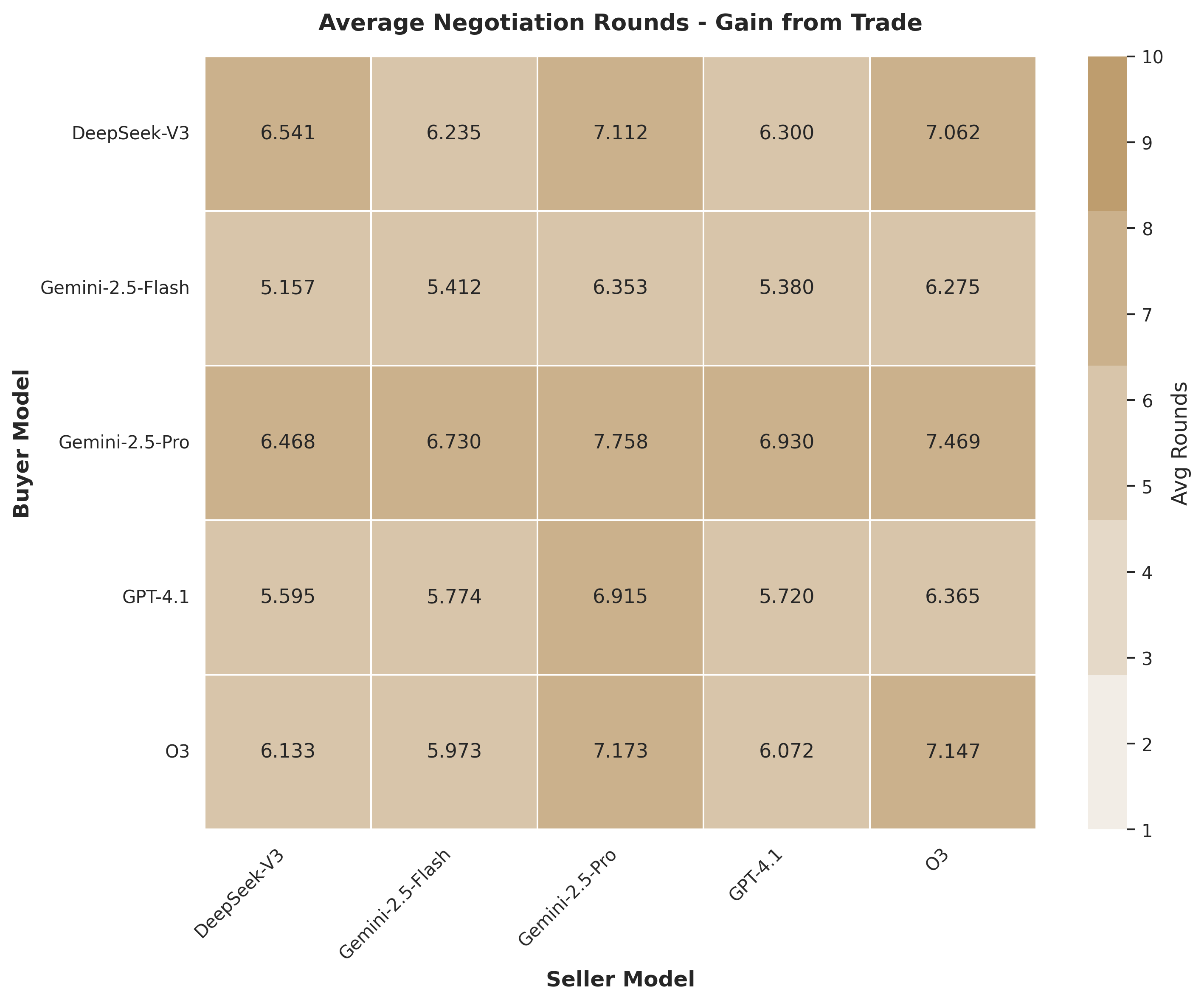}
\caption{Average negotiation rounds for all 25 buyer--seller pairings (GFT
scenarios). The protocol allows a maximum of 10 rounds.}
\label{fig:temporal_patience}
\end{figure}

\subsection{Price-Tier Decomposition}
\label{app:bracket_analysis}

This subsection provides price-tier decomposition results: aggregate tables
(\Cref{tab:frontier_bracket_surplus,tab:frontier_bracket_dealrate,%
tab:frontier_bracket_violation}) underlying \Cref{sec:benchmark_price_tier},
followed by per-opponent breakdowns. All tables report GFT scenarios with
reservation price quintiles (Q1\,=\,lowest, Q5\,=\,highest).

\subsubsection{Aggregate Bracket Tables}
\label{app:frontier_bracket_tables}

\begin{table}[H]
\centering
\small
\begin{tabular}{@{}l ccccc c ccccc c@{}}
\toprule
& \multicolumn{6}{c}{\textbf{Buyer Surplus by Buyer Value (\%)}} & \multicolumn{6}{c}{\textbf{Seller Surplus by Seller Cost (\%)}} \\
\cmidrule(lr){2-7} \cmidrule(lr){8-13}
\textbf{Model} & Q1 & Q2 & Q3 & Q4 & Q5 & Spread & Q1 & Q2 & Q3 & Q4 & Q5 & Spread \\
\midrule
o3             & 44.5 & 49.9 & 49.6 & 48.2 & 43.8 & 6.1 & 79.3 & 76.4 & 75.5 & 75.3 & 78.7 & 4.0 \\
Gem.-2.5-Pro   & 42.8 & 45.0 & 47.6 & 47.1 & 41.6 & 6.0 & 75.0 & 69.6 & 73.4 & 69.6 & 75.6 & 6.0 \\
GPT-4.1        & 26.1 & 34.1 & 34.5 & 38.6 & 36.9 & 12.5 & 59.7 & 55.9 & 56.3 & 58.2 & 58.5 & 3.8 \\
DeepSeek-V3    & 31.6 & 35.9 & 38.0 & 33.6 & 36.3 & 6.4 & 54.9 & 50.8 & 51.0 & 46.3 & 44.9 & 10.0 \\
Gem.-2.5-Flash & 16.7 & 23.0 & 25.6 & 26.8 & 29.5 & \textbf{12.8} & 59.6 & 56.5 & 54.8 & 55.9 & 61.4 & 6.6 \\
\bottomrule
\end{tabular}
\caption{Surplus share by reservation price quintile, averaged across all 5
opponents (GFT). Left panel: buyer surplus by buyer-value bracket. Right
panel: seller surplus by seller-cost bracket. Spread\,=\,max$-$min across
quintiles. Strong models (o3, Gem.-Pro) show 4--6\,pp spreads; weak buyers
(Gem.-Flash) show 13\,pp.}
\label{tab:frontier_bracket_surplus}
\end{table}

\begin{table}[H]
\centering
\small
\begin{tabular}{@{}l ccccc c ccccc c@{}}
\toprule
& \multicolumn{6}{c}{\textbf{Buyer Deal Rate by Buyer Value (\%)}} & \multicolumn{6}{c}{\textbf{Seller Deal Rate by Seller Cost (\%)}} \\
\cmidrule(lr){2-7} \cmidrule(lr){8-13}
\textbf{Model} & Q1 & Q2 & Q3 & Q4 & Q5 & Spread & Q1 & Q2 & Q3 & Q4 & Q5 & Spread \\
\midrule
o3             & 97.3 & 98.8 & 97.2 & 98.3 & 98.5 & 1.6 & 96.0 & 94.5 & 95.7 & 96.5 & 95.0 & 2.0 \\
Gem.-2.5-Pro   & 97.0 & 96.0 & 96.3 & 96.8 & 97.3 & 1.3 & 89.5 & 89.5 & 93.8 & 93.5 & 92.0 & 4.3 \\
GPT-4.1        & 94.0 & 91.3 & 94.2 & 94.5 & 95.8 & 4.5 & 92.3 & 92.5 & 97.0 & 97.0 & 97.0 & 4.7 \\
DeepSeek-V3    & 92.3 & 95.0 & 95.7 & 95.8 & 93.0 & 3.5 & 90.8 & 90.5 & 94.2 & 92.0 & 92.8 & 3.7 \\
Gem.-2.5-Flash & 87.5 & 79.8 & 87.8 & 87.5 & 89.5 & \textbf{9.7} & 92.0 & 94.3 & 95.0 & 96.8 & 96.8 & 4.8 \\
\bottomrule
\end{tabular}
\caption{Deal rates by reservation price quintile, averaged across all 5
opponents (GFT). Left panel: buyer deal rate by buyer-value bracket. Right
panel: seller deal rate by seller-cost bracket. Flash as buyer shows the
largest spread (9.7\,pp); all other models stay within 1--5\,pp.}
\label{tab:frontier_bracket_dealrate}
\end{table}

\begin{table}[H]
\centering
\small
\begin{tabular}{@{}l ccccc c ccccc c@{}}
\toprule
& \multicolumn{6}{c}{\textbf{As Buyer (\%)}} & \multicolumn{6}{c}{\textbf{As Seller (\%)}} \\
\cmidrule(lr){2-7} \cmidrule(lr){8-13}
\textbf{Model} & Q1 & Q2 & Q3 & Q4 & Q5 & Spread & Q1 & Q2 & Q3 & Q4 & Q5 & Spread \\
\midrule
o3             & 0.50 & 0.25 & 0.25 & 0.25 & 0.00 & 0.50 & 0.25 & 0.75 & 0.00 & 0.50 & 0.25 & 0.75 \\
Gem.-2.5-Pro   & 0.50 & 0.00 & 0.25 & 0.25 & 0.50 & 0.50 & 0.00 & 0.00 & 0.00 & 0.00 & 0.00 & \textbf{0.00} \\
GPT-4.1        & 0.50 & 0.00 & 0.50 & 0.75 & 0.25 & 0.75 & 0.50 & 0.25 & 0.25 & 0.25 & 0.25 & 0.25 \\
DeepSeek-V3    & 1.00 & 0.50 & 0.50 & 0.75 & 0.25 & 0.75 & 1.25 & 0.25 & 1.51 & 1.75 & 1.00 & 1.50 \\
Gem.-2.5-Flash & 1.00 & 0.50 & 0.00 & 0.75 & 0.00 & 1.00 & 0.25 & 0.50 & 0.00 & 0.25 & 0.00 & 0.50 \\
\bottomrule
\end{tabular}
\caption{Combined violation rates (either party) by reservation price
quintile, averaged across all 5 opponents (GFT). Left panel: by buyer-value
bracket (model as buyer). Right panel: by seller-cost bracket (model as
seller). All rates are below 2\% with no systematic bracket trend.
Gemini-2.5-Pro as seller: 0\% in every quintile.}
\label{tab:frontier_bracket_violation}
\end{table}

\subsubsection{Per-Opponent Buyer Surplus Share}
\label{app:bracket_surplus}

\begin{table}[H]
\centering
\small
\begin{tabular}{@{}l ccccc c@{}}
\toprule
\textbf{vs.\ Seller} & Q1 & Q2 & Q3 & Q4 & Q5 & Avg \\
\midrule
o3         & 26.8 & 37.1 & 35.0 & 34.5 & 28.0 & 32.3 \\
GPT-4.1    & 54.6 & 60.6 & 55.7 & 54.4 & 52.2 & 55.5 \\
Gem.-Pro   & 27.7 & 37.6 & 37.4 & 36.2 & 28.3 & 33.4 \\
Gem.-Flash & 59.7 & 52.0 & 55.3 & 52.5 & 45.4 & 53.0 \\
DeepSeek   & 53.7 & 62.2 & 64.5 & 63.6 & 65.2 & 61.8 \\
\bottomrule
\end{tabular}
\caption{o3 as buyer: surplus share (\%) by value quintile and opponent.}
\label{tab:bracket_o3_buyer}
\end{table}

\begin{table}[H]
\centering
\small
\begin{tabular}{@{}l ccccc c@{}}
\toprule
\textbf{vs.\ Seller} & Q1 & Q2 & Q3 & Q4 & Q5 & Avg \\
\midrule
o3         & 28.0 & 28.3 & 29.9 & 33.7 & 24.1 & 28.8 \\
GPT-4.1    & 53.6 & 52.3 & 49.2 & 49.6 & 44.8 & 49.9 \\
Gem.-Pro   & 30.4 & 38.1 & 42.0 & 40.8 & 31.6 & 36.6 \\
Gem.-Flash & 50.6 & 50.9 & 56.8 & 51.7 & 43.0 & 50.6 \\
DeepSeek   & 51.6 & 55.6 & 60.3 & 59.5 & 64.7 & 58.3 \\
\bottomrule
\end{tabular}
\caption{Gemini-2.5-Pro as buyer: surplus share (\%) by value quintile and opponent.}
\label{tab:bracket_pro_buyer}
\end{table}

\begin{table}[H]
\centering
\small
\begin{tabular}{@{}l ccccc c@{}}
\toprule
\textbf{vs.\ Seller} & Q1 & Q2 & Q3 & Q4 & Q5 & Avg \\
\midrule
o3         & 12.9 & 18.4 & 17.8 & 25.6 & 23.3 & 19.6 \\
GPT-4.1    & 32.1 & 38.3 & 42.2 & 39.8 & 42.2 & 38.9 \\
Gem.-Pro   & 12.8 & 23.7 & 25.3 & 28.8 & 26.3 & 23.4 \\
Gem.-Flash & 35.7 & 42.7 & 40.6 & 48.7 & 38.2 & 41.2 \\
DeepSeek   & 36.9 & 47.3 & 46.5 & 50.3 & 54.7 & 47.1 \\
\bottomrule
\end{tabular}
\caption{GPT-4.1 as buyer: surplus share (\%) by value quintile and opponent.}
\label{tab:bracket_gpt_buyer}
\end{table}

\begin{table}[H]
\centering
\small
\begin{tabular}{@{}l ccccc c@{}}
\toprule
\textbf{vs.\ Seller} & Q1 & Q2 & Q3 & Q4 & Q5 & Avg \\
\midrule
o3         & 19.6 & 20.8 & 26.1 & 23.5 & 22.7 & 22.5 \\
GPT-4.1    & 42.3 & 43.2 & 43.5 & 30.6 & 42.6 & 40.4 \\
Gem.-Pro   & 18.5 & 26.8 & 30.6 & 30.1 & 26.5 & 26.5 \\
Gem.-Flash & 33.8 & 41.0 & 40.6 & 36.1 & 34.9 & 37.3 \\
DeepSeek   & 43.6 & 47.5 & 49.1 & 47.5 & 54.9 & 48.5 \\
\bottomrule
\end{tabular}
\caption{DeepSeek-V3 as buyer: surplus share (\%) by value quintile and opponent.}
\label{tab:bracket_ds_buyer}
\end{table}

\begin{table}[H]
\centering
\small
\begin{tabular}{@{}l ccccc c@{}}
\toprule
\textbf{vs.\ Seller} & Q1 & Q2 & Q3 & Q4 & Q5 & Avg \\
\midrule
o3         &  6.0 &  9.9 & 10.7 & 14.0 & 17.5 & 11.6 \\
GPT-4.1    & 20.9 & 29.9 & 25.4 & 28.6 & 29.2 & 26.8 \\
Gem.-Pro   &  9.4 & 15.7 & 17.4 & 20.2 & 21.5 & 16.8 \\
Gem.-Flash & 28.1 & 26.3 & 35.8 & 28.4 & 30.3 & 29.8 \\
DeepSeek   & 19.0 & 33.3 & 38.9 & 42.7 & 48.8 & 36.5 \\
\bottomrule
\end{tabular}
\caption{Gemini-2.5-Flash as buyer: surplus share (\%) by value quintile and
opponent. Flash shows the strongest Q1$\to$Q5 increase, especially vs.\
DeepSeek (+29.8\,pp).}
\label{tab:bracket_flash_buyer}
\end{table}

\subsubsection{Per-Opponent Buyer Deal Rate}
\label{app:bracket_deal_rate}

\begin{table}[H]
\centering
\small
\begin{tabular}{@{}l ccccc c@{}}
\toprule
\textbf{vs.\ Seller} & Q1 & Q2 & Q3 & Q4 & Q5 & Avg \\
\midrule
o3         & 100.0 & 100.0 & 97.5 & 100.0 & 98.8 & 99.3 \\
GPT-4.1    & 98.8 & 98.8 & 97.5 & 98.8 & 97.5 & 98.3 \\
Gem.-Pro   & 96.3 & 100.0 & 100.0 & 98.8 & 100.0 & 99.0 \\
Gem.-Flash & 93.8 & 97.5 & 98.8 & 97.5 & 100.0 & 97.5 \\
DeepSeek   & 97.5 & 97.5 & 92.4 & 96.3 & 96.3 & 96.0 \\
\bottomrule
\end{tabular}
\caption{o3 as buyer: deal rate (\%) by value quintile and opponent.}
\label{tab:bracket_o3_buyer_dr}
\end{table}

\begin{table}[H]
\centering
\small
\begin{tabular}{@{}l ccccc c@{}}
\toprule
\textbf{vs.\ Seller} & Q1 & Q2 & Q3 & Q4 & Q5 & Avg \\
\midrule
o3         & 100.0 & 98.8 & 97.5 & 98.8 & 98.8 & 98.8 \\
GPT-4.1    & 98.8 & 97.5 & 98.8 & 100.0 & 97.5 & 98.5 \\
Gem.-Pro   & 95.0 & 98.8 & 98.8 & 96.3 & 93.8 & 96.5 \\
Gem.-Flash & 97.5 & 91.3 & 96.3 & 96.3 & 97.5 & 95.8 \\
DeepSeek   & 93.8 & 93.8 & 90.0 & 92.5 & 98.8 & 93.8 \\
\bottomrule
\end{tabular}
\caption{Gemini-2.5-Pro as buyer: deal rate (\%) by value quintile and opponent.}
\label{tab:bracket_pro_buyer_dr}
\end{table}

\begin{table}[H]
\centering
\small
\begin{tabular}{@{}l ccccc c@{}}
\toprule
\textbf{vs.\ Seller} & Q1 & Q2 & Q3 & Q4 & Q5 & Avg \\
\midrule
o3         & 98.8 & 91.3 & 96.3 & 97.5 & 100.0 & 96.8 \\
GPT-4.1    & 93.8 & 92.5 & 95.0 & 96.3 & 97.5 & 95.0 \\
Gem.-Pro   & 92.5 & 88.8 & 91.3 & 90.0 & 98.8 & 92.3 \\
Gem.-Flash & 96.3 & 96.3 & 94.9 & 96.3 & 93.8 & 95.5 \\
DeepSeek   & 88.8 & 87.5 & 93.8 & 92.5 & 88.8 & 90.3 \\
\bottomrule
\end{tabular}
\caption{GPT-4.1 as buyer: deal rate (\%) by value quintile and opponent.}
\label{tab:bracket_gpt_buyer_dr}
\end{table}

\begin{table}[H]
\centering
\small
\begin{tabular}{@{}l ccccc c@{}}
\toprule
\textbf{vs.\ Seller} & Q1 & Q2 & Q3 & Q4 & Q5 & Avg \\
\midrule
o3         & 90.0 & 91.3 & 95.0 & 95.0 & 86.3 & 91.5 \\
GPT-4.1    & 97.5 & 98.8 & 98.8 & 98.8 & 100.0 & 98.8 \\
Gem.-Pro   & 83.8 & 88.8 & 91.3 & 90.0 & 83.8 & 87.5 \\
Gem.-Flash & 93.8 & 97.5 & 95.0 & 98.8 & 100.0 & 97.0 \\
DeepSeek   & 96.3 & 98.8 & 98.7 & 96.3 & 95.0 & 97.0 \\
\bottomrule
\end{tabular}
\caption{DeepSeek-V3 as buyer: deal rate (\%) by value quintile and opponent.}
\label{tab:bracket_ds_buyer_dr}
\end{table}

\begin{table}[H]
\centering
\small
\begin{tabular}{@{}l ccccc c@{}}
\toprule
\textbf{vs.\ Seller} & Q1 & Q2 & Q3 & Q4 & Q5 & Avg \\
\midrule
o3         & 97.5 & 87.5 & 91.3 & 91.3 & 90.0 & 91.5 \\
GPT-4.1    & 78.8 & 78.8 & 87.5 & 87.5 & 93.8 & 85.3 \\
Gem.-Pro   & 88.8 & 70.0 & 82.5 & 82.5 & 91.3 & 83.0 \\
Gem.-Flash & 87.5 & 81.3 & 92.5 & 90.0 & 93.8 & 89.0 \\
DeepSeek   & 85.0 & 81.3 & 85.0 & 86.3 & 78.8 & 83.3 \\
\bottomrule
\end{tabular}
\caption{Gemini-2.5-Flash as buyer: deal rate (\%) by value quintile and
opponent. Flash shows the strongest bracket sensitivity, with Q1$\to$Q5
gains of up to 15\,pp (vs.\ GPT-4.1).}
\label{tab:bracket_flash_buyer_dr}
\end{table}

\subsubsection{Combined View vs.\ GPT-4.1}
\label{app:bracket_vs_gpt}

\begin{table}[H]
\centering
\small
\begin{tabular}{@{}l cccc cccc@{}}
\toprule
& \multicolumn{4}{c}{\textbf{Deal Rate (\%)}} & \multicolumn{4}{c}{\textbf{Buyer Surplus Share (\%)}} \\
\cmidrule(lr){2-5} \cmidrule(lr){6-9}
\textbf{Quintile} & o3 & Pro & DeepSeek & Flash & o3 & Pro & DeepSeek & Flash \\
\midrule
Q1 (lowest)  & 98.8 & 98.8 & 97.5 & 78.8 & 54.6 & 53.6 & 42.3 & 20.9 \\
Q2           & 98.8 & 97.5 & 98.8 & 78.8 & 60.6 & 52.3 & 43.2 & 29.9 \\
Q3           & 97.5 & 98.8 & 98.8 & 87.5 & 55.7 & 49.2 & 43.5 & 25.4 \\
Q4           & 98.8 & 100.0 & 98.8 & 87.5 & 54.4 & 49.6 & 30.6 & 28.6 \\
Q5 (highest) & 97.5 & 97.5 & 100.0 & 93.8 & 52.2 & 44.8 & 42.6 & 29.2 \\
\bottomrule
\end{tabular}
\caption{Per-model price-tier decomposition: buyer deal rate and buyer surplus
share vs.\ GPT-4.1 (GFT scenarios), partitioned by buyer value quintile.
Strong negotiators (o3, Pro) show flat performance across brackets; weaker
negotiators (Flash) show bracket-dependent deal rates.}
\label{tab:frontier_bracket}
\end{table}


\section{Training Results: Supplementary Tables}
\label{app:training_tables}

This appendix provides the full numerical tables underlying the figures in
\Cref{sec:training_results}.

\subsection{Violation Rates}
\label{app:training_viol_tables}

\begin{table}[H]
\centering
\small
\begin{tabular}{@{}lll cc cc cc@{}}
\toprule
& & & \multicolumn{2}{c}{\textbf{vs GPT-4.1 (\%)}} & \multicolumn{2}{c}{\textbf{vs o3-mini (\%)}} & \multicolumn{2}{c}{\textbf{Average (\%)}} \\
\cmidrule(lr){4-5} \cmidrule(lr){6-7} \cmidrule(lr){8-9}
\textbf{Model} & \textbf{Stage} & \textbf{Role} & Self & Opp. & Self & Opp. & Self & Opp. \\
\midrule
\multirow{6}{*}{Qwen3-8B}
  & \multirow{2}{*}{Base} & Buyer  & 0.50 & 0.00 & 1.00 & 0.00 & 0.75 & 0.00 \\
  &                       & Seller & \textbf{3.75} & 0.25 & 1.00 & 0.00 & \textbf{2.38} & 0.13 \\
  & \multirow{2}{*}{SFT}  & Buyer  & 0.00 & 0.50 & 1.25 & 0.75 & 0.63 & 0.63 \\
  &                       & Seller & \textbf{0.25} & 0.75 & 0.50 & 0.25 & \textbf{0.38} & 0.50 \\
  & \multirow{2}{*}{RL}   & Buyer  & 0.25 & 0.00 & 0.75 & 0.25 & 0.50 & 0.13 \\
  &                       & Seller & \textbf{3.25} & 0.50 & \textbf{2.50} & 0.00 & \textbf{2.88} & 0.25 \\
\midrule
\multirow{6}{*}{Qwen3-14B}
  & \multirow{2}{*}{Base} & Buyer  & 0.25 & 0.00 & 0.25 & 0.00 & 0.25 & 0.00 \\
  &                       & Seller & 1.00 & 0.50 & 0.25 & 0.25 & 0.63 & 0.38 \\
  & \multirow{2}{*}{SFT}  & Buyer  & 0.00 & 0.00 & 1.00 & 0.50 & 0.50 & 0.25 \\
  &                       & Seller & 0.00 & 0.50 & 0.50 & 0.00 & 0.25 & 0.25 \\
  & \multirow{2}{*}{RL}   & Buyer  & 0.25 & 0.00 & 0.25 & 0.00 & 0.25 & 0.00 \\
  &                       & Seller & 1.25 & 0.50 & 0.25 & 0.75 & 0.75 & 0.63 \\
\bottomrule
\end{tabular}
\caption{GFT violation rates across training stages. Self: the Qwen model
accepts a deal yielding negative utility. Opp.: the frontier counterpart
accepts such a deal when facing Qwen. GPT-4.1 is the RL training opponent
(in-distribution); o3-mini is unseen (out-of-distribution). Bold highlights
key self-violation values discussed in text.}
\label{tab:training_gft_viol}
\end{table}

\begin{table}[H]
\centering
\small
\begin{tabular}{@{}lll cc cc cc@{}}
\toprule
& & & \multicolumn{2}{c}{\textbf{vs GPT-4.1 (\%)}} & \multicolumn{2}{c}{\textbf{vs o3-mini (\%)}} & \multicolumn{2}{c}{\textbf{Average (\%)}} \\
\cmidrule(lr){4-5} \cmidrule(lr){6-7} \cmidrule(lr){8-9}
\textbf{Model} & \textbf{Stage} & \textbf{Role} & Self & Opp. & Self & Opp. & Self & Opp. \\
\midrule
\multirow{6}{*}{Qwen3-8B}
  & \multirow{2}{*}{Base} & Buyer  & 0.00 & 0.00 & 0.25 & 0.00 & 0.13 & 0.00 \\
  &                       & Seller & \textbf{4.00} & 0.00 & \textbf{4.00} & 0.00 & \textbf{4.00} & 0.00 \\
  & \multirow{2}{*}{SFT}  & Buyer  & 0.25 & 0.50 & 0.25 & 0.00 & 0.25 & 0.25 \\
  &                       & Seller & \textbf{0.50} & 0.00 & \textbf{0.25} & 0.00 & \textbf{0.38} & 0.00 \\
  & \multirow{2}{*}{RL}   & Buyer  & 0.00 & 0.00 & 0.00 & 0.00 & 0.00 & 0.00 \\
  &                       & Seller & \textbf{4.00} & 0.25 & \textbf{6.50} & 0.00 & \textbf{5.25} & 0.13 \\
\midrule
\multirow{6}{*}{Qwen3-14B}
  & \multirow{2}{*}{Base} & Buyer  & --- & --- & --- & --- & --- & --- \\
  &                       & Seller & 0.00 & 0.25 & 0.25 & 0.00 & 0.13 & 0.13 \\
  & \multirow{2}{*}{SFT}  & Buyer  & 0.00 & 0.00 & 0.00 & 0.00 & 0.00 & 0.00 \\
  &                       & Seller & 0.25 & 0.00 & 0.25 & 0.25 & 0.25 & 0.13 \\
  & \multirow{2}{*}{RL}   & Buyer  & 0.50 & 0.00 & 0.00 & 0.00 & 0.25 & 0.00 \\
  &                       & Seller & 0.25 & 0.00 & 0.00 & 0.00 & 0.13 & 0.00 \\
\bottomrule
\end{tabular}
\caption{NGFT violation rates across training stages. Self: the Qwen model
strikes a deal in a no-gains-from-trade scenario (irrational for at least
one party). Opp.: the frontier counterpart does so when facing Qwen.
Bold highlights key self-violation values discussed in text. Dash indicates
missing data.}
\label{tab:training_ngft_viol}
\end{table}

\subsection{Strategic Effectiveness and Allocative Efficiency}
\label{app:training_surplus_table}

\begin{table}[H]
\centering
\small
\begin{tabular}{@{}lll cc cc@{}}
\toprule
& & & \multicolumn{2}{c}{\textbf{Deal Rate (\%)}} & \multicolumn{2}{c}{\textbf{Surplus Share (\%)}} \\
\cmidrule(lr){4-5} \cmidrule(lr){6-7}
\textbf{Model} & \textbf{Stage} & \textbf{Role} & vs 4.1 & vs o3 & vs 4.1 & vs o3 \\
\midrule
\multirow{6}{*}{Qwen3-8B}
  & \multirow{2}{*}{Base} & Buyer  & 70.00 & 66.75 & 22.41 & 11.16 \\
  &                       & Seller & 80.00 & 70.25 & 32.53 & 24.80 \\
  & \multirow{2}{*}{SFT}  & Buyer  & 86.25 & 79.25 & 40.77 & 36.12 \\
  &                       & Seller & 70.25 & 56.50 & 60.56 & 64.79 \\
  & \multirow{2}{*}{RL}   & Buyer  & 69.75 & 69.50 & 20.39 & 11.60 \\
  &                       & Seller & 78.25 & 68.50 & 31.69 & 26.06 \\
\midrule
\multirow{6}{*}{Qwen3-14B}
  & \multirow{2}{*}{Base} & Buyer  & 97.25 & 95.25 & 23.31 & 12.96 \\
  &                       & Seller & 97.25 & 94.50 & 39.41 & 33.22 \\
  & \multirow{2}{*}{SFT}  & Buyer  & 71.75 & 64.25 & 44.77 & 36.83 \\
  &                       & Seller & 49.25 & 41.00 & 58.04 & 60.99 \\
  & \multirow{2}{*}{RL}   & Buyer  & 95.50 & 93.50 & 21.97 & 13.01 \\
  &                       & Seller & 99.00 & 95.50 & 40.18 & 37.77 \\
\bottomrule
\end{tabular}
\caption{Strategic effectiveness and allocative efficiency across training stages
(GFT scenarios). Deal rate: fraction of negotiations reaching agreement. Surplus
share: fraction of ZOPA captured conditional on agreement. GPT-4.1 is the RL
training opponent (in-distribution); o3-mini is unseen (out-of-distribution).}
\label{tab:training_surplus}
\end{table}

\subsection{Price-Tier Decomposition Tables}
\label{app:training_bracket_tables}

\begin{table}[H]
\centering
\scriptsize
\begin{tabular}{@{}ll cccccc cccccc@{}}
\toprule
& & \multicolumn{6}{c}{\textbf{Buyer Violation by Buyer Value (\%)}}
& \multicolumn{6}{c}{\textbf{Seller Violation by Seller Cost (\%)}} \\
\cmidrule(lr){3-8} \cmidrule(lr){9-14}
\textbf{Model--Stage} & \textbf{Opp.} & Q1 & Q2 & Q3 & Q4 & Q5 & Spread
                      & Q1 & Q2 & Q3 & Q4 & Q5 & Spread \\
\midrule
14B-Base & 4.1   & 0.00 & 0.00 & 1.25 & 0.00 & 0.00 & 1.25
                  & 2.50 & 0.00 & 2.50 & 1.25 & 1.25 & 2.50 \\
         & o3    & 0.00 & 0.00 & 1.25 & 0.00 & 0.00 & 1.25
                  & 1.25 & 0.00 & 1.25 & 0.00 & 0.00 & 1.25 \\
14B-SFT  & 4.1   & 0.00 & 0.00 & 0.00 & 0.00 & 0.00 & 0.00
                  & 0.00 & 1.25 & 0.00 & 0.00 & 1.25 & 1.25 \\
         & o3    & 1.25 & 1.25 & 3.75 & 0.00 & 1.25 & 3.75
                  & 1.25 & 1.25 & 0.00 & 0.00 & 0.00 & 1.25 \\
14B-RL   & 4.1   & 0.00 & 0.00 & 0.00 & 1.25 & 0.00 & 1.25
                  & 2.50 & 1.25 & 1.25 & 3.75 & 0.00 & 3.75 \\
         & o3    & 1.25 & 0.00 & 0.00 & 0.00 & 0.00 & 1.25
                  & 0.00 & 0.00 & 2.50 & 2.50 & 0.00 & 2.50 \\
\midrule
8B-Base  & 4.1   & 1.25 & 0.00 & 1.25 & 0.00 & 0.00 & 1.25
                  & 2.50 & 2.50 & \textbf{7.50} & \textbf{7.50} & 0.00 & \textbf{7.50} \\
         & o3    & 1.25 & 1.25 & 1.25 & 1.25 & 0.00 & 1.25
                  & 0.00 & 1.25 & 1.25 & 0.00 & 2.50 & 2.50 \\
8B-SFT   & 4.1   & 0.00 & 0.00 & 0.00 & 0.00 & 2.50 & 2.50
                  & 1.25 & 1.25 & 0.00 & 1.25 & 1.25 & 1.25 \\
         & o3    & 2.50 & 0.00 & 1.25 & 5.00 & 1.25 & 5.00
                  & 0.00 & 1.25 & 0.00 & 3.75 & 2.50 & 3.75 \\
8B-RL    & 4.1   & 1.25 & 0.00 & 0.00 & 0.00 & 0.00 & 1.25
                  & 2.50 & 1.25 & \textbf{5.00} & \textbf{6.25} & 3.75 & \textbf{5.00} \\
         & o3    & 2.50 & 0.00 & 1.25 & 0.00 & 1.25 & 2.50
                  & 0.00 & 3.75 & 2.50 & 3.75 & 2.50 & 3.75 \\
\bottomrule
\end{tabular}
\caption{Combined violation rate by reservation-price quintile and training
stage, split by opponent. Bold highlights 8B seller violations exceeding 5\%.}
\label{tab:training_bracket_violation}
\end{table}

\begin{table}[H]
\centering
\scriptsize
\begin{tabular}{@{}ll cccccc cccccc@{}}
\toprule
& & \multicolumn{6}{c}{\textbf{Buyer Surplus by Buyer Value (\%)}}
& \multicolumn{6}{c}{\textbf{Seller Surplus by Seller Cost (\%)}} \\
\cmidrule(lr){3-8} \cmidrule(lr){9-14}
\textbf{Model--Stage} & \textbf{Opp.} & Q1 & Q2 & Q3 & Q4 & Q5 & Spread
                      & Q1 & Q2 & Q3 & Q4 & Q5 & Spread \\
\midrule
14B-Base & 4.1   & 18.9 & 24.3 & 18.7 & 20.0 & 34.1 & 15.4
                  & 42.8 & 38.4 & 39.0 & 37.0 & 40.0 & 5.8 \\
         & o3    & 4.7 & 7.9 & 12.4 & 14.0 & 26.5 & 21.8
                  & 35.0 & 32.1 & 31.3 & 37.5 & 30.2 & 7.3 \\
14B-SFT  & 4.1   & \textbf{49.3} & \textbf{48.9} & \textbf{40.7} & \textbf{43.3} & \textbf{42.5} & \textbf{8.6}
                  & \textbf{64.2} & \textbf{51.4} & \textbf{58.1} & \textbf{63.1} & \textbf{56.0} & 12.8 \\
         & o3    & \textbf{30.7} & \textbf{37.7} & \textbf{38.7} & \textbf{37.7} & \textbf{39.3} & \textbf{8.6}
                  & \textbf{60.1} & \textbf{66.4} & \textbf{59.4} & \textbf{56.8} & \textbf{61.6} & 9.6 \\
14B-RL   & 4.1   & 14.0 & 20.3 & 18.1 & 22.9 & 34.7 & 20.7
                  & 42.2 & 37.4 & 44.0 & 33.8 & 43.4 & 10.2 \\
         & o3    & 2.9 & 9.4 & 11.1 & 15.0 & 25.3 & 22.4
                  & 36.5 & 42.3 & 35.8 & 36.8 & 37.5 & 6.5 \\
\midrule
8B-Base  & 4.1   & 12.0 & 22.7 & 18.7 & 21.6 & 36.6 & 24.6
                  & 32.9 & 31.7 & 32.1 & 35.2 & 31.1 & 4.1 \\
         & o3    & 1.8 & 4.3 & 9.0 & 11.7 & 26.3 & 24.5
                  & 22.2 & 30.3 & 19.8 & 26.2 & 25.0 & 10.5 \\
8B-SFT   & 4.1   & \textbf{37.9} & \textbf{44.5} & \textbf{43.5} & \textbf{36.9} & \textbf{41.0} & \textbf{7.6}
                  & \textbf{64.5} & \textbf{56.2} & \textbf{58.7} & \textbf{57.4} & \textbf{66.1} & 9.9 \\
         & o3    & \textbf{31.5} & \textbf{34.2} & \textbf{34.7} & \textbf{39.4} & \textbf{40.3} & \textbf{8.8}
                  & \textbf{65.4} & \textbf{63.2} & \textbf{66.1} & \textbf{68.7} & \textbf{61.3} & 7.4 \\
8B-RL    & 4.1   & 11.8 & 22.7 & 16.7 & 20.9 & 28.9 & 17.1
                  & 32.2 & 28.8 & 35.3 & 31.1 & 31.5 & 6.5 \\
         & o3    & 6.9 & 5.7 & 6.5 & 10.8 & 28.3 & 22.6
                  & 23.9 & 29.4 & 20.0 & 32.8 & 25.4 & 12.8 \\
\bottomrule
\end{tabular}
\caption{Surplus share by reservation-price quintile and training stage,
split by opponent. SFT compresses buyer-side spreads to 7--9pp; Base and RL
show 15--25pp Q1$\to$Q5 gradients.}
\label{tab:training_bracket_surplus}
\end{table}

\begin{table}[H]
\centering
\scriptsize
\begin{tabular}{@{}ll cccccc cccccc@{}}
\toprule
& & \multicolumn{6}{c}{\textbf{Buyer Deal Rate by Buyer Value (\%)}}
& \multicolumn{6}{c}{\textbf{Seller Deal Rate by Seller Cost (\%)}} \\
\cmidrule(lr){3-8} \cmidrule(lr){9-14}
\textbf{Model--Stage} & \textbf{Opp.} & Q1 & Q2 & Q3 & Q4 & Q5 & Spread
                      & Q1 & Q2 & Q3 & Q4 & Q5 & Spread \\
\midrule
14B-Base & 4.1   & 97.5 & 97.5 & 96.3 & 95.0 & 100.0 & 5.0
                  & 96.3 & 97.5 & 97.5 & 98.8 & 96.3 & 2.5 \\
         & o3    & 96.3 & 97.5 & 96.3 & 95.0 & 91.3 & 6.2
                  & 93.8 & 93.8 & 95.0 & 93.8 & 96.3 & 2.5 \\
14B-SFT  & 4.1   & 73.8 & 62.5 & 77.5 & 71.3 & 73.8 & 15.0
                  & 48.8 & 56.3 & 45.0 & 35.0 & 61.3 & 26.3 \\
         & o3    & 62.5 & 72.5 & 55.0 & 62.5 & 68.8 & 17.5
                  & 36.3 & 42.5 & 45.0 & 36.3 & 45.0 & 8.7 \\
14B-RL   & 4.1   & 96.3 & 96.3 & 96.3 & 93.8 & 95.0 & 2.5
                  & 100.0 & 98.8 & 100.0 & 97.5 & 98.8 & 2.5 \\
         & o3    & 87.5 & 92.5 & 96.3 & 95.0 & 96.3 & 8.8
                  & 96.3 & 96.3 & 93.8 & 96.3 & 95.0 & 2.5 \\
\midrule
8B-Base  & 4.1   & 68.8 & 71.3 & 73.8 & 67.5 & 68.8 & 6.3
                  & 77.5 & 82.5 & 77.5 & 77.5 & 85.0 & 7.5 \\
         & o3    & 66.3 & 55.0 & 67.5 & 72.5 & 72.5 & 17.5
                  & 68.8 & 75.0 & 68.8 & 66.3 & 72.5 & 8.7 \\
8B-SFT   & 4.1   & 83.8 & 88.8 & 88.8 & 90.0 & 80.0 & 10.0
                  & 70.0 & 75.0 & 67.5 & 67.5 & 71.3 & 7.5 \\
         & o3    & 73.8 & 75.0 & 82.5 & 82.5 & 82.5 & 8.7
                  & 57.5 & 58.8 & 60.0 & 46.3 & 60.0 & 13.7 \\
8B-RL    & 4.1   & 68.8 & 67.5 & 72.5 & 63.8 & 76.3 & 12.5
                  & 76.3 & 78.8 & 73.8 & 80.0 & 82.5 & 8.7 \\
         & o3    & 66.3 & 68.8 & 71.3 & 73.8 & 67.5 & 7.5
                  & 73.8 & 70.0 & 68.8 & 60.0 & 70.0 & 13.8 \\
\bottomrule
\end{tabular}
\caption{Deal rate by reservation-price quintile and training stage,
split by opponent. 14B-SFT as seller shows the largest bracket spread
(26.3pp vs.\ GPT-4.1).}
\label{tab:training_bracket_dealrate}
\end{table}